\documentclass[apj]{emulateapj}
\usepackage{amsmath,graphicx,epsfig,multirow,natbib,rotating,tabularx,eqnarray}
\usepackage{hyperref}
\usepackage{cleveref}
\usepackage[usenames]{color}
\shorttitle{SuGOHI II: Lens Environments and Lines of Sight to $z \sim 0.8$}
\shortauthors{K. C. WONG et al.}

\begin{document}

\newcommand{\comment}[1]{\textcolor{red}{\textbf{#1}}}
\newcommand{\nnew}{41}
\newcommand{\nmain}{87}
\newcommand{\ncont}{100}
\newcommand{\zlavg}{0.514}

\title{Survey of Gravitationally-lensed Objects in HSC Imaging (SuGOHI). II. Environments and Line-of-Sight Structure of Strong Gravitational Lens Galaxies to $z \sim 0.8$}
\author{
Kenneth C. Wong\altaffilmark{1,14},
Alessandro Sonnenfeld\altaffilmark{2},
James H. H. Chan\altaffilmark{3,4,5},
Cristian E. Rusu\altaffilmark{6,15},
Masayuki Tanaka\altaffilmark{1},
Anton T. Jaelani\altaffilmark{7},
Chien-Hsiu Lee\altaffilmark{6},
Anupreeta More\altaffilmark{2,8},
Masamune Oguri\altaffilmark{2,9,10},
Sherry H. Suyu\altaffilmark{11,5,12},
and Yutaka Komiyama\altaffilmark{1,13}}
\altaffiltext{1}{National Astronomical Observatory of Japan, 2-21-1 Osawa, Mitaka, Tokyo 181-8588, Japan}
\altaffiltext{2}{Kavli IPMU (WPI), UTIAS, The University of Tokyo, Kashiwa, Chiba 277-8583, Japan}
\altaffiltext{3}{Laboratoire d'Astrophysique, Ecole Polytechnique F{\'e}d{\'e}rale de Lausanne (EPFL), Observatoire de Sauverny, CH-1290 Versoix, Switzerland}
\altaffiltext{4}{Department of Physics, National Taiwan University, Taipei 10617, Taiwan}
\altaffiltext{5}{Institute of Astronomy and Astrophysics, Academia Sinica, 11F of ASMAB, No.1, Section 4, Roosevelt Road, Taipei 10617, Taiwan}
\altaffiltext{6}{Subaru Telescope, National Astronomical Observatory of Japan, 650 N Aohoku Pl., Hilo, HI 96720, USA}
\altaffiltext{7}{Graduate School of Science, Tohoku University, 980-8578, 6-3 Aramaki Aoba, Aoba-ku, Sendai, Miyagi, Japan}
\altaffiltext{8}{Inter-University Centre for Astronomy and Astrophysics, Post Bag 4, Ganeshkhind, Pune 411 007, India}
\altaffiltext{9}{Department of Physics, The University of Tokyo, 7-3-1 Hongo, Bunkyo-ku, Tokyo 113-0033, Japan}
\altaffiltext{10}{Research Center for the Early Universe, University of Tokyo, Tokyo 113-0033, Japan}
\altaffiltext{11}{Max-Planck-Institut f{\"u}r Astrophysik, Karl-Schwarzschild-Str.~1, 85748 Garching, Germany}
\altaffiltext{12}{Physik-Department, Technische Universit\"at M\"unchen, James-Franck-Stra\ss{}e~1, 85748 Garching, Germany}
\altaffiltext{13}{Department of Astronomy, School of Science, Graduate University for Advanced Studies (SOKENDAI), 2-21-1 Osawa, Mitaka, Tokyo 181-8588, Japan}
\altaffiltext{14}{EACOA Fellow}
\altaffiltext{15}{Subaru Fellow}

\begin{abstract}
We investigate the local and line-of-sight overdensities of strong gravitational lens galaxies using wide-area multiband imaging from the Hyper Suprime-Cam Subaru Strategic Program.  We present \nnew~new definite or probable lens candidates discovered in Data Release 2 of the survey.  Using a combined sample of \nmain~galaxy-scale lenses out to a lens redshift of $z_{\mathrm{L}} \sim 0.8$, we compare galaxy number counts in lines of sight toward known and newly-discovered lenses in the survey to those of a control sample consisting of random lines of sight.  We also compare the local overdensity of lens galaxies to a sample of ``twin" galaxies that have a similar redshift and velocity dispersion to test whether lenses lie in different environments from similar non-lens galaxies.  We find that lens fields contain higher number counts of galaxies compared to the control fields, but this effect arises from the local environment of the lens.  Once galaxies in the lens plane are removed, the lens lines of sight are consistent with the control sample.  The local environments of the lenses are overdense compared to the control sample, and are slightly overdense compared to those of the twin sample, although the significance is marginal.  There is no significant evidence of the evolution of the local overdensity of lens environments with redshift.
\end{abstract}


\section{Introduction} \label{sec:intro}
Strong gravitational lensing is a useful tool for studying the distribution of matter in the universe.  The lensing effect arises from the total mass distribution along the line of sight (LOS) -- including both baryonic and dark matter -- and is generally dominated by the primary lensing mass.  Thus, lensing is a unique probe of the mass structure of lens galaxies \citep[e.g.,][]{treu+2006,koopmans+2009,auger+2010,sonnenfeld+2013b}.  Strongly-lensed quasars can also be used to probe the Hubble constant ($H_{0}$) through a measurement of their time delays \citep{refsdal1964}, independent of and complementary to other probes such as the cosmological distance ladder \citep[e.g.,][]{freedman+2012,riess+2016}.  The latest constraints from time-delay lenses \citep{suyu+2017,wong+2017,bonvin+2017} show a higher $H_{0}$ value than {\it Planck} CMB measurements, highlighting the importance of this independent probe.

However, lenses do not exist in isolation, but are embedded within the large-scale structure of the Universe, including both the local lens environment and uncorrelated structure in projection.  The lensing signal is perturbed by the integrated effect of the LOS mass distribution from the observer back to the source redshift, $z_{\mathrm{S}}$.  Most of these perturbations are small and can be either ignored or approximated as a tidal perturbation \citep[c.f.,][]{mccully+2017}, but the presence of significant perturbers such as groups and clusters \citep[e.g.,][]{momcheva+2006,wong+2011,wilson+2016,wilson+2017,sluse+2017} or the combined effect of multiple perturbers can lead to biases in lens modeling if unaccounted for.

If strong lenses are randomly distributed on the sky, these effects should average out over a sufficiently large sample. On the other hand, these LOS structures could potentially boost the lensing efficiency and cause lenses to lie along biased lines of sight \citep[e.g.,][]{oguri+2005}.  In addition, the mass surface density of perturbers along the LOS, which we refer to as external convergence ($\kappa_{\mathrm{ext}}$), is unconstrained from lens modeling due to the mass-sheet degeneracy \citep[e.g.,][]{falco+1985,gorenstein+1988,saha2000,schneider+2013,xu+2016}.  $\kappa_{\mathrm{ext}}$ does not average out, but instead leads to a bias on the inferred $H_{0}$ from time-delay lensed quasars if unaccounted for \citep[e.g.,][]{collett+2013,greene+2013,mccully+2014,mccully+2017}.

In order to use strong lens galaxies to infer the generic properties of typical galaxies, it is necessary to ascertain whether lenses are in unbiased lines of sight and have environments that are representative of the underlying galaxy population.  The Sloan Lens ACS Survey \citep[SLACS;][]{bolton+2006} studied over 100 galaxy-scale strong lenses and found that their structural properties are consistent with that of similar early-type non-lens galaxies \citep{bolton+2008} and reside in similar local environments \citep{treu+2009}.  However, the SLACS sample is primarily at low redshift ($z_{\mathrm{L}} \sim 0.2$) and is selected through spectroscopic identification of background sources.  It is not known whether this holds true for lenses selected by broadband imaging at higher redshifts, where the clustering and bias properties of galaxies may differ from those selected by SLACS.

Using {\it Hubble Space Telescope} ({\it HST}) imaging, \citet{fassnacht+2011} studied the lines of sight toward 20 known strong lens galaxies, using galaxy number counts to evaluate the LOS overdensity relative to a control sample of randomly chosen lines of sight.  This study found that the LOS toward lenses tends to be overdense, but much of this effect arises from the local environment of the lens galaxies, which are biased toward massive elliptical galaxies and are known to cluster \citep[e.g.,][]{dressler1980}.  Once the local lens environment is corrected for, the relative densities of the lens LOS are not significantly different from the random sample, which is consistent with results from cosmological simulations \citep{hilbert+2007}.  However, this study had several limitations, including the small sample size of lenses and the relatively small field of view of the {\it HST} Advanced Camera for Surveys, which only allowed for a maximum aperture radius of 45\arcsec~over which to evaluate the LOS.  In addition, there were uncertainties arising from cosmic variance due to a small control sample of truly random fields, as a larger contiguous control field was itself found to be biased.  Furthermore, this study lacked galaxy redshifts to distinguish the contribution of the local lens environment from projected structure along the LOS, instead relying on a statistical correction based on galaxy clustering results.

Using deep multiband imaging data from the ongoing Hyper Suprime Cam Subaru Strategic Program \citep[HSC SSP;][]{aihara+2018a}, we have recently discovered dozens of new gravitational lenses as part of the Survey of Gravitationally-lensed Objects in HSC Imaging \citep[SuGOHI;][]{sonnenfeld+2018} project.  With the sample of SuGOHI galaxy-scale lenses (hereafter ``SuGOHI-g"), we use the high-quality imaging from the HSC SSP to overcome the technical limitations of the \citet{fassnacht+2011} analysis and study lines of sight toward strong lenses in much more detail than has been possible in the past. The deep multiband imaging provides accurate photometric redshifts \citep{tanaka+2018} for distinguishing galaxies in the local lens environment from those projected along the LOS.  In addition, the SuGOHI-g sample includes lenses up to $z \sim 0.8$, allowing us to study lens environments to higher redshifts than was previously possible with the SLACS sample, and to compare them to non-lens galaxies of similar redshifts and masses.  Finally, detailed analyses of individual SuGOHI-g lenses may require ancillary data, and it will be useful to consider which lenses are likely to be impacted by overdense environments or lines of sight when deciding which systems to prioritize for such follow-up observations.

This paper is organized as follows.  We describe the HSC SSP imaging data and BOSS spectroscopic data used in this study in Section~\ref{sec:data}.  In Section~\ref{sec:dr2_lenses}, we present newly-discovered lens candidates in Data Release 2 of the HSC SSP.  We describe our sample selection of lenses for this study, as well as our selection of twin galaxies and control fields in Section~\ref{sec:sample}.  In Section~\ref{sec:los}, we describe our methodology for calculating the density of the local environment and the LOS.  We present our main results in Section~\ref{sec:results} and summarize our conclusions in Section~\ref{sec:conclusion}.  Throughout this paper, we assume $\Omega_{m} = 0.3$, $\Omega_{\Lambda} = 0.7$, and $h = 0.7$.  All magnitudes given are on the AB system.

\section{Data} \label{sec:data}

\subsection{Hyper Suprime-Cam Imaging Data} \label{subsec:hsc}
The HSC SSP \citep{aihara+2018a} is an ongoing imaging survey with the Hyper Suprime-Cam \citep[HSC;][]{miyazaki+2012,miyazaki+2018,furusawa+2018,kawanomoto+2018,komiyama+2018} on the Subaru Telescope.  The Wide component of the HSC SSP will observe a $\sim1400$ deg$^{2}$ area in the $grizy$ bands to a depth of $i = 26.2$.  The data used in this study are taken from Data Release 2 of the HSC SSP, which comprises data taken through the S17A semester and covers 776 deg$^{2}$ in all bands, including 289 deg$^{2}$ to the full depth.  The median $i$-band seeing of the data is $\sim 0\farcs6$.  The data are reduced with {\tt hscPipe} version 5.4.0 \citep{bosch+2018}, which is based on the Large Synoptic Survey Telescope (LSST) pipeline \citep{axelrod+2010,juric+2015}.

The photometric redshifts used in this analysis are determined using the {\sc mizuki} algorithm \citep{tanaka2015}.  A description of its application to the HSC SSP data is presented in \citet{tanaka+2018}.  The robustness of the photometric redshifts is a function of galaxy redshift and brightness, and are quantified in terms of $\Delta z / (1+z_{\mathrm{ref}})$, where $\Delta z \equiv |z - z_{\mathrm{ref}}|$ and $z_{\mathrm{ref}}$ is a reference redshift.  The galaxies that we use to quantify the environment and LOS effects are brighter than $i = 24$.  Roughly 97\% of these galaxies have a photometric redshift of $z \leq 2$, which is approximately the median source redshift expected for lenses discovered in similar imaging surveys \citep{collett2015}.  For HSC SSP galaxies that are brighter than $i = 24$ and that are assigned a photometric redshift $z_{\mathrm{phot}} \leq 2$, {\sc mizuki} has a photometric redshift scatter of $\sigma_{\mathrm{conv}} = 0.047$ and a catastrophic outlier rate (defined as the fraction of objects such that $\Delta z / (1+z_{\mathrm{true}}) > 0.15$) of $9.0\%$.

\subsection{BOSS Spectroscopic Data} \label{subsec:boss}
Our sample of strong lenses using various search algorithms (Section~\ref{subsec:lens}) is based on a pre-selection of candidate lens galaxies.  These candidates are taken from the Baryon Oscillation Spectroscopic Survey \citep[BOSS;][]{dawson+2013}, a component of the Sloan Digital Sky Survey III \citep[SDSS-III;][]{eisenstein+2011}.  The BOSS sample consists of the LOWZ subsample (primarily $z < 0.4$ LRGs) and the CMASS subsample (primarily $0.4 < z < 0.7$ LRGs).  The initial sample of SuGOHI-g lenses used the BOSS Data Release 12 \citep{alam+2015}.  The lens galaxy velocity dispersions used in this study come from Data Release 14 \citep{abolfathi+2017}.


\section{New Lens Candidates in Data Release 2 of the HSC SSP} \label{sec:dr2_lenses}
The SuGOHI-g sample of lenses are found through a combination of several lens search methodologies, which are described in \citet{sonnenfeld+2018}.
The primary search method is the {\sc YattaLens} algorithm, which searches for arc-like features around massive galaxies and fits simple lens models to evaluate the likelihood of those systems being lenses.  The new lens candidates discovered in Data Release 2 of the HSC SSP and presented here are found with {\sc YattaLens}.

We run {\sc YattaLens} on 31,286 massive galaxies in BOSS and find 772 lens candidates, 131 of which are in the LOWZ subsample and 641 of which are in the CMASS subsample.  These candidates are visually inspected by the coauthors and individually graded using the following scheme:
\begin{itemize}
\item Grade A: definite lenses 
\item Grade B: probable lenses 
\item Grade C: possible lenses 
\item Grade D: non-lenses 
\end{itemize}

\begin{figure*}
\plotone{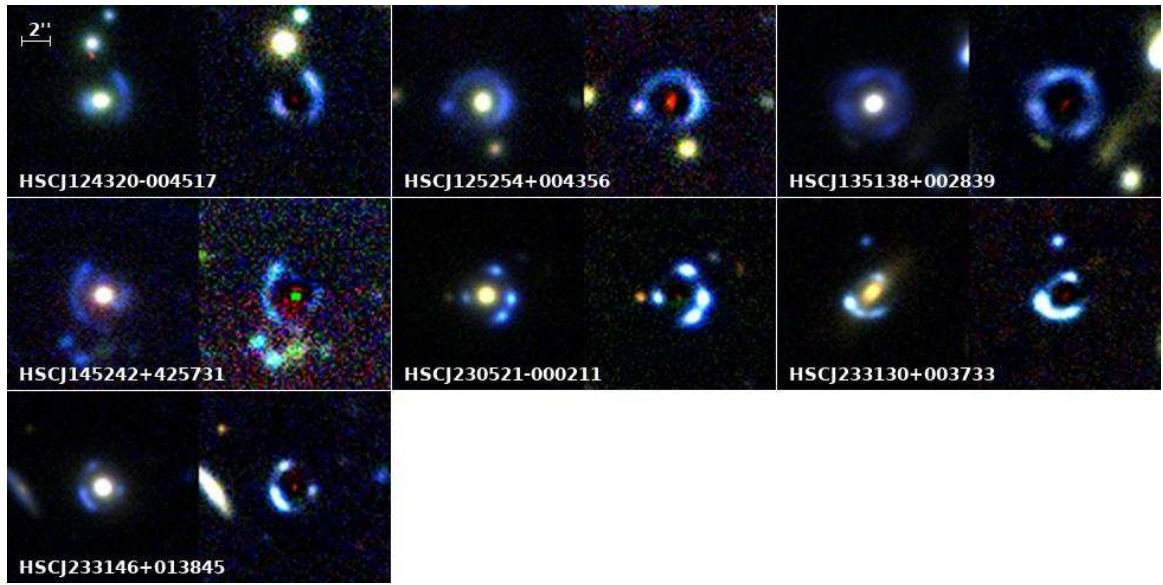}
\caption{
Grade A lenses. For each system, the left-hand panel is a color-composite image in g, r, and i bands, while the right-hand panel is a lens-subtracted version of the image.  A higher contrast is used in the right-hand panel to enhance the images of the lensed source.
\label{fig:Alenses}}
\end{figure*}

In total, we find 7 new lenses with an average grade of A, 34 with an average grade of B, and 159 with an average grade of C.  The full list of candidates, including grade C systems, can be found online\footnote{http://hsc.mtk.nao.ac.jp/ssp/science/strong-lensing}.  We present newly-discovered lenses with an average grade of A or B in Table~\ref{tab:s17a_lenses}.  We show images of the grade A lenses in Figure~\ref{fig:Alenses} and the grade B lens candidates in Figure~\ref{fig:Blenses}.  The BOSS spectra are visually inspected to identify emission lines from the lensed source, and thus measure the source redshift.  However, we are only able to measure the source redshift for one system in the new sample of grade A or B lenses.

\begin{figure*}
\plotone{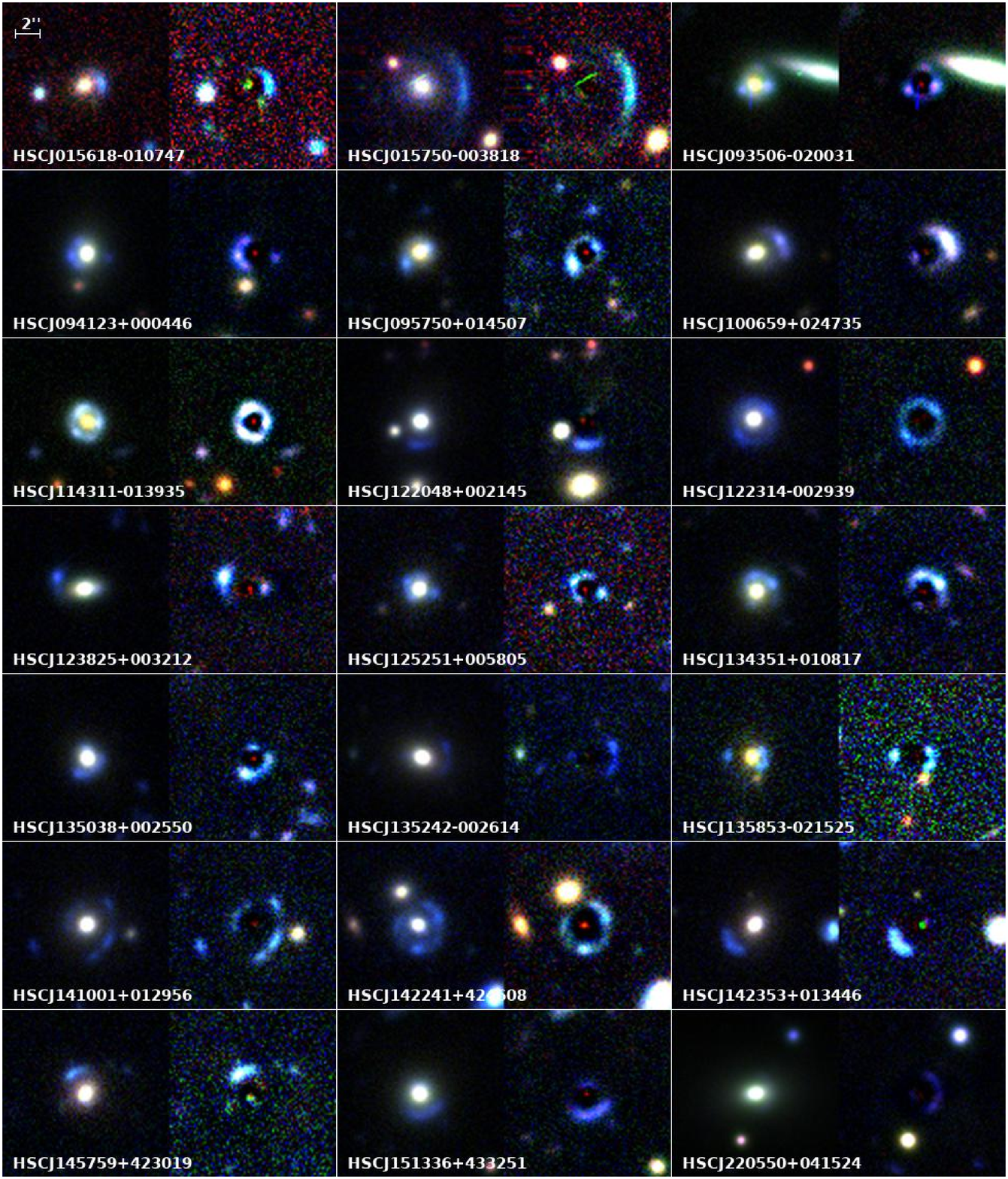}
\caption{
Grade B lens candidates. For each system, the left-hand panel is a color-composite image in g, r, and i bands, while the right-hand panel is a lens-subtracted version of the image.  A higher contrast is used in the right-hand panel to enhance the images of the lensed source.
\label{fig:Blenses}}
\end{figure*}

\addtocounter{figure}{-1}
\begin{figure*}
\plotone{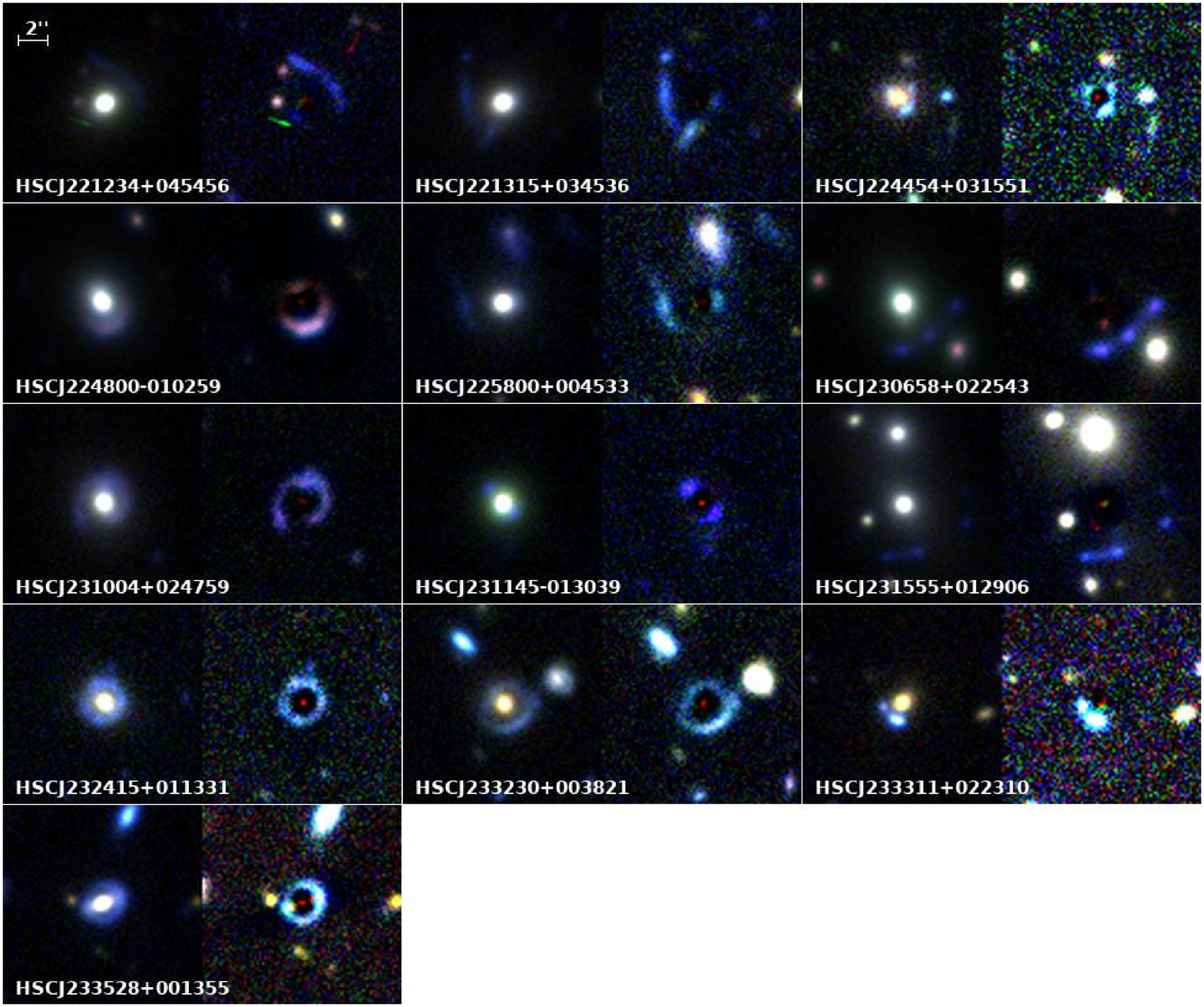}
\caption{
(continued)
}
\end{figure*}

\renewcommand*\arraystretch{1.2}
\begin{table*}
\caption{Grade A Lenses and Grade B Candidates in Data Release 2 of the HSC SSP\label{tab:s17a_lenses}}
\begin{ruledtabular}
\begin{tabular}{lcccccc}
Name & 
$\alpha$ (J2000) & 
$\delta$ (J2000) & 
$z_{\mathrm{L}}$ & 
$z_{\mathrm{S}}$ & 
Subsample & 
Grade
\\
\tableline
HSCJ015618$-$010747 & 
 29.0755 & 
 $-$1.1298 & 
0.542 & 
1.170 & 
CMASS & 
B
\\
HSCJ015750$-$003818 & 
 29.4599 & 
 $-$0.6385 & 
0.520 & 
--- & 
CMASS & 
B
\\
HSCJ093506$-$020031 & 
143.7761 & 
 $-$2.0089 & 
0.531 & 
--- & 
CMASS & 
B
\\
HSCJ094123+000446 & 
145.3460 & 
  0.0796 & 
0.486 & 
--- & 
CMASS & 
B
\\
HSCJ095750+014507 & 
149.4587 & 
  1.7521 & 
0.574 & 
--- & 
CMASS & 
B
\\
HSCJ100659+024735 & 
151.7470 & 
  2.7931 & 
0.482 & 
--- & 
CMASS & 
B
\\
HSCJ114311$-$013935 & 
175.7970 & 
 $-$1.6598 & 
0.644 & 
--- & 
CMASS & 
B
\\
HSCJ122048+002145 & 
185.2014 & 
  0.3628 & 
0.407 & 
--- & 
LOWZ & 
B
\\
HSCJ122314$-$002939 & 
185.8114 & 
 $-$0.4944 & 
0.547 & 
--- & 
CMASS & 
B
\\
HSCJ123825+003212 & 
189.6043 & 
  0.5368 & 
0.478 & 
--- & 
CMASS & 
B
\\
HSCJ124320$-$004517 & 
190.8365 & 
 $-$0.7550 & 
0.654 & 
--- & 
CMASS & 
A
\\
HSCJ125251+005805 & 
193.2132 & 
  0.9683 & 
0.540 & 
--- & 
CMASS & 
B
\\
HSCJ125254+004356 & 
193.2275 & 
  0.7323 & 
0.649 & 
--- & 
CMASS & 
A
\\
HSCJ134351+010817 & 
205.9665 & 
  1.1382 & 
0.567 & 
--- & 
CMASS & 
B
\\
HSCJ135038+002550 & 
207.6591 & 
  0.4306 & 
0.526 & 
--- & 
CMASS & 
B
\\
HSCJ135138+002839 & 
207.9122 & 
  0.4778 & 
0.461 & 
--- & 
CMASS & 
A
\\
HSCJ135242$-$002614 & 
208.1791 & 
 $-$0.4372 & 
0.508 & 
--- & 
CMASS & 
B
\\
HSCJ135853$-$021525 & 
209.7230 & 
 $-$2.2571 & 
0.737 & 
--- & 
CMASS & 
B
\\
HSCJ141001+012956 & 
212.5044 & 
  1.4992 & 
0.541 & 
--- & 
CMASS & 
B
\\
HSCJ142241+424608 & 
215.6716 & 
 42.7689 & 
0.605 & 
--- & 
CMASS & 
B
\\
HSCJ142353+013446 & 
215.9711 & 
  1.5797 & 
0.519 & 
--- & 
CMASS & 
B
\\
HSCJ145242+425731 & 
223.1789 & 
 42.9589 & 
0.718 & 
--- & 
CMASS & 
A
\\
HSCJ145759+423019 & 
224.4966 & 
 42.5053 & 
0.607 & 
--- & 
CMASS & 
B
\\
HSCJ151336+433251 & 
228.4032 & 
 43.5475 & 
0.487 & 
--- & 
CMASS & 
B
\\
HSCJ220550+041524 & 
331.4591 & 
  4.2569 & 
0.268 & 
--- & 
LOWZ & 
B
\\
HSCJ221234+045456 & 
333.1438 & 
  4.9158 & 
0.262 & 
--- & 
LOWZ & 
B
\\
HSCJ221315+034536 & 
333.3137 & 
  3.7602 & 
0.446 & 
--- & 
LOWZ & 
B
\\
HSCJ224454+031551 & 
341.2286 & 
  3.2642 & 
0.791 & 
--- & 
CMASS & 
B
\\
HSCJ224800$-$010259 & 
342.0030 & 
 $-$1.0499 & 
0.277 & 
--- & 
LOWZ & 
B
\\
HSCJ225800+004533 & 
344.5026 & 
  0.7594 & 
0.576 & 
--- & 
CMASS & 
B
\\
HSCJ230521$-$000211 & 
346.3403 & 
 $-$0.0366 & 
0.492 & 
--- & 
CMASS & 
A
\\
HSCJ230658+022543 & 
346.7428 & 
  2.4286 & 
0.362 & 
--- & 
LOWZ & 
B
\\
HSCJ231004+024759 & 
347.5196 & 
  2.7999 & 
0.390 & 
--- & 
LOWZ & 
B
\\
HSCJ231145$-$013039 & 
347.9379 & 
 $-$1.5109 & 
0.400 & 
--- & 
LOWZ & 
B
\\
HSCJ231555+012906 & 
348.9800 & 
  1.4850 & 
0.424 & 
--- & 
LOWZ & 
B
\\
HSCJ232415+011331 & 
351.0652 & 
  1.2255 & 
0.592 & 
--- & 
CMASS & 
B
\\
HSCJ233130+003733 & 
352.8770 & 
  0.6259 & 
0.552 & 
--- & 
CMASS & 
A
\\
HSCJ233146+013845 & 
352.9434 & 
  1.6460 & 
0.476 & 
--- & 
CMASS & 
A
\\
HSCJ233230+003821 & 
353.1289 & 
  0.6394 & 
0.623 & 
--- & 
CMASS & 
B
\\
HSCJ233311+022310 & 
353.2963 & 
  2.3864 & 
0.472 & 
--- & 
CMASS & 
B
\\
HSCJ233528+001355 & 
353.8677 & 
  0.2322 & 
0.568 & 
--- & 
CMASS & 
B
\\
\end{tabular}
\end{ruledtabular}
\tablecomments{New lens candidates in Data Release 2 of the HSC SSP.  Columns 4 and 5 list the lens and source redshift (when measured), respectively.  Column 6 lists which subsample of BOSS LRGs the lens galaxy belongs to.  Column 7 indicates the grade of the candidate.}
\end{table*}

\renewcommand*\arraystretch{1.0}

\section{Sample Selection for Environment and Line-of-Sight Analysis} \label{sec:sample}

\subsection{Lens Sample} \label{subsec:lens}
The sample of lenses used in this study includes the previously discovered candidates from \citet{sonnenfeld+2018} along with new candidates discovered in Data Release 2 (Section~\ref{sec:dr2_lenses}).  We only use lenses whose average grades are A or B (definite or probable lenses).  To increase our sample size of lenses, we also include previously-known lenses in our sample.  We search the Master Lens Database\footnote{http://masterlens.astro.utah.edu/} (Moustakas et al. in preparation), selecting for systems with a grade of A or B where the lensing object is a galaxy (as opposed to a galaxy group or cluster).  The Master Lens Database uses a grading scheme qualitatively similar to our own, and many of the lenses have follow-up {\it HST} imaging and/or spectroscopically-confirmed source redshifts, so we expect these to be real lenses with high confidence.  We add two additional known lenses in the survey area, HSCJ142449-005321 \citep[the ``Eye of Horus";][]{tanaka+2016} and H-ATLAS J090740.0-004200 \citep[``SDP.9";][]{negrello+2010}.  We further include two new grade A and B galaxy-scale lenses that have subsequently been discovered by the {\sc Chitah} algorithm \citep{chan+2015}.  These were originally classified as grade C lenses by \citet{sonnenfeld+2018}, but were given higher grades by the classifiers in Chan et al. (in preparation).

We manually remove systems where there is clearly more than one galaxy contributing to the strong lensing (i.e., there are multiple galaxies within the region enclosed by the lensed images).  Some lenses outside of the SuGOHI-g sample do not have velocity dispersion measurements from BOSS, while others have measured velocity dispersions smaller than 100 km s$^{-1}$ or larger than 400 km s$^{-1}$, indicative of significant errors in its measurement.  We do not include these lenses in our analysis since we cannot build a sample of twin galaxies for them, but we can still calculate their local and LOS environments.  Our final lens sample, which we refer to as the ``main sample", contains \nmain~lenses with a mean lens redshift of $\langle z_{\mathrm{L}} \rangle = \zlavg$ and is presented in Table~\ref{tab:lens_summary}.  For completeness, we show the results for individual lenses not in the main sample in Appendix~\ref{app:badvd}.

\renewcommand*\arraystretch{0.6}
\begin{table*}
\caption{Main Lens Sample\label{tab:lens_summary}}
\begin{ruledtabular}
\begin{tabular}{l|cccccc}
Lens &
$\alpha$ (J2000) &
$\delta$ (J2000) &
$z_{\mathrm{L}}$ &
$z_{\mathrm{S}}$ &
$\sigma$\tablenotemark{a} (km s$^{-1}$) &
Reference\tablenotemark{b}
\\
\tableline
HSCJ015731$-$033057 & 
 29.3812 & 
 $-$3.5160 & 
0.621 & 
$-$ &
190 $\pm$ 56 & 
SuGOHI I 
\\
HSCJ015756$-$021809 & 
 29.4859 & 
 $-$2.3028 & 
0.372 & 
$-$ &
244 $\pm$ 17 & 
SuGOHI I 
\\
SDSSJ0157$-$0056 & 
 29.4956 & 
 $-$0.9406 & 
0.513 & 
0.924 &
238 $\pm$ 35 & 
\citet{bolton+2008} 
\\
HSCJ020141$-$030946 & 
 30.4249 & 
 $-$3.1628 & 
0.362 & 
$-$ &
272 $\pm$ 32 & 
SuGOHI I 
\\
HSCJ020241$-$064611 & 
 30.6725 & 
 $-$6.7698 & 
0.502 & 
2.75 &
155 $\pm$ 27 & 
SuGOHI I 
\\
HSCJ020846$-$032727 & 
 32.1952 & 
 $-$3.4577 & 
0.618 & 
$-$ &
140 $\pm$ 39 & 
SuGOHI I 
\\
SL2SJ021737$-$051329 & 
 34.4048 & 
 $-$5.2248 & 
0.644 & 
1.847 &
305 $\pm$ 60 & 
\citet{sonnenfeld+2013} 
\\
HSCJ022140$-$021020 & 
 35.4172 & 
 $-$2.1723 & 
0.708 & 
$-$ &
289 $\pm$ 140 & 
SuGOHI I 
\\
SL2SJ022315$-$062906 & 
 35.8142 & 
 $-$6.4851 & 
0.550 & 
$-$ &
305 $\pm$ 43 & 
\citet{more+2012} 
\\
SL2SJ022346$-$053418 & 
 35.9423 & 
 $-$5.5718 & 
0.499 & 
1.44 &
258 $\pm$ 27 & 
\citet{sonnenfeld+2013} 
\\
SL2SJ022357$-$065142 & 
 35.9914 & 
 $-$6.8618 & 
0.473 & 
$-$ &
116 $\pm$ 35 & 
\citet{sonnenfeld+2013} 
\\
SL2SJ022439$-$040045 & 
 36.1628 & 
 $-$4.0125 & 
0.430 & 
$-$ &
191 $\pm$ 36 & 
\citet{more+2012} 
\\
SL2SJ022511$-$045433 & 
 36.2960 & 
 $-$4.9093 & 
0.238 & 
1.199 &
249 $\pm$ 13 & 
\citet{sonnenfeld+2013} 
\\
SL2SJ022610$-$042011 & 
 36.5444 & 
 $-$4.3366 & 
0.494 & 
1.232 &
233 $\pm$ 39 & 
\citet{sonnenfeld+2013} 
\\
HSCJ023217$-$021703 & 
 38.0724 & 
 $-$2.2844 & 
0.508 & 
$-$ &
263 $\pm$ 127 & 
SuGOHI I 
\\
SL2SJ023307$-$043838 & 
 38.2795 & 
 $-$4.6439 & 
0.671 & 
1.87 &
325 $\pm$ 123 & 
\citet{more+2012} 
\\
HSCJ023538$-$063406 & 
 38.9093 & 
 $-$6.5684 & 
0.181 & 
$-$ &
241 $\pm$ 11 & 
SuGOHI I 
\\
HSCJ023637$-$033220 & 
 39.1554 & 
 $-$3.5389 & 
0.270 & 
$-$ &
263 $\pm$ 19 & 
SuGOHI I 
\\
HSCJ023655$-$023656 & 
 39.2303 & 
 $-$2.6156 & 
0.562 & 
$-$ &
184 $\pm$ 41 & 
SuGOHI I 
\\
HSCJ083943+004740 & 
129.9293 & 
  0.7947 & 
0.621 & 
$-$ &
302 $\pm$ 37 & 
SuGOHI I 
\\
HSCJ085855$-$010208 & 
134.7333 & 
 $-$1.0357 & 
0.468 & 
1.42 &
167 $\pm$ 27 & 
SuGOHI I 
\\
HSCJ090507$-$001030 & 
136.2806 & 
 $-$0.1750 & 
0.494 & 
$-$ &
162 $\pm$ 24 & 
SuGOHI I 
\\
HSCJ090709+005648 & 
136.7904 & 
  0.9468 & 
0.478 & 
$-$ &
246 $\pm$ 37 & 
SuGOHI I 
\\
SDSSJ0915$-$0055 & 
138.8192 & 
 $-$0.9168 & 
0.402 & 
1.1705 &
210 $\pm$ 15 & 
\citet{brownstein+2012} 
\\
HSCJ091608+034710 & 
139.0355 & 
  3.7862 & 
0.531 & 
$-$ &
299 $\pm$ 33 & 
Chan et al. (in preparation) 
\\
HSCJ091904+033638 & 
139.7692 & 
  3.6107 & 
0.444 & 
$-$ &
248 $\pm$ 25 & 
SuGOHI I 
\\
HSCJ092101+035521 & 
140.2565 & 
  3.9228 & 
0.472 & 
$-$ &
175 $\pm$ 26 & 
Chan et al. (in preparation) 
\\
HSCJ093506$-$020031 & 
143.7761 & 
 $-$2.0089 & 
0.531 & 
$-$ &
235 $\pm$ 38 & 
SuGOHI II 
\\
HSCJ094123+000446 & 
145.3460 & 
  0.0796 & 
0.486 & 
$-$ &
255 $\pm$ 29 & 
SuGOHI II 
\\
SDSSJ0944$-$0147 & 
146.1145 & 
 $-$1.7951 & 
0.539 & 
1.179 &
199 $\pm$ 38 & 
\citet{brownstein+2012} 
\\
HSCJ095750+014507 & 
149.4587 & 
  1.7521 & 
0.574 & 
$-$ &
161 $\pm$ 36 & 
SuGOHI II 
\\
COSMOS0013+2249 & 
150.0579 & 
  2.3803 & 
0.346 & 
$-$ &
234 $\pm$ 40 & 
\citet{faure+2011} 
\\
COSMOS0056+1226 & 
150.2366 & 
  2.2072 & 
0.361 & 
0.808 &
241 $\pm$ 29 & 
\citet{faure+2011} 
\\
HSCJ100659+024735 & 
151.7470 & 
  2.7931 & 
0.482 & 
$-$ &
246 $\pm$ 40 & 
SuGOHI II 
\\
HSCJ114311$-$013935 & 
175.7970 & 
 $-$1.6598 & 
0.644 & 
$-$ &
226 $\pm$ 38 & 
SuGOHI II 
\\
HSCJ115653$-$003948 & 
179.2210 & 
 $-$0.6635 & 
0.508 & 
$-$ &
187 $\pm$ 39 & 
SuGOHI I 
\\
HSCJ120623+001507 & 
181.5994 & 
  0.2520 & 
0.563 & 
3.12 &
266 $\pm$ 48 & 
SuGOHI I 
\\
HSCJ121052$-$011905 & 
182.7187 & 
 $-$1.3181 & 
0.700 & 
$-$ &
285 $\pm$ 71 & 
SuGOHI I 
\\
HSCJ122048+002145 & 
185.2014 & 
  0.3628 & 
0.407 & 
$-$ &
285 $\pm$ 24 & 
SuGOHI II 
\\
HSCJ122314$-$002939 & 
185.8114 & 
 $-$0.4944 & 
0.547 & 
$-$ &
276 $\pm$ 34 & 
SuGOHI II 
\\
HSCJ123825+003212 & 
189.6043 & 
  0.5368 & 
0.478 & 
$-$ &
259 $\pm$ 45 & 
SuGOHI II 
\\
HSCJ124320$-$004517 & 
190.8365 & 
 $-$0.7550 & 
0.654 & 
$-$ &
365 $\pm$ 65 & 
SuGOHI II 
\\
HSCJ125254+004356 & 
193.2275 & 
  0.7323 & 
0.649 & 
$-$ &
388 $\pm$ 59 & 
SuGOHI II 
\\
HSCJ134351+010817 & 
205.9665 & 
  1.1382 & 
0.567 & 
$-$ &
260 $\pm$ 63 & 
SuGOHI II 
\\
HSCJ135038+002550 & 
207.6591 & 
  0.4306 & 
0.526 & 
$-$ &
210 $\pm$ 47 & 
SuGOHI II 
\\
HSCJ135138+002839 & 
207.9122 & 
  0.4778 & 
0.461 & 
$-$ &
355 $\pm$ 56 & 
SuGOHI II 
\\
HSCJ135242$-$002614 & 
208.1791 & 
 $-$0.4372 & 
0.508 & 
$-$ &
364 $\pm$ 45 & 
SuGOHI II 
\\
HSCJ135853$-$021525 & 
209.7230 & 
 $-$2.2571 & 
0.737 & 
$-$ &
225 $\pm$ 44 & 
SuGOHI II 
\\
HSCJ140929$-$011410 & 
212.3738 & 
 $-$1.2363 & 
0.584 & 
$-$ &
204 $\pm$ 39 & 
SuGOHI I 
\\
HSCJ141001+012956 & 
212.5044 & 
  1.4992 & 
0.541 & 
$-$ &
257 $\pm$ 34 & 
SuGOHI II 
\\
HSCJ141300$-$012608 & 
213.2503 & 
 $-$1.4356 & 
0.749 & 
$-$ &
248 $\pm$ 65 & 
SuGOHI I 
\\
HSCJ141831$-$000052 & 
214.6309 & 
 $-$0.0146 & 
0.263 & 
$-$ &
260 $\pm$ 13 & 
SuGOHI I 
\\
HSCJ142353+013446 & 
215.9711 & 
  1.5797 & 
0.519 & 
$-$ &
374 $\pm$ 43 & 
SuGOHI II 
\\
HSCJ142449$-$005321 & 
216.2042 & 
 $-$0.8893 & 
0.795 & 
$z_{\mathrm{S1}}=1.302$; $z_{\mathrm{S2}}=1.988$ & 
246 $\pm$ 94 & 
\citet{tanaka+2016} 
\\
HSCJ142720+001916 & 
216.8356 & 
  0.3211 & 
0.551 & 
$-$ &
250 $\pm$ 32 & 
SuGOHI I 
\\
HSCJ142748+000958 & 
216.9515 & 
  0.1663 & 
0.589 & 
$-$ &
281 $\pm$ 72 & 
SuGOHI I 
\\
HSCJ144307$-$004056 & 
220.7798 & 
 $-$0.6823 & 
0.500 & 
1.07 &
251 $\pm$ 35 & 
SuGOHI I 
\\
HSCJ144428$-$005142 & 
221.1198 & 
 $-$0.8618 & 
0.575 & 
$-$ &
199 $\pm$ 27 & 
SuGOHI I 
\\
HSCJ145236$-$002142 & 
223.1527 & 
 $-$0.3617 & 
0.733 & 
$-$ &
325 $\pm$ 47 & 
SuGOHI I 
\\
HSCJ145732$-$015917 & 
224.3858 & 
 $-$1.9882 & 
0.526 & 
$-$ &
238 $\pm$ 40 & 
SuGOHI I 
\\
HSCJ145759+423019 & 
224.4966 & 
 42.5053 & 
0.607 & 
$-$ &
272 $\pm$ 68 & 
SuGOHI II 
\\
HSCJ145902$-$012351 & 
224.7613 & 
 $-$1.3975 & 
0.482 & 
$-$ &
268 $\pm$ 40 & 
SuGOHI I 
\\
HSCJ151336+433251 & 
228.4032 & 
 43.5475 & 
0.487 & 
$-$ &
204 $\pm$ 26 & 
SuGOHI II 
\\
HSCJ155826+432830 & 
239.6111 & 
 43.4752 & 
0.444 & 
$-$ &
292 $\pm$ 30 & 
SuGOHI I 
\\
SL2SJ220202+014710 & 
330.5069 & 
  1.7860 & 
0.300 & 
$-$ &
213 $\pm$ 8 & 
\citet{more+2012} 
\\
SL2SJ220506+014703 & 
331.2788 & 
  1.7844 & 
0.476 & 
2.53 &
269 $\pm$ 46 & 
\citet{sonnenfeld+2013} 
\\
HSCJ220550+041524 & 
331.4591 & 
  4.2569 & 
0.268 & 
$-$ &
221 $\pm$ 15 & 
SuGOHI II 
\\
SL2SJ220642+041131 & 
331.6751 & 
  4.1919 & 
0.620 & 
$-$ &
226 $\pm$ 38 & 
\citet{more+2012} 
\\
HSCJ221726+000350 & 
334.3602 & 
  0.0640 & 
0.398 & 
$-$ &
216 $\pm$ 20 & 
SuGOHI I 
\\
SL2SJ221852+014038 & 
334.7193 & 
  1.6775 & 
0.564 & 
$-$ &
264 $\pm$ 44 & 
\citet{sonnenfeld+2013} 
\\
HSCJ222801+012805 & 
337.0082 & 
  1.4683 & 
0.647 & 
$-$ &
336 $\pm$ 49 & 
SuGOHI I 
\\
HSCJ223518$-$004747 & 
338.8263 & 
 $-$0.7965 & 
0.640 & 
$-$ &
196 $\pm$ 54 & 
SuGOHI I 
\\
HSCJ223733+005015 & 
339.3897 & 
  0.8377 & 
0.604 & 
$-$ &
256 $\pm$ 42 & 
SuGOHI I 
\\
HSCJ224201+022810 & 
340.5049 & 
  2.4696 & 
0.443 & 
$-$ &
189 $\pm$ 49 & 
SuGOHI I 
\\
HSCJ224221+001144 & 
340.5899 & 
  0.1958 & 
0.385 & 
$-$ &
256 $\pm$ 20 & 
SuGOHI I 
\\
HSCJ224454+031551 & 
341.2286 & 
  3.2642 & 
0.791 & 
$-$ &
277 $\pm$ 68 & 
SuGOHI II 
\\
HSCJ224800$-$010259 & 
342.0030 & 
 $-$1.0499 & 
0.277 & 
$-$ &
245 $\pm$ 13 & 
SuGOHI II 
\\
HSCJ225800+004533 & 
344.5026 & 
  0.7594 & 
0.576 & 
$-$ &
280 $\pm$ 45 & 
SuGOHI II 
\\
SDSSJ2303+0037 & 
345.8965 & 
  0.6176 & 
0.458 & 
0.936 &
269 $\pm$ 35 & 
\citet{brownstein+2012} 
\\
HSCJ230521$-$000211 & 
346.3403 & 
 $-$0.0366 & 
0.492 & 
$-$ &
185 $\pm$ 28 & 
SuGOHI II 
\\
HSCJ231004+024759 & 
347.5196 & 
  2.7999 & 
0.390 & 
$-$ &
270 $\pm$ 28 & 
SuGOHI II 
\\
HSCJ231145$-$013039 & 
347.9379 & 
 $-$1.5109 & 
0.400 & 
$-$ &
241 $\pm$ 31 & 
SuGOHI II 
\\
HSCJ232415+011331 & 
351.0652 & 
  1.2255 & 
0.592 & 
$-$ &
296 $\pm$ 63 & 
SuGOHI II 
\\
HSCJ233130+003733 & 
352.8770 & 
  0.6259 & 
0.552 & 
$-$ &
168 $\pm$ 41 & 
SuGOHI II 
\\
HSCJ233146+013845 & 
352.9434 & 
  1.6460 & 
0.476 & 
$-$ &
188 $\pm$ 62 & 
SuGOHI II 
\\
HSCJ233311+022310 & 
353.2963 & 
  2.3864 & 
0.472 & 
$-$ &
119 $\pm$ 47 & 
SuGOHI II 
\\
HSCJ233528+001355 & 
353.8677 & 
  0.2322 & 
0.568 & 
$-$ &
230 $\pm$ 43 & 
SuGOHI II 
\\
\end{tabular}
\end{ruledtabular}
\tablenotetext{1}{Velocity dispersions from SDSS DR14}
\tablenotetext{2}{SuGOHI I = \citet{sonnenfeld+2018}; SuGOHI II = this work}
\end{table*}

\renewcommand*\arraystretch{1.0}

\subsection{Twin Sample} \label{subsec:twin}
We construct a sample of ``twin" galaxies for comparison with the lens sample in a manner similar to that of \citet{treu+2009}.  We select galaxies from the BOSS DR14 using the same selection criteria as for the pre-selected galaxies used as targets for the {\sc YattaLens} algorithm, excluding the galaxies in the lens sample itself.  We then construct the twin sample based on galaxy redshift and velocity dispersion.

The strong lensing cross section of a galaxy is a steep function of its velocity dispersion ($\propto \sigma^{4}$ for an isothermal mass profile).  Given this fact, a lensing-selected sample effectively selects galaxies based on their true velocity dispersion, which could result in an Eddington bias by which the measured velocity dispersions of our lens sample from BOSS are systematically low.  This would result in our selection of the twin sample to select less massive galaxies than the lenses if we did a simple matching based on measure velocity dispersion.  The expected velocity dispersion of a galaxy (assuming a singular isothermal sphere profile) is 
\begin{equation} \label{eq:rein_sis}
\sigma_{\mathrm{SIS}} = c \sqrt{\frac{\theta_{\mathrm{E}}}{4 \pi} \frac{D_{\mathrm{S}}}{D_{\mathrm{LS}}}}.
\end{equation}
We can then compare the expected velocity dispersion to the lenses' measured velocity dispersions from BOSS.  We have a subsample of 10 lens systems in which we have measured source redshifts and Einstein radii from the literature, and thus can compare their expected $\sigma_{\mathrm{SIS}}$ to their measured velocity dispersion from BOSS.  Of these 10 systems, nine of them have a $\sigma_{\mathrm{SIS}}$ larger than their measured $\sigma$ with a mean difference between the two quantities of $\sim15\%$.  While this subsample is small, it suggests that this bias is real.  For the SLACS sample, \citet{bolton+2008} find that $\sigma / \sigma_{\mathrm{SIS}} \approx 0.948$ before correcting for differences in the physical sizes of the SDSS fiber aperture at the lens redshift.  This likely has a smaller influence in our sample due to our higher mean lens redshift, and the magnitude of the effect does not seem large enough to explain this bias.  Thus, we must account for this effect when selecting the twin sample.

For each lens galaxy, we assume the probability distribution of its velocity dispersion is a Gaussian with a mean and scatter given by the value and uncertainty from BOSS.  We then weight this distribution by $\sigma^{4}$ and re-normalize, taking the mean of the new distribution to be the ``corrected" velocity dispersion.  On average, the corrected lens galaxy velocity dispersions are $\sim13\%$ higher than the measured values.  For each lens, we then randomly pick ten twin galaxies that have a spectroscopic redshift such that $\Delta z/(1+z_{\mathrm{L}}) \leq 0.01$ and a velocity dispersion that is within 10\% of the lens galaxy's corrected velocity dispersion.  We also enforce a condition that half of the twins have a smaller velocity dispersion and half have a larger velocity dispersion than that of its corresponding lens to minimize bias from the slope of the galaxy velocity dispersion function.  The same galaxy can be selected more than once if it is a ``twin" of more than one lens galaxy.  Our final twin sample contains 870 galaxies, of which 855 are unique.

\subsection{Control Fields} \label{subsec:control}
We select random control fields from the HSC SSP by randomly selecting patches ($12\arcmin \times 12\arcmin$ square regions) from the survey patch catalog, then selecting random coordinates within that patch.  The chosen locations are taken to be the center of each control field.  Patches that have {\tt imag\_psf\_depth} $< 24$ are excluded in order to remove fields that do not have a similar depth to the lens fields.  We also exclude patches that do not have a defined PSF magnitude depth in the other four filters, indicating that it lacks coverage in all five bands.  The control sample consists of \ncont~random fields that meet these criteria, which is a large enough sample to obtain good statistics within a reasonable computation time.  In principle, the randomly chosen control fields can overlap with each other or with the lens and twin fields, but we verify that none of the fields overlap to within a 120\arcsec~radius.

\subsection{Correction for Survey Gaps} \label{subsec:geo}
The HSC SSP area has gaps due to survey edges, chip gaps, bright star masks, etc.  To account for this, we use the HSC SSP random point catalog \citep{coupon+2018}, which contains randomly sampled points in the survey region that are flagged in the same way as the objects.  The random points are drawn with a density of 100 points per arcmin$^{2}$.  When evaluating galaxy number counts within our selected apertures, we use the random point catalog to correct the raw counts, $N$, by a scaling factor to account for occulted area in the aperture.  We define the effective number counts as
\begin{equation} \label{eq:n_eff}
N_{\mathrm{eff}} \equiv N \times \frac{100 \pi r_{\mathrm{ap}}^{2}}{N_{\mathrm{rand}}},
\end{equation}
where $r_{\mathrm{ap}}$ is the radius of the chosen aperture and $N_{\mathrm{rand}}$ is the number of points from the random catalog in the aperture.  For small apertures and/or bright magnitude limits where the galaxy density is more sparse, this correction can lead to a bias toward higher $N_{\mathrm{eff}}$ in individual fields if most of them do not contain galaxies in the survey gaps \citep{cooper+2005}, but the ensemble results over the entire sample should be unbiased.

If $N_{\mathrm{rand}} / (100 \pi r_{\mathrm{ap}}^{2}) < 0.75$, we exclude that field from the calculation for that particular aperture and lens redshift.  \citet{rusu+2017} similarly excluded fields from their analysis of the field of the time-delay lens HE 0435-1223, testing a cut of both 0.5 and 0.75, and finding that the results were insensitive to the choice of masked fraction threshold.

\section{Quantifying Line-of-Sight and Local Environment Overdensity} \label{sec:los}

\subsection{Galaxy Selection} \label{subsec:galselect}
To evaluate local and line-of-sight densities, we select galaxies from the HSC SSP catalog that satisfy the following criteria, which are defined in \citet{bosch+2018}:
\begin{itemize}
\item{{\tt i\_extendedness\_value} $> 0$}
\item{{\tt i\_cmodel\_mag} $\leq 24.0$}
\item{{\tt merge\_peak\_i} = True}
\item{{\tt isprimary} = True}
\item{{\tt i\_cmodel\_flag} = False}
\item{{\tt i\_pixelflags\_edge} = False}
\item{{\tt i\_pixelflags\_bad} = False}
\item{{\tt i\_pixelflags\_saturatedcenter} = False}
\item{{\tt i\_pixelflags\_crcenter} = False}
\end{itemize}

\begin{figure*}
\plotone{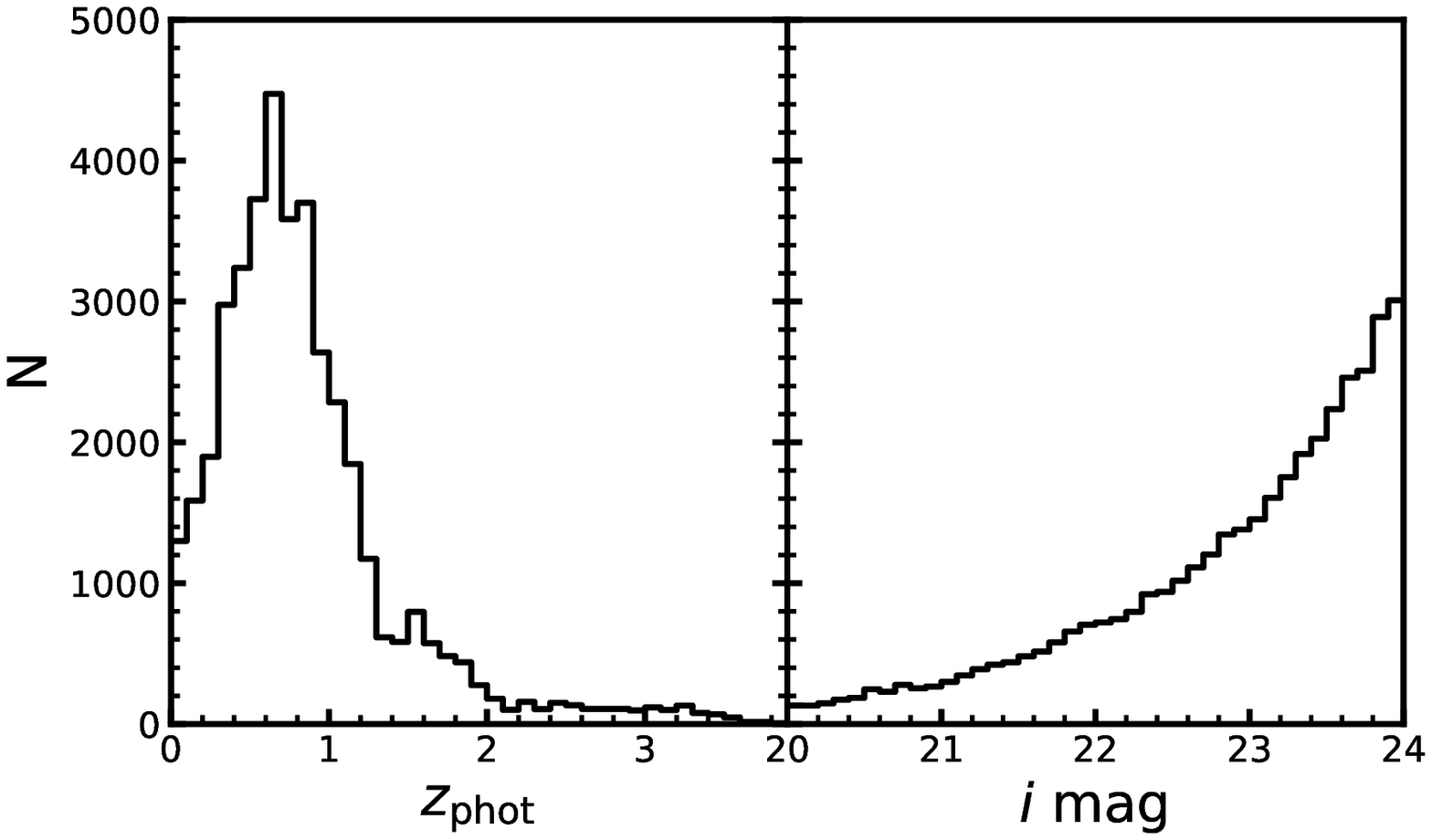}
\caption{
Distribution of photometric redshift (left panel) and $i$ magnitude (right panel) for galaxies satisfying our selection criteria within $r \leq 120\arcsec$ of all control fields.
\label{fig:z_mag_hist}}
\end{figure*}

We further require that galaxies lie in patches that have observations in all five bands, even if not to the full survey depth, as the photometric redshifts are less reliable without measurements in all filters.  The galaxy magnitudes used in this analysis are cModel magnitudes, which are measured by fitting galaxy models convolved with the point spread function (PSF) to the light profile of the object \citep{abazajian+2004}.  The photometric redshift and $i$-band magnitude distributions of galaxies selected by these criteria within $r \leq 120\arcsec$ of all control fields are shown in Figure~\ref{fig:z_mag_hist}.

For our calculation of $N_{\mathrm{eff}}$, we select points from the random point catalog using the following criteria:
\begin{itemize}
\item{{\tt isprimary} = True}
\item{{\tt i\_pixelflags\_edge} = False}
\item{{\tt i\_pixelflags\_bad} = False}
\item{{\tt i\_pixelflags\_saturatedcenter} = False}
\item{{\tt i\_pixelflags\_crcenter} = False}
\end{itemize}

\subsection{Line of Sight Structure} \label{subsec:los}
We quantify the density of a line of sight by the effective galaxy number counts, $N_{\mathrm{eff}}$, brighter than a limiting $i$-band magnitude and within a projected radial distance $r_{\mathrm{ap}}$ of each lens.  We exclude galaxies that are within $2\farcs5$ of the lens and twin samples to remove any background galaxies strongly affected by magnification bias \citep[e.g.,][]{fassnacht+2011}, and exclude a similar region from the centers of the control fields so that we are comparing similar volumes.  We test aperture sizes of $r_{\mathrm{ap}} = \{30\arcsec, 60\arcsec, 90\arcsec, 120\arcsec\}$ and limiting magnitudes of $i \leq \{21.0, 22.0, 23.0, 24.0\}$.  \citet{collett+2013} suggest that galaxies projected further than 120\arcsec~from a lens galaxy and fainter than $i$ = 24 are unlikely to make a significant contribution to the external convergence.  These limits also encompass the aperture sizes and limiting magnitudes tested by \citet{rusu+2017} in evaluating the external convergence in HE 0435-1223.

To separate the contribution of the local lens environment from the LOS, we remove galaxies that are close to the lens in redshift.  We define $N_{\mathrm{eff}}^{\mathrm{LOS}}$ to be $N_{\mathrm{eff}}$ calculated without counting the galaxies that have photometric redshifts such that $\Delta z/(1+z_{\mathrm{L}}) \leq 0.05$.  The threshold of 0.05 is taken to be a conservative estimate of the typical scatter of the {\sc mizuki} photometric redshift algorithm out to $i = 24$ when applied to HSC SSP data \citep{tanaka+2018}.

We can calculate the fractional overdensity of a lens LOS relative to a random LOS by normalizing $N_{\mathrm{eff}}$ by a factor of $\langle N_{\mathrm{eff}}^{\mathrm{cont}} \rangle$, which is the mean $N_{\mathrm{eff}}$ across all control fields for the same aperture size and limiting magnitude.  To ensure that we are comparing equivalent volumes, we exclude a redshift slice of width $\Delta z/(1+z_{\mathrm{L,rand}}) \leq 0.05$ centered on a randomly drawn lens redshift for each control field when calculating $\langle N_{\mathrm{eff}}^{\mathrm{cont}} \rangle$.  The uncertainties are taken to be the combination of Poisson error on $N$ and $N_{\mathrm{rand}}$ in the calculation of $N_{\mathrm{eff}}$, along with the error on the mean of $N_{\mathrm{eff}}^{\mathrm{cont}}$.  This is a conservative estimate, as Poisson error on $N$ dominates the uncertainty budget.  For certain applications, it may be useful to have the uncertainties that do not include Poisson error on $N$ \citep[e.g.,][]{rusu+2017}, so we show these values in Appendix~\ref{app:no_poisson}.  We ignore Poisson uncertainties in the $N_{\mathrm{eff}}$ calculations for individual control fields, as it is negligible in comparison to the sample variance.

\subsection{Local Lens Environment} \label{subsec:lensenv}
To evaluate the local overdensity in the environment of a lens or twin galaxy, we use a standard measure of environment, $\Sigma_{10}$ \citep[e.g.,][]{dressler1980,cooper+2006,treu+2009}, defined as
\begin{equation} \label{eq:sig_10}
\Sigma_{10} = \frac{10}{\pi D_{p,10}^{2}},
\end{equation}
where $D_{p,10}$ is the projected distance in Mpc to the tenth nearest neighbor that is brighter than $i = 24$ and has a redshift within $\Delta z/(1+z_{\mathrm{L}}) \leq 0.05$ of the lens redshift.  \citet{cooper+2005} find that $D_{p,10}$ is a reasonable estimator of local galaxy environments over a broad range of scales and is reasonably robust to survey edge effects.  We correct the $\Sigma_{10}$ values for the survey geometry using the random object catalog to calculate a scaling factor as in Equation~\ref{eq:n_eff}, using $r_{\mathrm{ap}} = D_{p,10}$.


The absolute $\Sigma_{10}$ values depend on the chosen magnitude cut and lens redshift, so in practice, we use a normalized density, $\Sigma_{10} / \langle \Sigma_{10}^{\mathrm{cont}} \rangle$, where $\langle \Sigma_{10}^{\mathrm{cont}} \rangle$ is the mean $\Sigma_{10}$ evaluated at the center of each of the control fields at the same redshift as the lens being considered.  We take the uncertainty on this relative density measure to be the combination of Poisson error on $n=10$ in the calculation of $\Sigma_{10}$ and the error on the mean of $\Sigma_{10}^{\mathrm{cont}}$, similar to \citet{treu+2009}.

\section{Results and Discussion} \label{sec:results}

\subsection{Lens Lines of Sight} \label{sec:los_results}
\begin{figure*}
\plotone{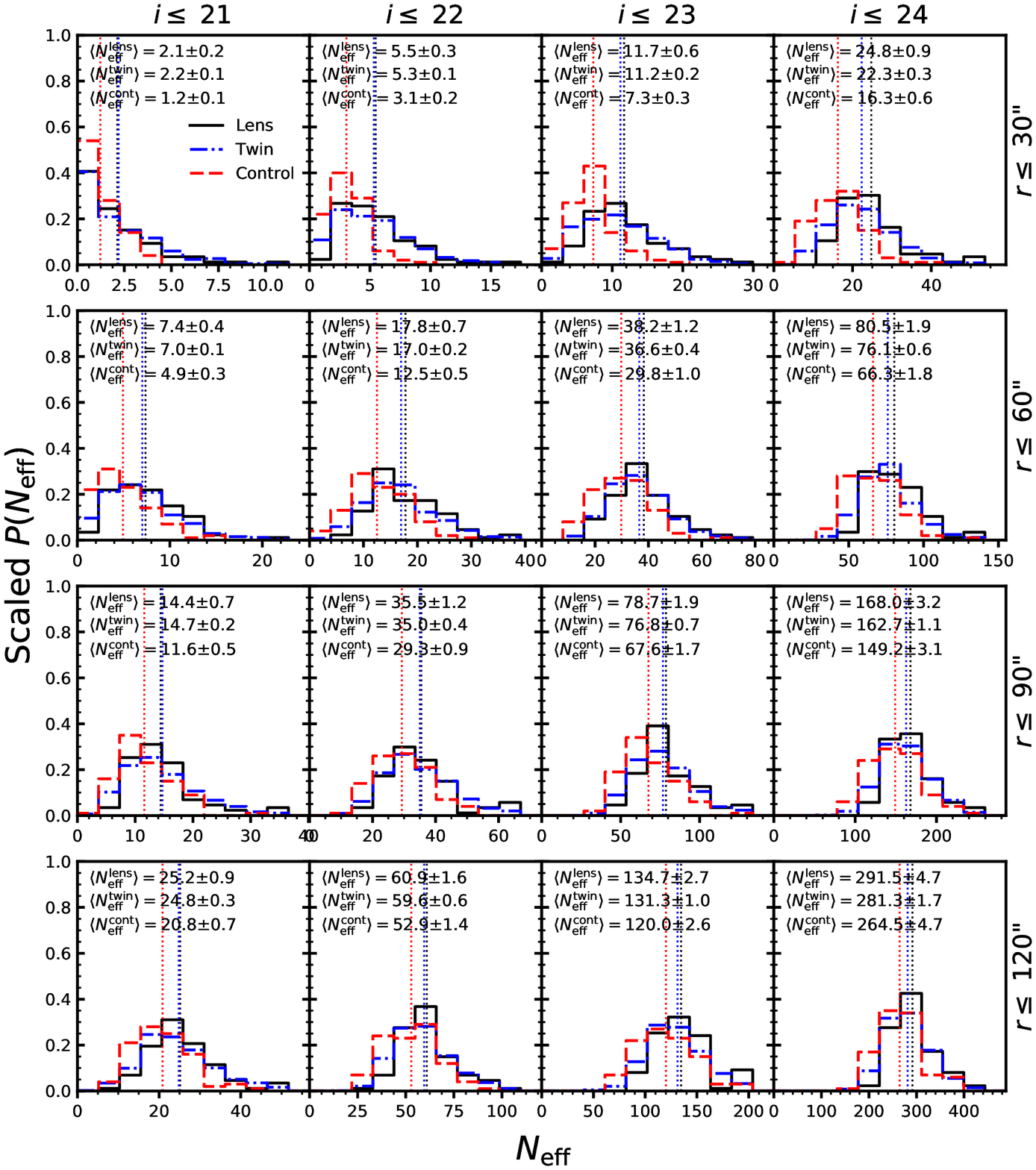}
\caption{
Scaled histogram of $N_{\mathrm{eff}}$ values for the lens (black solid line), twin (blue dot-dashed line), and control (red dashed line) samples.  The columns represent different limiting magnitudes, and the rows represent different aperture radii.  The dotted lines of the corresponding color indicate the mean of the distributions.  The mean and the standard error on the mean are given for the different samples within each panel.  The lens and twin samples have higher $N_{\mathrm{eff}}$ in general than the control sample.
\label{fig:nhist_all}}
\end{figure*}

We show normalized histograms of $N_{\mathrm{eff}}$ for the lens, twin, and control lines of sight in Figure~\ref{fig:nhist_all}.  Each panel represents a different aperture size and magnitude cut.  In general, the lens and twin lines of sight have higher $N_{\mathrm{eff}}$ than the control sample.  We quantify the significance of these differences using a 2-sample Anderson-Darling test, from which we calculate $P_{\mathrm{AD}}$, the significance level at which the null hypothesis that the two samples are drawn from the same parent distribution can be rejected (i.e., a lower $P_{\mathrm{AD}}$ means that it is less likely that the null hypothesis is true).  Commonly accepted significance levels are $P_{\mathrm{AD}} < 0.05$ ($2\sigma$) or $P_{\mathrm{AD}} < 0.003$ ($3\sigma$).  We perform this test in a pairwise manner among the lens, twin, and control samples for each of the aperture sizes and magnitude cuts.  These results are shown in Table~\ref{tab:pad_all}.

\renewcommand*\arraystretch{1.2}
\begin{table}
\caption{$2$-sample Anderson-Darling Test Significance Levels - Full LOS \label{tab:pad_all}}
\begin{ruledtabular}
\begin{tabular}{l|cccc}
\multicolumn{5}{c}{Lens-Control}
\\
r (arcsec) &
$i \leq 21$ &
$i \leq 22$ &
$i \leq 23$ &
$i \leq 24$
\\
\tableline
30 &
\bf{7.2e-04} &
\bf{8.1e-06} &
\bf{1.3e-05} &
\bf{1.7e-04}
\\
60 &
\bf{2.2e-05} &
\bf{8.8e-06} &
\bf{1.6e-05} &
\bf{1.1e-05}
\\
90 &
\bf{1.2e-03} &
\bf{1.7e-04} &
\bf{1.0e-04} &
\bf{1.0e-04}
\\
120 &
\bf{4.1e-04} &
\bf{4.5e-04} &
\bf{3.1e-04} &
\bf{1.7e-04}
\\
\tableline
\multicolumn{5}{c}{Lens-Twin}
\\
r (arcsec) &
$i \leq 21$ &
$i \leq 22$ &
$i \leq 23$ &
$i \leq 24$
\\
\tableline
30 &
8.9e-01 &
6.2e-01 &
2.2e-01 &
\emph{2.0e-02}
\\
60 &
2.3e-01 &
2.7e-01 &
2.2e-01 &
5.9e-02
\\
90 &
4.2e-01 &
4.9e-01 &
1.7e-01 &
1.6e-01
\\
120 &
3.6e-01 &
2.4e-01 &
7.4e-02 &
\emph{3.5e-02}
\\
\tableline
\multicolumn{5}{c}{Twin-Control}
\\
r (arcsec) &
$i \leq 21$ &
$i \leq 22$ &
$i \leq 23$ &
$i \leq 24$
\\
\tableline
30 &
\bf{1.7e-05} &
\bf{1.9e-04} &
\bf{3.0e-04} &
\bf{6.6e-04}
\\
60 &
\bf{1.3e-05} &
\bf{1.1e-05} &
\bf{8.9e-06} &
\bf{1.0e-05}
\\
90 &
\bf{3.7e-05} &
\bf{2.7e-05} &
\bf{7.0e-05} &
\bf{1.5e-04}
\\
120 &
\bf{1.7e-04} &
\bf{5.3e-04} &
\bf{7.4e-04} &
\bf{1.5e-03}
\\
\end{tabular}
\end{ruledtabular}
\tablecomments{The values represent $P_{\mathrm{AD}}$, the significance level at which the null hypothesis that the samples are drawn from the same parent distribution can be rejected.  Values in italics are significant at greater than the $2\sigma$ level, while values in bold are significant at greater than the $3\sigma$ level.}
\end{table}

\renewcommand*\arraystretch{1.0}

Our results show that the lens and twin samples are consistent with each other and any differences are of marginal significance, as would be expected if lenses are representative of the underlying population of galaxies of similar masses and redshifts.  Furthermore, the lens fields (as well as the twin fields) show a distinct difference in their $N_{\mathrm{eff}}$ distributions compared to the control fields, with the lens fields skewed toward higher density lines of sight.  This is consistent with the results of \citet{fassnacht+2011}.  The overdensities of each lens's line of sight relative to the control sample is given in Table~\ref{tab:overdens}.

\renewcommand*\arraystretch{0.7}
\begin{table*}
\caption{Local and LOS Overdensity of Individual Lens Fields\label{tab:overdens}}
\begin{ruledtabular}
\begin{tabular}{l|cccc|c}
\multirow{2}{*}{Lens} &
$N_{\mathrm{eff}}/\langle N_{\mathrm{eff}}^{\mathrm{cont}} \rangle$ & 
$N_{\mathrm{eff}}/\langle N_{\mathrm{eff}}^{\mathrm{cont}} \rangle$ & 
$N_{\mathrm{eff}}/\langle N_{\mathrm{eff}}^{\mathrm{cont}} \rangle$ & 
$N_{\mathrm{eff}}/\langle N_{\mathrm{eff}}^{\mathrm{cont}} \rangle$ & 
\multirow{2}{*}{$\Sigma_{10}/\langle \Sigma_{10}^{\mathrm{cont}} \rangle$}
\\
&
($r \leq 30\arcsec$) & 
($r \leq 60\arcsec$) & 
($r \leq 90\arcsec$) & 
($r \leq 120\arcsec$) & 
\\
\tableline
HSCJ015731$-$033057 & 
1.33 $\pm$ 0.33 & 
1.48 $\pm$ 0.18 & 
1.45 $\pm$ 0.12 & 
1.26 $\pm$ 0.08 & 
2.37 $\pm$ 0.80
\\
HSCJ015756$-$021809 & 
1.68 $\pm$ 0.38 & 
1.10 $\pm$ 0.15 & 
1.02 $\pm$ 0.09 & 
0.95 $\pm$ 0.07 & 
0.87 $\pm$ 0.28
\\
SDSSJ0157$-$0056 & 
0.75 $\pm$ 0.23 & 
1.00 $\pm$ 0.14 & 
1.12 $\pm$ 0.10 & 
1.20 $\pm$ 0.08 & 
1.32 $\pm$ 0.43
\\
HSCJ020141$-$030946 & 
1.54 $\pm$ 0.36 & 
1.15 $\pm$ 0.15 & 
0.96 $\pm$ 0.09 & 
1.02 $\pm$ 0.07 & 
2.61 $\pm$ 0.87
\\
HSCJ020241$-$064611 & 
1.28 $\pm$ 0.32 & 
0.95 $\pm$ 0.13 & 
1.03 $\pm$ 0.09 & 
1.13 $\pm$ 0.08 & 
1.05 $\pm$ 0.34
\\
HSCJ020846$-$032727 & 
1.71 $\pm$ 0.38 & 
1.36 $\pm$ 0.17 & 
1.30 $\pm$ 0.11 & 
1.33 $\pm$ 0.08 & 
3.42 $\pm$ 1.17
\\
SL2SJ021737$-$051329 & 
1.90 $\pm$ 0.41 & 
1.53 $\pm$ 0.18 & 
1.19 $\pm$ 0.10 & 
1.10 $\pm$ 0.07 & 
4.49 $\pm$ 1.55
\\
HSCJ022140$-$021020 & 
1.26 $\pm$ 0.31 & 
0.82 $\pm$ 0.12 & 
0.86 $\pm$ 0.08 & 
0.86 $\pm$ 0.06 & 
5.02 $\pm$ 1.76
\\
SL2SJ022315$-$062906 & 
1.90 $\pm$ 0.43 & 
1.16 $\pm$ 0.15 & 
1.01 $\pm$ 0.09 & 
0.91 $\pm$ 0.07 & 
4.24 $\pm$ 1.45
\\
SL2SJ022346$-$053418 & 
1.80 $\pm$ 0.40 & 
1.16 $\pm$ 0.15 & 
1.06 $\pm$ 0.10 & 
1.09 $\pm$ 0.07 & 
2.21 $\pm$ 0.73
\\
SL2SJ022357$-$065142 & 
1.29 $\pm$ 0.31 & 
0.98 $\pm$ 0.14 & 
1.05 $\pm$ 0.09 & 
1.04 $\pm$ 0.07 & 
1.37 $\pm$ 0.45
\\
SL2SJ022439$-$040045 & 
1.35 $\pm$ 0.34 & 
0.87 $\pm$ 0.13 & 
0.96 $\pm$ 0.09 & 
1.05 $\pm$ 0.07 & 
2.06 $\pm$ 0.68
\\
SL2SJ022511$-$045433 & 
1.40 $\pm$ 0.33 & 
1.20 $\pm$ 0.15 & 
1.15 $\pm$ 0.10 & 
1.10 $\pm$ 0.07 & 
1.05 $\pm$ 0.34
\\
SL2SJ022610$-$042011 & 
1.56 $\pm$ 0.36 & 
0.93 $\pm$ 0.13 & 
0.91 $\pm$ 0.09 & 
0.97 $\pm$ 0.07 & 
1.13 $\pm$ 0.37
\\
HSCJ023217$-$021703 & 
1.21 $\pm$ 0.31 & 
1.14 $\pm$ 0.15 & 
1.15 $\pm$ 0.10 & 
1.02 $\pm$ 0.07 & 
1.11 $\pm$ 0.36
\\
SL2SJ023307$-$043838 & 
1.41 $\pm$ 0.34 & 
1.22 $\pm$ 0.16 & 
1.19 $\pm$ 0.10 & 
1.11 $\pm$ 0.07 & 
1.33 $\pm$ 0.44
\\
HSCJ023538$-$063406 & 
1.57 $\pm$ 0.36 & 
1.06 $\pm$ 0.14 & 
0.93 $\pm$ 0.09 & 
0.87 $\pm$ 0.06 & 
1.05 $\pm$ 0.34
\\
HSCJ023637$-$033220 & 
1.86 $\pm$ 0.40 & 
1.57 $\pm$ 0.18 & 
1.49 $\pm$ 0.12 & 
1.36 $\pm$ 0.08 & 
4.08 $\pm$ 1.35
\\
HSCJ023655$-$023656 & 
1.04 $\pm$ 0.28 & 
0.97 $\pm$ 0.14 & 
0.94 $\pm$ 0.09 & 
0.97 $\pm$ 0.07 & 
2.03 $\pm$ 0.67
\\
HSCJ083943+004740 & 
1.05 $\pm$ 0.28 & 
1.03 $\pm$ 0.14 & 
1.01 $\pm$ 0.09 & 
1.10 $\pm$ 0.07 & 
1.48 $\pm$ 0.49
\\
HSCJ085855$-$010208 & 
2.35 $\pm$ 0.48 & 
1.22 $\pm$ 0.16 & 
1.16 $\pm$ 0.10 & 
1.01 $\pm$ 0.07 & 
1.20 $\pm$ 0.39
\\
HSCJ090507$-$001030 & 
1.26 $\pm$ 0.32 & 
0.96 $\pm$ 0.13 & 
0.89 $\pm$ 0.09 & 
0.93 $\pm$ 0.07 & 
1.48 $\pm$ 0.49
\\
HSCJ090709+005648 & 
1.45 $\pm$ 0.34 & 
0.92 $\pm$ 0.13 & 
0.98 $\pm$ 0.09 & 
0.99 $\pm$ 0.07 & 
1.21 $\pm$ 0.40
\\
SDSSJ0915$-$0055 & 
2.61 $\pm$ 0.53 & 
1.96 $\pm$ 0.21 & 
1.52 $\pm$ 0.12 & 
1.38 $\pm$ 0.09 & 
3.22 $\pm$ 1.08
\\
HSCJ091608+034710 & 
1.13 $\pm$ 0.30 & 
1.21 $\pm$ 0.16 & 
1.09 $\pm$ 0.10 & 
1.06 $\pm$ 0.07 & 
3.57 $\pm$ 1.21
\\
HSCJ091904+033638 & 
1.47 $\pm$ 0.35 & 
1.24 $\pm$ 0.16 & 
1.09 $\pm$ 0.10 & 
1.01 $\pm$ 0.07 & 
4.57 $\pm$ 1.56
\\
HSCJ092101+035521 & 
1.59 $\pm$ 0.37 & 
1.00 $\pm$ 0.14 & 
0.83 $\pm$ 0.08 & 
0.81 $\pm$ 0.06 & 
1.64 $\pm$ 0.54
\\
HSCJ093506$-$020031 & 
1.42 $\pm$ 0.35 & 
0.85 $\pm$ 0.13 & 
0.72 $\pm$ 0.08 & 
0.76 $\pm$ 0.06 & 
1.21 $\pm$ 0.40
\\
HSCJ094123+000446 & 
1.32 $\pm$ 0.33 & 
1.00 $\pm$ 0.14 & 
0.94 $\pm$ 0.09 & 
0.88 $\pm$ 0.06 & 
1.80 $\pm$ 0.59
\\
SDSSJ0944$-$0147 & 
1.79 $\pm$ 0.40 & 
1.64 $\pm$ 0.19 & 
1.30 $\pm$ 0.11 & 
1.12 $\pm$ 0.07 & 
2.01 $\pm$ 0.66
\\
HSCJ095750+014507 & 
0.97 $\pm$ 0.27 & 
0.86 $\pm$ 0.13 & 
0.87 $\pm$ 0.09 & 
0.91 $\pm$ 0.07 & 
1.10 $\pm$ 0.36
\\
COSMOS0013+2249 & 
0.93 $\pm$ 0.26 & 
1.09 $\pm$ 0.15 & 
1.12 $\pm$ 0.10 & 
1.15 $\pm$ 0.08 & 
4.10 $\pm$ 1.38
\\
COSMOS0056+1226 & 
1.12 $\pm$ 0.29 & 
0.99 $\pm$ 0.14 & 
1.00 $\pm$ 0.09 & 
0.99 $\pm$ 0.07 & 
2.16 $\pm$ 0.71
\\
HSCJ100659+024735 & 
1.95 $\pm$ 0.42 & 
1.28 $\pm$ 0.16 & 
1.11 $\pm$ 0.10 & 
1.00 $\pm$ 0.07 & 
7.93 $\pm$ 2.83
\\
HSCJ114311$-$013935 & 
0.92 $\pm$ 0.26 & 
0.96 $\pm$ 0.13 & 
1.07 $\pm$ 0.10 & 
1.14 $\pm$ 0.08 & 
1.47 $\pm$ 0.49
\\
HSCJ115653$-$003948 & 
1.19 $\pm$ 0.30 & 
1.36 $\pm$ 0.17 & 
1.38 $\pm$ 0.11 & 
1.43 $\pm$ 0.09 & 
1.18 $\pm$ 0.38
\\
HSCJ120623+001507 & 
1.41 $\pm$ 0.34 & 
1.13 $\pm$ 0.15 & 
1.14 $\pm$ 0.10 & 
1.22 $\pm$ 0.08 & 
3.66 $\pm$ 1.24
\\
HSCJ121052$-$011905 & 
0.85 $\pm$ 0.25 & 
1.06 $\pm$ 0.14 & 
1.03 $\pm$ 0.09 & 
1.19 $\pm$ 0.08 & 
1.49 $\pm$ 0.49
\\
HSCJ122048+002145 & 
3.11 $\pm$ 0.58 & 
2.06 $\pm$ 0.22 & 
1.62 $\pm$ 0.13 & 
1.52 $\pm$ 0.09 & 
5.71 $\pm$ 1.97
\\
HSCJ122314$-$002939 & 
1.25 $\pm$ 0.33 & 
1.32 $\pm$ 0.16 & 
1.19 $\pm$ 0.10 & 
1.07 $\pm$ 0.07 & 
1.66 $\pm$ 0.55
\\
HSCJ123825+003212 & 
2.00 $\pm$ 0.43 & 
1.37 $\pm$ 0.17 & 
1.26 $\pm$ 0.11 & 
1.18 $\pm$ 0.08 & 
2.56 $\pm$ 0.85
\\
HSCJ124320$-$004517 & 
2.10 $\pm$ 0.45 & 
1.40 $\pm$ 0.17 & 
1.27 $\pm$ 0.11 & 
1.14 $\pm$ 0.08 & 
4.09 $\pm$ 1.41
\\
HSCJ125254+004356 & 
1.57 $\pm$ 0.37 & 
1.43 $\pm$ 0.17 & 
1.25 $\pm$ 0.11 & 
1.08 $\pm$ 0.07 & 
2.43 $\pm$ 0.82
\\
HSCJ134351+010817 & 
1.67 $\pm$ 0.38 & 
1.51 $\pm$ 0.18 & 
1.33 $\pm$ 0.11 & 
1.18 $\pm$ 0.08 & 
2.76 $\pm$ 0.93
\\
HSCJ135038+002550 & 
1.49 $\pm$ 0.36 & 
1.40 $\pm$ 0.17 & 
1.19 $\pm$ 0.10 & 
1.03 $\pm$ 0.07 & 
2.32 $\pm$ 0.77
\\
HSCJ135138+002839 & 
2.71 $\pm$ 0.54 & 
1.46 $\pm$ 0.18 & 
1.46 $\pm$ 0.12 & 
1.35 $\pm$ 0.08 & 
8.16 $\pm$ 2.92
\\
HSCJ135242$-$002614 & 
0.70 $\pm$ 0.22 & 
0.77 $\pm$ 0.12 & 
0.84 $\pm$ 0.08 & 
0.93 $\pm$ 0.07 & 
0.75 $\pm$ 0.24
\\
HSCJ135853$-$021525 & 
2.03 $\pm$ 0.44 & 
1.37 $\pm$ 0.17 & 
1.21 $\pm$ 0.10 & 
1.22 $\pm$ 0.08 & 
3.60 $\pm$ 1.23
\\
HSCJ140929$-$011410 & 
1.36 $\pm$ 0.33 & 
1.66 $\pm$ 0.19 & 
1.34 $\pm$ 0.11 & 
1.27 $\pm$ 0.08 & 
2.28 $\pm$ 0.76
\\
HSCJ141001+012956 & 
1.19 $\pm$ 0.30 & 
1.12 $\pm$ 0.15 & 
1.09 $\pm$ 0.10 & 
1.17 $\pm$ 0.08 & 
1.82 $\pm$ 0.60
\\
HSCJ141300$-$012608 & 
1.10 $\pm$ 0.29 & 
0.93 $\pm$ 0.13 & 
0.92 $\pm$ 0.09 & 
0.97 $\pm$ 0.07 & 
0.74 $\pm$ 0.24
\\
HSCJ141831$-$000052 & 
1.63 $\pm$ 0.37 & 
1.59 $\pm$ 0.18 & 
1.15 $\pm$ 0.10 & 
1.15 $\pm$ 0.08 & 
2.44 $\pm$ 0.80
\\
HSCJ142353+013446 & 
1.30 $\pm$ 0.32 & 
1.31 $\pm$ 0.16 & 
1.08 $\pm$ 0.10 & 
1.11 $\pm$ 0.07 & 
2.19 $\pm$ 0.73
\\
HSCJ142449$-$005321 & 
3.20 $\pm$ 0.61 & 
1.79 $\pm$ 0.20 & 
1.74 $\pm$ 0.13 & 
1.68 $\pm$ 0.10 & 
8.54 $\pm$ 3.16
\\
HSCJ142720+001916 & 
1.55 $\pm$ 0.36 & 
1.49 $\pm$ 0.18 & 
1.38 $\pm$ 0.11 & 
1.31 $\pm$ 0.08 & 
3.21 $\pm$ 1.08
\\
HSCJ142748+000958 & 
1.30 $\pm$ 0.32 & 
1.36 $\pm$ 0.17 & 
1.17 $\pm$ 0.10 & 
1.27 $\pm$ 0.08 & 
2.05 $\pm$ 0.68
\\
HSCJ144307$-$004056 & 
0.76 $\pm$ 0.24 & 
1.06 $\pm$ 0.14 & 
0.99 $\pm$ 0.09 & 
0.96 $\pm$ 0.07 & 
1.34 $\pm$ 0.44
\\
HSCJ144428$-$005142 & 
1.96 $\pm$ 0.42 & 
1.26 $\pm$ 0.16 & 
1.09 $\pm$ 0.10 & 
1.09 $\pm$ 0.07 & 
8.53 $\pm$ 3.10
\\
HSCJ145236$-$002142 & 
2.63 $\pm$ 0.52 & 
1.61 $\pm$ 0.19 & 
1.35 $\pm$ 0.11 & 
1.18 $\pm$ 0.08 & 
6.60 $\pm$ 2.37
\\
HSCJ145732$-$015917 & 
2.61 $\pm$ 0.53 & 
1.48 $\pm$ 0.18 & 
1.34 $\pm$ 0.11 & 
1.16 $\pm$ 0.08 & 
4.93 $\pm$ 1.71
\\
HSCJ145759+423019 & 
1.56 $\pm$ 0.35 & 
1.10 $\pm$ 0.14 & 
1.13 $\pm$ 0.10 & 
1.11 $\pm$ 0.07 & 
4.11 $\pm$ 1.42
\\
HSCJ145902$-$012351 & 
1.00 $\pm$ 0.27 & 
1.01 $\pm$ 0.14 & 
0.88 $\pm$ 0.09 & 
1.00 $\pm$ 0.07 & 
1.38 $\pm$ 0.45
\\
HSCJ151336+433251 & 
1.12 $\pm$ 0.29 & 
1.40 $\pm$ 0.17 & 
1.21 $\pm$ 0.10 & 
1.13 $\pm$ 0.08 & 
2.63 $\pm$ 0.88
\\
HSCJ155826+432830 & 
1.35 $\pm$ 0.34 & 
1.07 $\pm$ 0.14 & 
0.93 $\pm$ 0.09 & 
1.02 $\pm$ 0.07 & 
1.04 $\pm$ 0.34
\\
SL2SJ220202+014710 & 
0.78 $\pm$ 0.24 & 
0.67 $\pm$ 0.11 & 
0.84 $\pm$ 0.08 & 
0.88 $\pm$ 0.06 & 
0.86 $\pm$ 0.28
\\
SL2SJ220506+014703 & 
2.22 $\pm$ 0.47 & 
1.69 $\pm$ 0.20 & 
1.45 $\pm$ 0.12 & 
1.35 $\pm$ 0.09 & 
1.83 $\pm$ 0.60
\\
HSCJ220550+041524 & 
1.81 $\pm$ 0.40 & 
1.38 $\pm$ 0.17 & 
1.27 $\pm$ 0.11 & 
1.20 $\pm$ 0.08 & 
2.07 $\pm$ 0.67
\\
SL2SJ220642+041131 & 
1.79 $\pm$ 0.40 & 
1.34 $\pm$ 0.16 & 
1.07 $\pm$ 0.10 & 
1.00 $\pm$ 0.07 & 
5.17 $\pm$ 1.82
\\
HSCJ221726+000350 & 
1.16 $\pm$ 0.30 & 
1.09 $\pm$ 0.15 & 
1.01 $\pm$ 0.09 & 
0.91 $\pm$ 0.07 & 
7.10 $\pm$ 2.48
\\
SL2SJ221852+014038 & 
1.06 $\pm$ 0.29 & 
1.04 $\pm$ 0.14 & 
0.93 $\pm$ 0.09 & 
1.00 $\pm$ 0.07 & 
1.62 $\pm$ 0.53
\\
HSCJ222801+012805 & 
1.10 $\pm$ 0.29 & 
1.00 $\pm$ 0.14 & 
1.02 $\pm$ 0.09 & 
1.11 $\pm$ 0.07 & 
1.36 $\pm$ 0.45
\\
HSCJ223518$-$004747 & 
1.16 $\pm$ 0.30 & 
1.14 $\pm$ 0.15 & 
1.13 $\pm$ 0.10 & 
1.09 $\pm$ 0.07 & 
4.38 $\pm$ 1.52
\\
HSCJ223733+005015 & 
0.83 $\pm$ 0.25 & 
1.06 $\pm$ 0.14 & 
0.95 $\pm$ 0.09 & 
0.96 $\pm$ 0.07 & 
1.33 $\pm$ 0.44
\\
HSCJ224201+022810 & 
1.51 $\pm$ 0.36 & 
0.98 $\pm$ 0.14 & 
0.90 $\pm$ 0.09 & 
1.05 $\pm$ 0.07 & 
1.28 $\pm$ 0.42
\\
HSCJ224221+001144 & 
1.14 $\pm$ 0.30 & 
0.95 $\pm$ 0.13 & 
1.21 $\pm$ 0.10 & 
1.18 $\pm$ 0.08 & 
2.81 $\pm$ 0.93
\\
HSCJ224454+031551 & 
1.29 $\pm$ 0.32 & 
1.11 $\pm$ 0.15 & 
1.01 $\pm$ 0.09 & 
1.03 $\pm$ 0.07 & 
1.39 $\pm$ 0.46
\\
HSCJ224800$-$010259 & 
1.34 $\pm$ 0.33 & 
1.19 $\pm$ 0.15 & 
1.22 $\pm$ 0.11 & 
1.04 $\pm$ 0.07 & 
1.64 $\pm$ 0.53
\\
HSCJ225800+004533 & 
1.02 $\pm$ 0.28 & 
1.30 $\pm$ 0.16 & 
1.19 $\pm$ 0.10 & 
1.14 $\pm$ 0.08 & 
1.93 $\pm$ 0.64
\\
SDSSJ2303+0037 & 
1.96 $\pm$ 0.43 & 
1.30 $\pm$ 0.16 & 
1.16 $\pm$ 0.10 & 
1.11 $\pm$ 0.07 & 
3.17 $\pm$ 1.06
\\
HSCJ230521$-$000211 & 
1.59 $\pm$ 0.36 & 
1.08 $\pm$ 0.14 & 
0.89 $\pm$ 0.09 & 
0.87 $\pm$ 0.06 & 
1.42 $\pm$ 0.46
\\
HSCJ231004+024759 & 
3.30 $\pm$ 0.63 & 
2.00 $\pm$ 0.22 & 
1.71 $\pm$ 0.13 & 
1.50 $\pm$ 0.09 & 
10.29 $\pm$ 3.68
\\
HSCJ231145$-$013039 & 
1.61 $\pm$ 0.36 & 
1.14 $\pm$ 0.15 & 
1.02 $\pm$ 0.09 & 
1.02 $\pm$ 0.07 & 
1.99 $\pm$ 0.66
\\
HSCJ232415+011331 & 
1.04 $\pm$ 0.28 & 
0.95 $\pm$ 0.13 & 
1.02 $\pm$ 0.09 & 
1.16 $\pm$ 0.08 & 
1.80 $\pm$ 0.59
\\
HSCJ233130+003733 & 
1.39 $\pm$ 0.33 & 
1.14 $\pm$ 0.15 & 
1.33 $\pm$ 0.11 & 
1.39 $\pm$ 0.09 & 
1.41 $\pm$ 0.46
\\
HSCJ233146+013845 & 
1.44 $\pm$ 0.34 & 
1.15 $\pm$ 0.15 & 
1.11 $\pm$ 0.10 & 
1.05 $\pm$ 0.07 & 
3.48 $\pm$ 1.17
\\
HSCJ233311+022310 & 
1.06 $\pm$ 0.28 & 
0.91 $\pm$ 0.13 & 
0.98 $\pm$ 0.09 & 
0.89 $\pm$ 0.07 & 
1.07 $\pm$ 0.35
\\
HSCJ233528+001355 & 
1.75 $\pm$ 0.39 & 
1.32 $\pm$ 0.16 & 
1.15 $\pm$ 0.10 & 
1.24 $\pm$ 0.08 & 
4.01 $\pm$ 1.37
\\
\end{tabular}
\end{ruledtabular}
\tablecomments{Relative LOS overdensities are calculated for a magnitude limit of $i \leq 24$.  Overdensities are normalized by the mean values across all control fields.  The uncertainties include Poisson uncertainties in the calculation of lens field quantites and error on the mean in the control fields.}
\end{table*}

\renewcommand*\arraystretch{1.0}

\begin{figure*}
\plotone{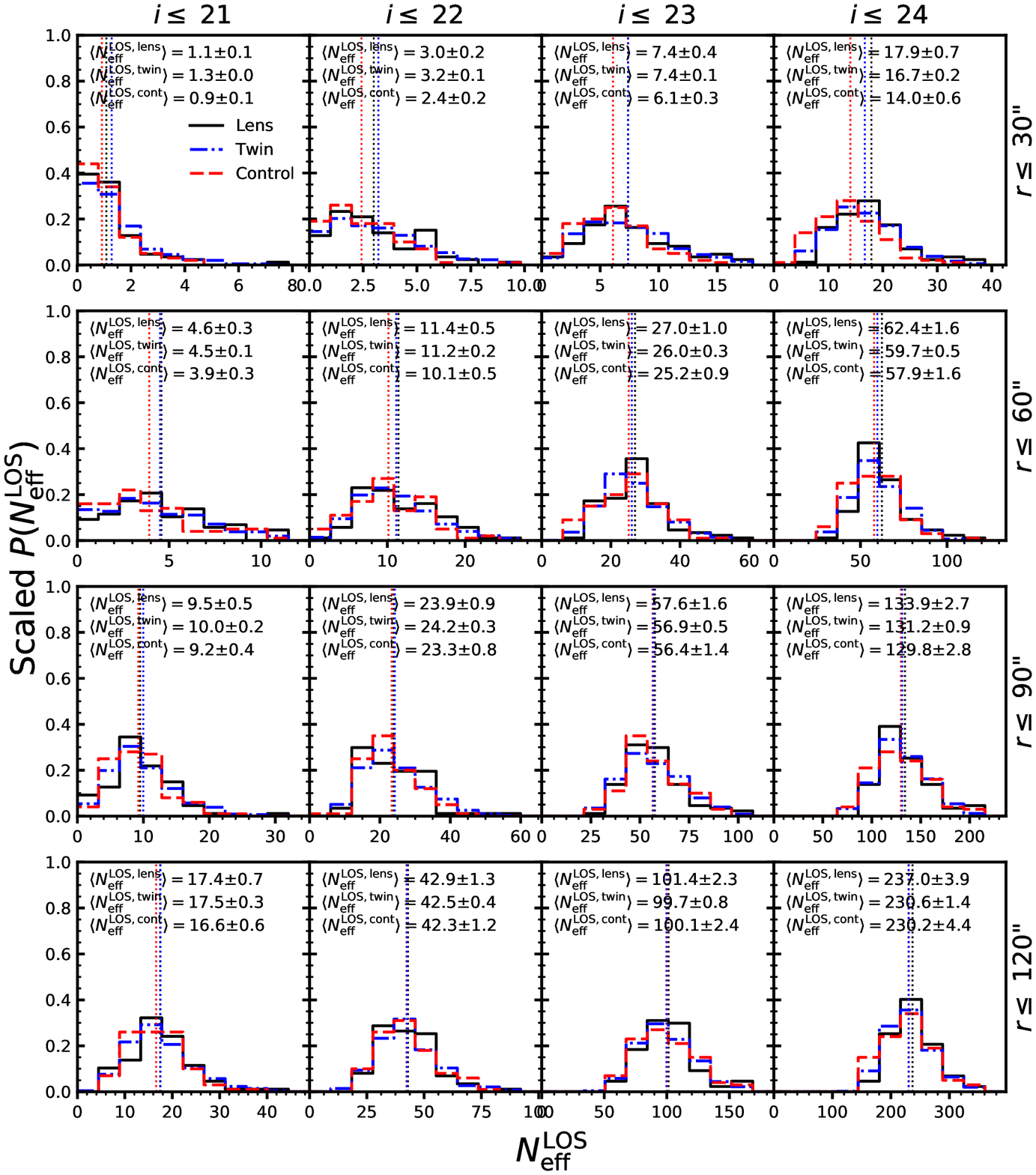}
\caption{
Same as Figure~\ref{fig:nhist_all}, but with galaxies in the lens plane removed for the lens and twin samples, and an equivalent cut made for the control sample by randomly drawing lens redshifts for each control field.  The three samples are now generally consistent, indicating that lenses do not lie along biased lines of sight once the local environment has been accounted for.  There is still an excess in the lens and twin samples in the smallest $r_{\mathrm{ap}}$ bin.  We discuss a possible origin of this effect in Section~\ref{subsubsec:photoz_outlier}.
\label{fig:nhist_los}}
\end{figure*}

We then exclude galaxies close to the lens and twin galaxies in redshift space to examine the LOS effects alone.  As mentioned in Section~\ref{subsec:los}, we randomly draw a lens redshift from our sample of lens galaxies and exclude objects within a redshift slice centered on that redshift.  This way, we are comparing fair volumes across samples in our calculation.  Figure~\ref{fig:nhist_los} shows the $N_{\mathrm{eff}}$ histograms excluding those within our chosen redshift cuts.  Table~\ref{tab:pad_los} shows the corresponding $P_{\mathrm{AD}}$ values.
 
\renewcommand*\arraystretch{1.2}
\begin{table}
\caption{$2$-sample Anderson-Darling Test Significance Levels - Lens Plane Removed \label{tab:pad_los}}
\begin{ruledtabular}
\begin{tabular}{l|ccccc}
\multicolumn{5}{c}{Lens-Control}
\\
r (arcsec) &
$i \leq 21$ &
$i \leq 22$ &
$i \leq 23$ &
$i \leq 24$
\\
\tableline
30 &
7.0e-01 &
5.2e-02 &
\emph{3.0e-02} &
\bf{8.5e-05}
\\
60 &
8.6e-02 &
1.8e-01 &
3.6e-01 &
\emph{2.0e-02}
\\
90 &
8.0e-01 &
9.1e-01 &
7.3e-01 &
1.7e-01
\\
120 &
3.5e-01 &
7.1e-01 &
2.0e-01 &
1.4e-01
\\
\tableline
\multicolumn{5}{c}{Lens-Twin}
\\
r (arcsec) &
$i \leq 21$ &
$i \leq 22$ &
$i \leq 23$ &
$i \leq 24$
\\
\tableline
30 &
4.1e-01 &
7.1e-01 &
9.8e-01 &
1.8e-01
\\
60 &
6.0e-01 &
8.4e-01 &
5.0e-01 &
1.4e-01
\\
90 &
5.0e-01 &
5.5e-01 &
4.6e-01 &
4.5e-01
\\
120 &
6.4e-01 &
6.8e-01 &
2.5e-01 &
1.3e-01
\\
\tableline
\multicolumn{5}{c}{Twin-Control}
\\
r (arcsec) &
$i \leq 21$ &
$i \leq 22$ &
$i \leq 23$ &
$i \leq 24$
\\
\tableline
30 &
\emph{2.2e-02} &
\bf{2.2e-03} &
\bf{2.2e-03} &
\bf{9.6e-05}
\\
60 &
1.2e-01 &
1.2e-01 &
4.0e-01 &
1.5e-01
\\
90 &
2.5e-01 &
4.9e-01 &
9.0e-01 &
4.7e-01
\\
120 &
3.4e-01 &
1.0e+00 &
7.6e-01 &
7.4e-01
\\
\end{tabular}
\end{ruledtabular}
\tablecomments{The values represent $P_{\mathrm{AD}}$, the significance level at which the null hypothesis that the samples are drawn from the same parent distribution can be rejected.  Values in italics are significant at greater than the $2\sigma$ level, while values in bold are significant at greater than the $3\sigma$ level.}
\end{table}

\renewcommand*\arraystretch{1.0}

Once the local environment of the lens galaxies are excluded, they are generally consistent with the control sample, particularly for larger aperture sizes where shot noise has a smaller effect.  This result affirms that of \citet{fassnacht+2011}, showing that lenses do not lie along biased lines of sight.  This is also consistent with the findings of \citet{collettcunnington2016} that show that lensed quasars have external convergence contributions of $\kappa_{\mathrm{ext}} < 0.01$ on average when considering just the projected LOS structure.  There does appear to be a slight excess in the lens and twin samples, particularly at small $r_{\mathrm{ap}}$ and/or bright limiting magnitudes where Poisson uncertainty due to small number statistics can be an issue.  These excesses are not statistically significant in most of the individual panels of Figure~\ref{fig:nhist_los} but appear consistently.  We discuss a possible origin of this effect in Section~\ref{subsubsec:photoz_outlier}.

\subsection{Local Lens Environments} \label{sec:env_results}
\begin{figure}
\plotone{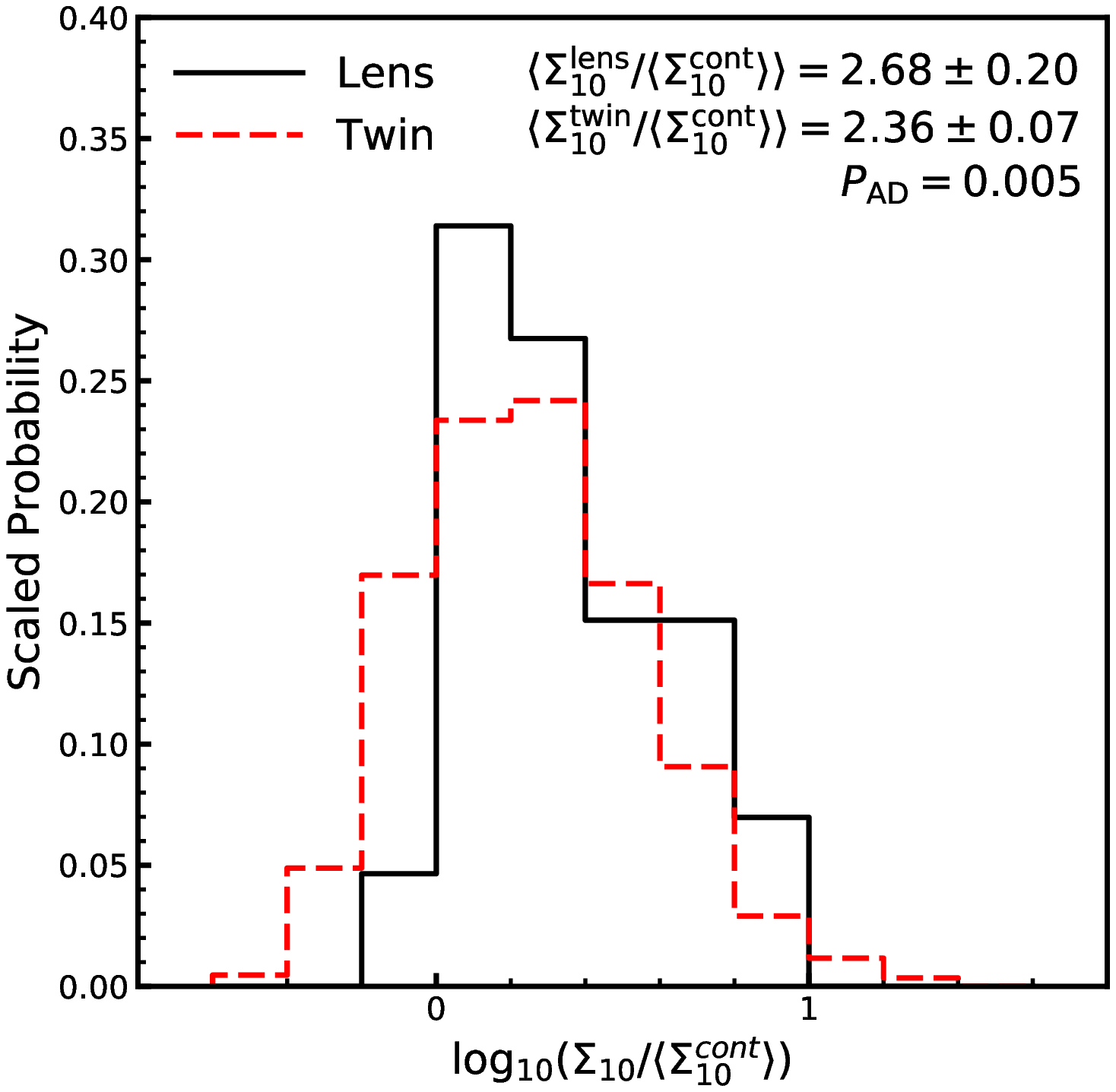}
\caption{
Normalized histogram of $\Sigma_{10}$ distributions for the lens (black solid line) and twin (red dashed line) samples.  The mean and standard error of each distribution is given in the upper right corner, along with the $P_{\mathrm{AD}}$ value.  The lens galaxies have a slightly higher local overdensity than the twin galaxies, although the significance is marginal.
\label{fig:sig10_hist}}
\end{figure}

We compare the distribution of normalized $\Sigma_{10}$ values of the lens and twin samples in Figure~\ref{fig:sig10_hist}.  The mean $\Sigma_{10} / \langle \Sigma_{10}^{\mathrm{cont}} \rangle$ for the lens and twin samples relative to the mean of the control fields are $2.68 \pm 0.20$ and $2.36 \pm 0.07$, respectively.  This shows that the local environments of both lenses and their twins are more overdense than random, as expected.  This also suggests a slight enhancement in the local density of lens galaxies than twins.  A 2-sample Anderson-Darling test of the two samples shows a $P_{\mathrm{AD}} = 0.005$, which is marginally significant but cannot conclusively rule out lenses not exhibiting a particular bias in their local environment.
\citet{treu+2009} also find a slightly higher $\Sigma_{10}/\langle \Sigma_{10}^{\mathrm{cont}} \rangle$ for a sample of lower-redshift SLACS lenses compared to their twin sample, but with larger uncertainties and a lower significance.  The relative overdensities of each lens are given in Table~\ref{tab:overdens}.

\begin{figure}
\plotone{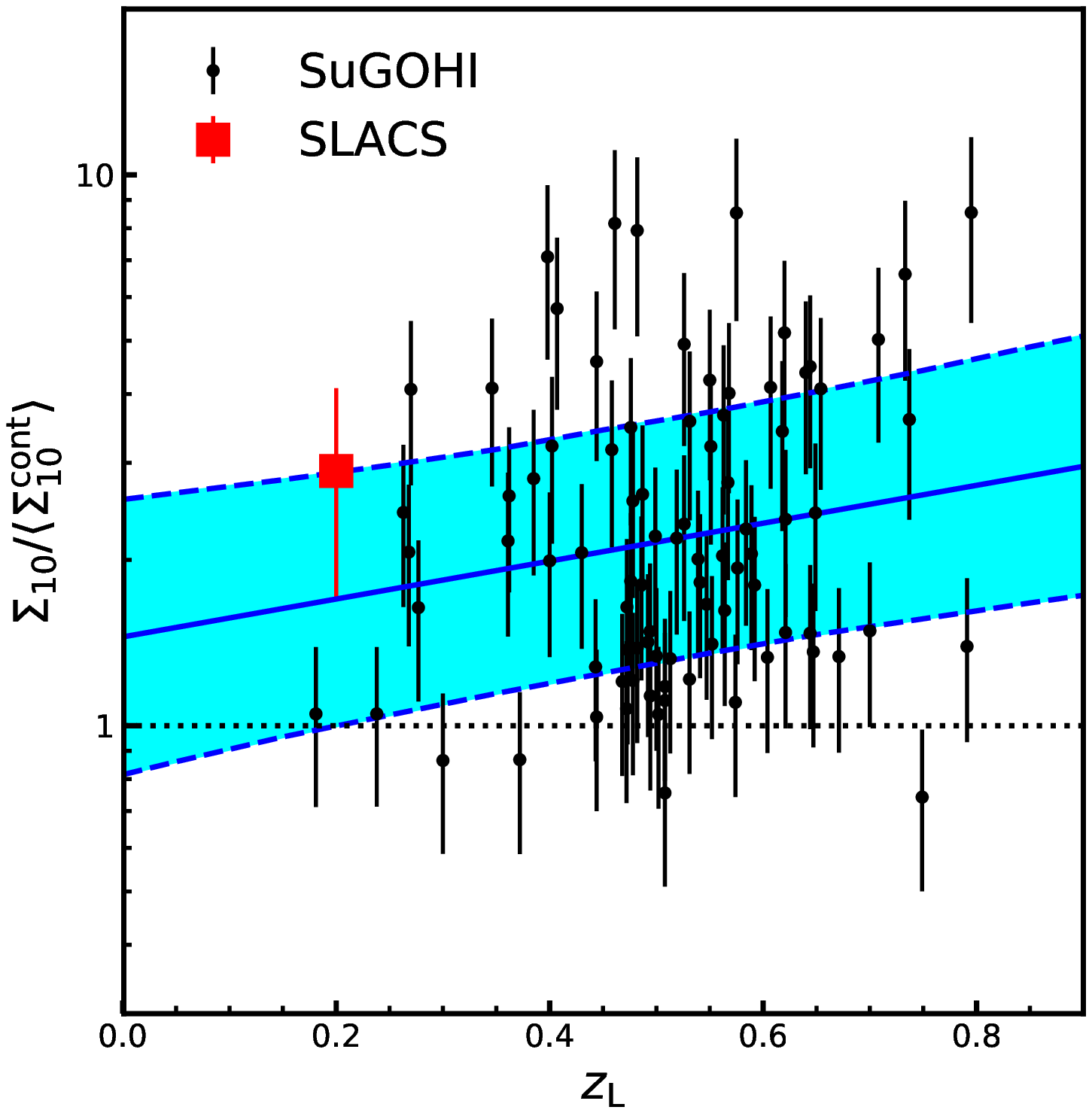}
\caption{
Normalized $\Sigma_{10}$ distributions for the individual SuGOHI-g lenses (black points) as a function of lens redshift.  The black dotted line represents an density equal to that of a random field.  For comparison, we plot the SLACS result of \citet{treu+2009} at the mean redshift of their sample (red square), although we note that they use a variable magnitude cut, so their results may not be exactly analogous to ours.  The blue line and shaded region represents the best-fit power-law relation and $68\%$ pointwise confidence interval using the methodology of \citet{kelly2007}.  We cannot rule out weak evolution, but the slope is still consistent with no redshift evolution in the relation.
\label{fig:z_od}}
\end{figure}

With the longer redshift baseline of the SuGOHI-g sample compared to previous lens samples, we investigate whether the relative overdensity of lens environments shows any trend with redshift.  In Figure~\ref{fig:z_od}, we plot the normalized $\Sigma_{10}$ values as a function of lens redshift for the individual systems.  There is considerable scatter, but we fit a power law to the points using the methodology of \citet{kelly2007} to account for intrinsic scatter.  The solid blue line is the best fit relation, with the blue band representing the $68\%$ pointwise confidence interval.  We cannot rule out weak evolution, but the slope is still consistent with no redshift evolution in the relation.

\subsection{Possible Sources of Systematic Error} \label{subsec:sys}

\subsubsection{Contamination by Stars} \label{subsubsec:stargal}
\begin{figure}
\plotone{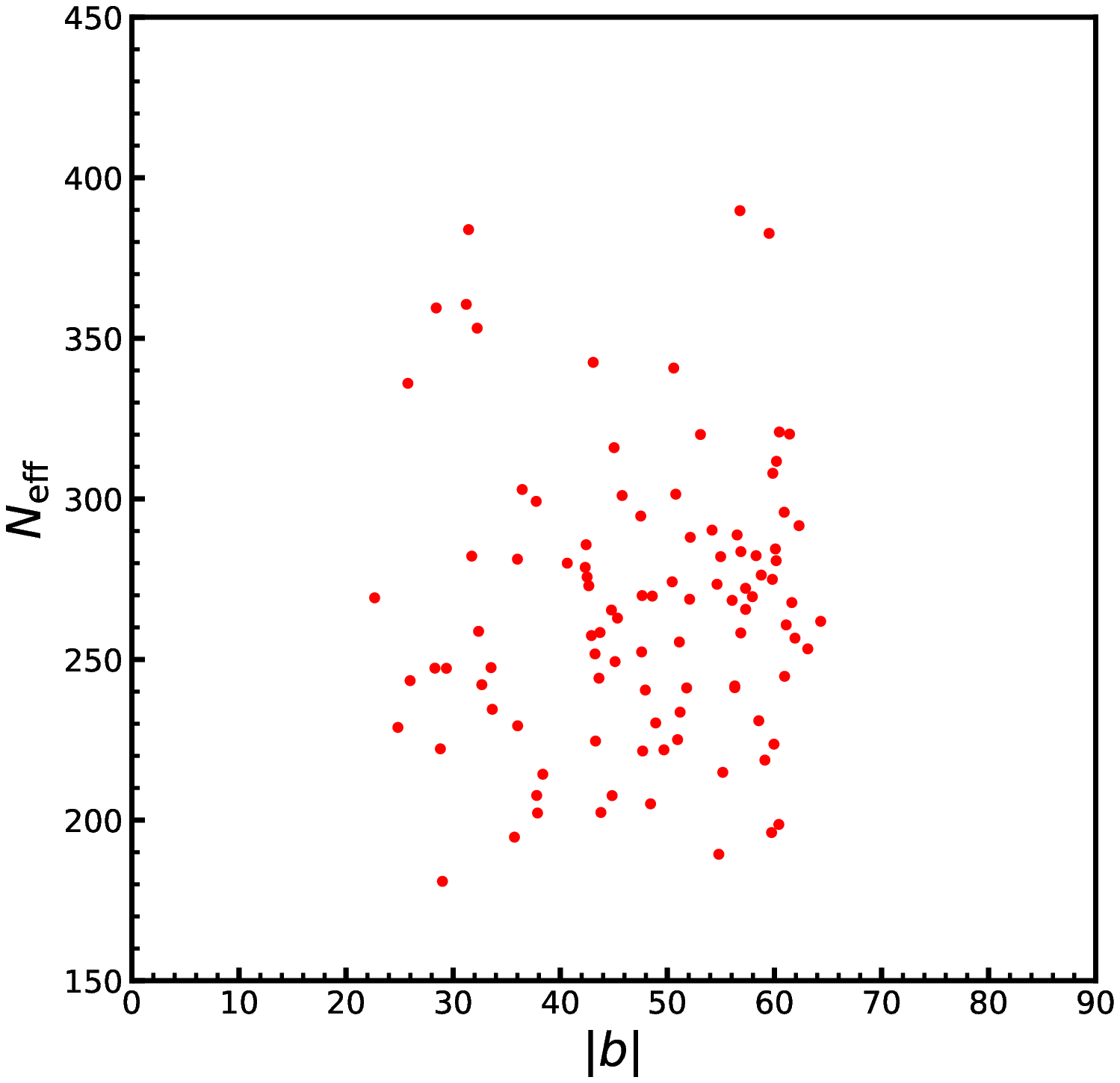}
\caption{
Mean $N_{\mathrm{eff}}$ for a limiting magnitude of $i = 24$ within $r \leq 120\arcsec$ of each control field as a function of distance from the Galactic equator, $|b|$.  The effective number counts are not systematically higher at low Galactic latitudes, suggesting that our extendedness cut is adequately selecting galaxies.
\label{fig:cont_gal_lat}}
\end{figure}

Although the star/galaxy classification should be reliable to $i \sim 24$ \citep{aihara+2018b}, we check for possible contamination of our galaxy sample by interloping Milky Way stars that may not have been excluded by our extendedness cut.  To do this, we plot the mean $N_{\mathrm{eff}}$ for a limiting magnitude of $i = 24$ within $r \leq 120\arcsec$ of each control field as a function of distance from the Galactic equator, $|b|$, in Figure~\ref{fig:cont_gal_lat}.  If there is contamination from stars, there should be a higher galaxy density at lower Galactic latitude.  In fact, there is no such effect, suggesting that our extendedness cut is adequately selecting galaxies.

\subsubsection{Magnification Bias} \label{subsubsec:mag_bias}
One might suspect that since lenses reside in locally overdense environments, there may be enhancement or depletion of LOS galaxy number counts due to lensing magnification effects.  The direction and size of the effect depends on the logarithmic slope of the cumulative number counts of background galaxies at the limiting magnitude being considered.  For a log slope steeper than the lensing-invariant slope of $s = 0.4$, there should be an enhancement, while for a log slope shallower than $s = 0.4$, there should be a depletion \citep[e.g.,][]{chiu+2016}.

\begin{figure}
\plotone{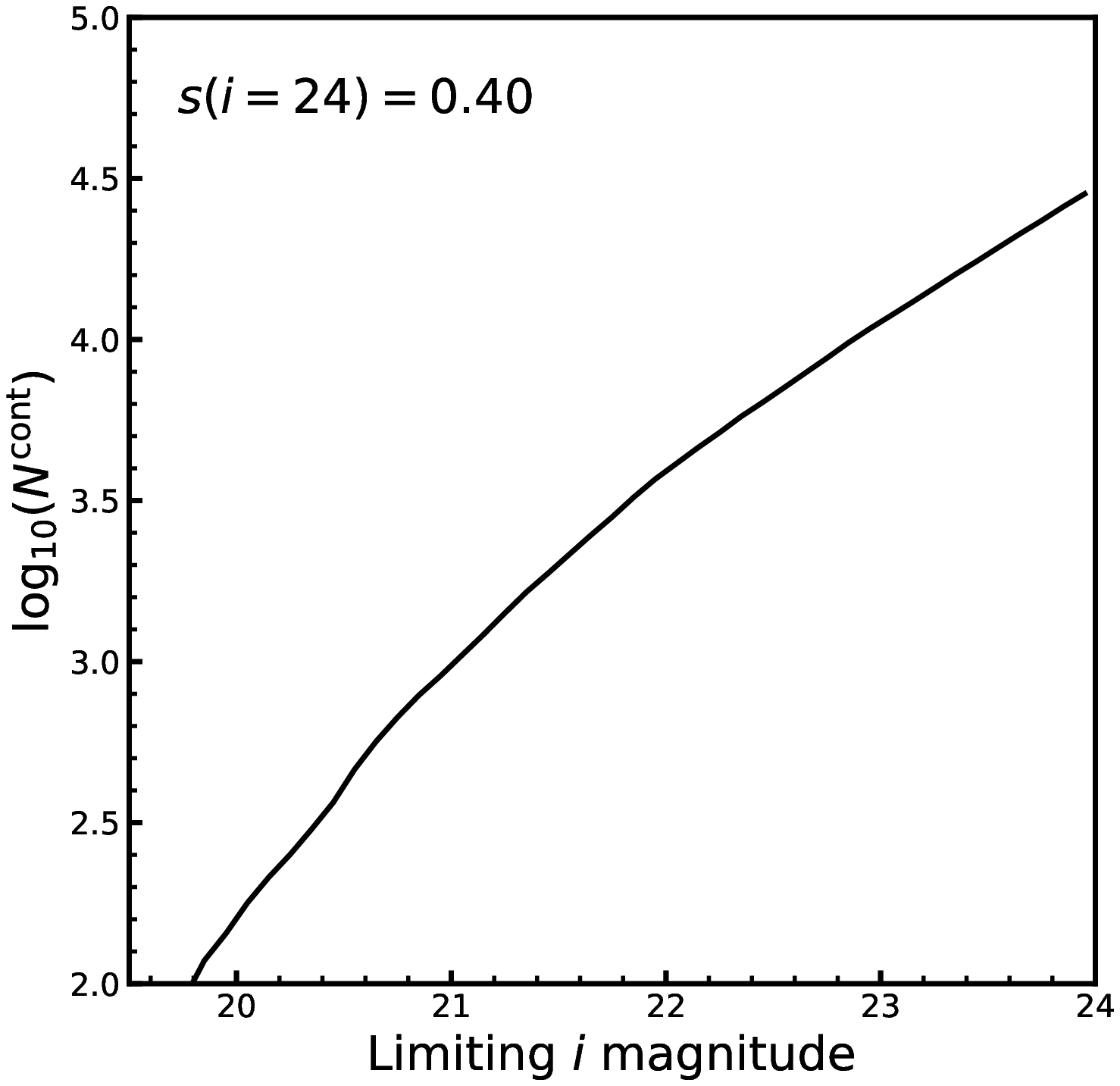}
\caption{
Cumulative number counts as a function of limiting magnitude for all galaxies within an aperture of $r \leq 120\arcsec$ centered on the control fields that have a photometric redshift higher the mean redshift of the lens sample, $z \geq \zlavg$.  The logarithmic slope of the cumulative number counts at $i=24$ is $s = 0.40$.  This matches the lensing-invariant slope of $s = 0.4$, so there magnification bias arising from overdensities associated with the lens environment should have a minimal effect.
\label{fig:ncum_maglim}}
\end{figure}

In Figure~\ref{fig:ncum_maglim}, we plot the cumulative number counts as a function of limiting magnitude for all galaxies within an aperture of $r \leq 120\arcsec$ centered on the control fields that have a photometric redshift higher than the mean redshift of the lenses in our sample.  We find that the log slope at $i = 24$ is $s = 0.40$, which matches the lensing-invariant value, implying that magnification bias is not a significant effect.  The slope steepens toward brighter limiting magnitudes, so in principle there could be a slight enhancement in the galaxy number counts at these brighter limits due to magnification effects.  However, this will be diluted by lower-redshift galaxies that are either unaffected by magnification from the lens galaxy and its local environment (since they are in the foreground) or have a low lensing efficiency (since $D_{\mathrm{LS}} / D_{\mathrm{S}}$ is small).  This effect is also offset by the fact that samples of brighter objects will contain a higher fraction of objects at lower redshift.

We run a simple test to determine how large an effect lensing magnification could have.  Using the {\sc Gravlens} software \citep{keeton2001}, we generate a mass model consisting of an NFW halo \citep{navarro+1997} at $z = \zlavg$, which is the mean redshift of our main lens sample.  The halo is given a mass of $10^{14}~\mathrm{M_{\odot}}$, which would represent a lens environment that is more massive than typical, but is still plausible given the velocity dispersion and stellar mass range of our lens sample.  The halo concentration parameter for this halo mass is taken from the results of simulations by \citet{zhao+2009}, and we assume an ellipticity of $\epsilon = 0.3$.  We calculate a source-plane magnification map centered on this model for a source redshift of $z_{\mathrm{S}} = 1.0$, which is close to the mean photometric redshift of the galaxies in the control sample that are at higher redshift than our assumed lens redshift.  We then place galaxies from the control sample, which should be unbiased on average, at random locations on the source plane between $2\farcs5 \leq r \leq 120\arcsec$ and calculate their apparent magnitude after being magnified.  We compare the number density of galaxies in the unmagnified and magnified cases after accounting for depletion of source area in the fixed aperture due to lensing.  We find that even in this aggressive scenario that represents a rich lens environment, the typical effect is a $< 1\%$ change in number counts at $i \leq 23$, with a likely comparable change at $i \leq 24$ given the relatively constant slope in Figure~\ref{fig:ncum_maglim}.  We repeat this test for a $10^{15}~\mathrm{M_{\odot}}$ halo, which represents an extreme cluster environment, and find a similarly small enhancement.  We therefore conclude that the effect of magnification bias on galaxy number counts is negligible in the context of this study, although we cannot necessarily extend this to deeper surveys or lens samples with a significantly different redshift distribution.

\subsubsection{Effect of Photometric Redshift Outliers} \label{subsubsec:photoz_outlier}
The photometric redshifts used in our analysis have a scatter and a possibility of catastrophic outliers, as described in Section~\ref{subsec:hsc}.  Since lenses reside in locally overdense environments, we intuitively expect that redshift outliers will lead to an underestimate of the local overdensity of lens galaxies since more galaxies will scatter out of the lens plane than will scatter into it.  To demonstrate this, we show a simple model.

For a given lens at a redshift $z_{\mathrm{L}}$, let $N_{z = z_\mathrm{L}}^{\mathrm{lens}}$ and $N_{z \ne z_\mathrm{L}}^{\mathrm{lens}}$ be the number of objects in the lens plane and along the LOS, respectively, in the lens field.  Similarly, let $N_{z = z_\mathrm{L}}^{\mathrm{cont}}$ and $N_{z \ne z_\mathrm{L}}^{\mathrm{cont}}$ be the number of objects in the lens plane and along the LOS, respectively, in a control field.  The total number counts in the lens field is then $N^{\mathrm{lens}} = N_{z = z_\mathrm{L}}^{\mathrm{lens}} + N_{z \ne z_\mathrm{L}}^{\mathrm{lens}}$ and the total number counts in the control field is $N^{\mathrm{cont}} = N_{z = z_\mathrm{L}}^{\mathrm{cont}} + N_{z \ne z_\mathrm{L}}^{\mathrm{cont}}$.

We then allow for photometric redshift outliers and assume that the global outlier rate applies to both the lens and control fields.  Let $f_{\mathrm{in}}$ be the fraction of LOS galaxies that scatter into the lens plane, and $f_{\mathrm{out}}$ be the fraction of galaxies in the lens plane that scatter out of the lens plane and into the LOS.  Our intuitive conjecture is that the measured relative overdensity is smaller than the true overdensity.  In other words, we expect that
\begin{equation} \label{eq:true_meas_od}
\frac{(1-f_{\mathrm{out}}) N_{z = z_\mathrm{L}}^{\mathrm{lens}} + f_{\mathrm{in}} N_{z \ne z_\mathrm{L}}^{\mathrm{lens}}}{(1-f_{\mathrm{out}}) N_{z = z_\mathrm{L}}^{\mathrm{cont}} + f_{\mathrm{in}} N_{z \ne z_\mathrm{L}}^{\mathrm{cont}}} < \frac{N_{z = z_{\mathrm{L}}}^{\mathrm{lens}}}{N_{z = z_{\mathrm{L}}}^{\mathrm{cont}}}.
\end{equation}
Multiplying through Equation~\ref{eq:true_meas_od} by the denominators of both sides and eliminating common factors, this simplifies to
\begin{equation} \label{eq:od_simp1}
N_{z \ne z_\mathrm{L}}^{\mathrm{lens}} N_{z = z_\mathrm{L}}^{\mathrm{cont}} < N_{z \ne z_\mathrm{L}}^{\mathrm{cont}} N_{z = z_\mathrm{L}}^{\mathrm{lens}}.
\end{equation}
Rearranging to separate the lens field and control field quantities, we get
\begin{equation} \label{eq:od_simp2}
\frac{N_{z = z_\mathrm{L}}^{\mathrm{cont}}}{N_{z \ne z_\mathrm{L}}^{\mathrm{cont}}} < \frac{N_{z = z_\mathrm{L}}^{\mathrm{lens}}}{N_{z \ne z_\mathrm{L}}^{\mathrm{lens}}}.
\end{equation}
Equation~\ref{eq:od_simp2} is simply the statement that lenses lie in locally overdense redshift planes relative to a random control field, which our results have already demonstrated.  Therefore, our conjecture that photometric redshift outliers will lead to an underestimate of the relative local overdensity of lenses is true, making our results conservative.

Conversely, this means that there will be a bias in the calculation of $N_{\mathrm{eff}}^{\mathrm{LOS}}$ such that lens and twin fields will have slightly higher measured values compared to the control sample, i.e., the fact that Equation~\ref{eq:true_meas_od} is true necessarily means that
\begin{equation} \label{eq:true_meas_nefflos}
\frac{(1-f_{\mathrm{in}}) N_{z \ne z_\mathrm{L}}^{\mathrm{lens}} + f_{\mathrm{out}} N_{z = z_\mathrm{L}}^{\mathrm{lens}}}{(1-f_{\mathrm{in}}) N_{z \ne z_\mathrm{L}}^{\mathrm{cont}} + f_{\mathrm{out}} N_{z = z_\mathrm{L}}^{\mathrm{cont}}} > \frac{N_{z \ne z_{\mathrm{L}}}^{\mathrm{lens}}}{N_{z \ne z_{\mathrm{L}}}^{\mathrm{cont}}}.
\end{equation}

To estimate the magnitude of this effect, we first define $y$ to be the true overdensity in the lens plane,
\begin{equation} \label{eq:y}
y \equiv \frac{N_{z=z_{\mathrm{L}}}^{\mathrm{lens}}}{N_{z=z_{\mathrm{L}}}^{\mathrm{cont}}}.
\end{equation}
For small $\Delta z / (1+z_{\mathrm{L}})$, we can make the approximation $f_{\mathrm{in}} \approx 0$ because the total outlier fraction for LOS galaxies is small (of order $\sim0.1$) and the likelihood of an outlier scattering into the lens plane is small (of order $\sim \Delta z  / (1+z_{\mathrm{L}}) \sim 0.1$), so their product is small.  We can then define $x$ to be the measured LOS overdensity (the left-hand side of Equation~\ref{eq:true_meas_nefflos}) under this approximation,
\begin{align}
x &\equiv \frac{N_{z \ne z_\mathrm{L}}^{\mathrm{lens}} + f_{\mathrm{out}} N_{z = z_\mathrm{L}}^{\mathrm{lens}}}{N_{z \ne z_\mathrm{L}}^{\mathrm{cont}} + f_{\mathrm{out}} N_{z = z_\mathrm{L}}^{\mathrm{cont}}}\\
&= \frac{N_{z \ne z_\mathrm{L}}^{\mathrm{lens}} + y f_{\mathrm{out}} N_{z = z_\mathrm{L}}^{\mathrm{cont}}}{N_{z \ne z_\mathrm{L}}^{\mathrm{cont}} + f_{\mathrm{out}} N_{z = z_\mathrm{L}}^{\mathrm{cont}}}.\label{eq:x}
\end{align}
We can then rearrange Equation~\ref{eq:x} to calculate the true LOS overdensity,
\begin{equation}
\frac{N_{z \ne z_{\mathrm{L}}}^{\mathrm{lens}}}{N_{z \ne z_{\mathrm{L}}}^{\mathrm{cont}}} = x - f_{\mathrm{out}} \frac{N_{z = z_{\mathrm{L}}}^{\mathrm{cont}}}{N_{z \ne z_{\mathrm{L}}}^{\mathrm{cont}}} (y-x).
\end{equation}
$f_{\mathrm{out}}$ should be similar to the global value of galaxies that scatter beyond $\Delta z  / (1+z_{\mathrm{L}}) = 0.05$ and, given the typical $1\sigma$ redshift scatter, is approximately $f_{\mathrm{out}} \approx 0.3$.  $N_{z = z_{\mathrm{L}}}^{\mathrm{cont}} / N_{z \ne z_{\mathrm{L}}}^{\mathrm{cont}}$ is the ratio of number counts in the lens plane to number counts in the LOS for the control field and can be estimated from the fractional difference in the distributions for a given magnitude and aperture cut between Figures~\ref{fig:nhist_all} and~\ref{fig:nhist_los}.  This roughly gives $N_{z = z_{\mathrm{L}}}^{\mathrm{cont}} / N_{z \ne z_{\mathrm{L}}}^{\mathrm{cont}} \approx 0.2$.  $y$ can be estimated from Figures~\ref{fig:nhist_all} and~\ref{fig:nhist_los} for a given magnitude and aperture cut and is generally in the range $1.5 \lesssim y \lesssim 2$, and $x \approx 1$, thus $0.5 \lesssim (y-x) \lesssim 1$.  Therefore, the bias due to photometric redshift outliers is a few percent, which may explain the statistically similar but consistently larger LOS overdensities in the lens fields relative to the control fields in Figure~\ref{fig:nhist_los}.

\subsubsection{Magnitude-dependent Clustering Effects} \label{subsubsec:clustering}
Our adoption of a simple magnitude cut may lead to a systematic effect in our calculation of the relative overdensities of the local lens environments.  Although our normalization of $\Sigma_{10}$ by the average density of the control fields for an equivalent redshift slice should mitigate this, there could still be a second-order effect since galaxy clustering strength is known to be stronger for brighter and more massive galaxies \citep[e.g.,][]{zehavi+2005}.

\begin{figure}
\plotone{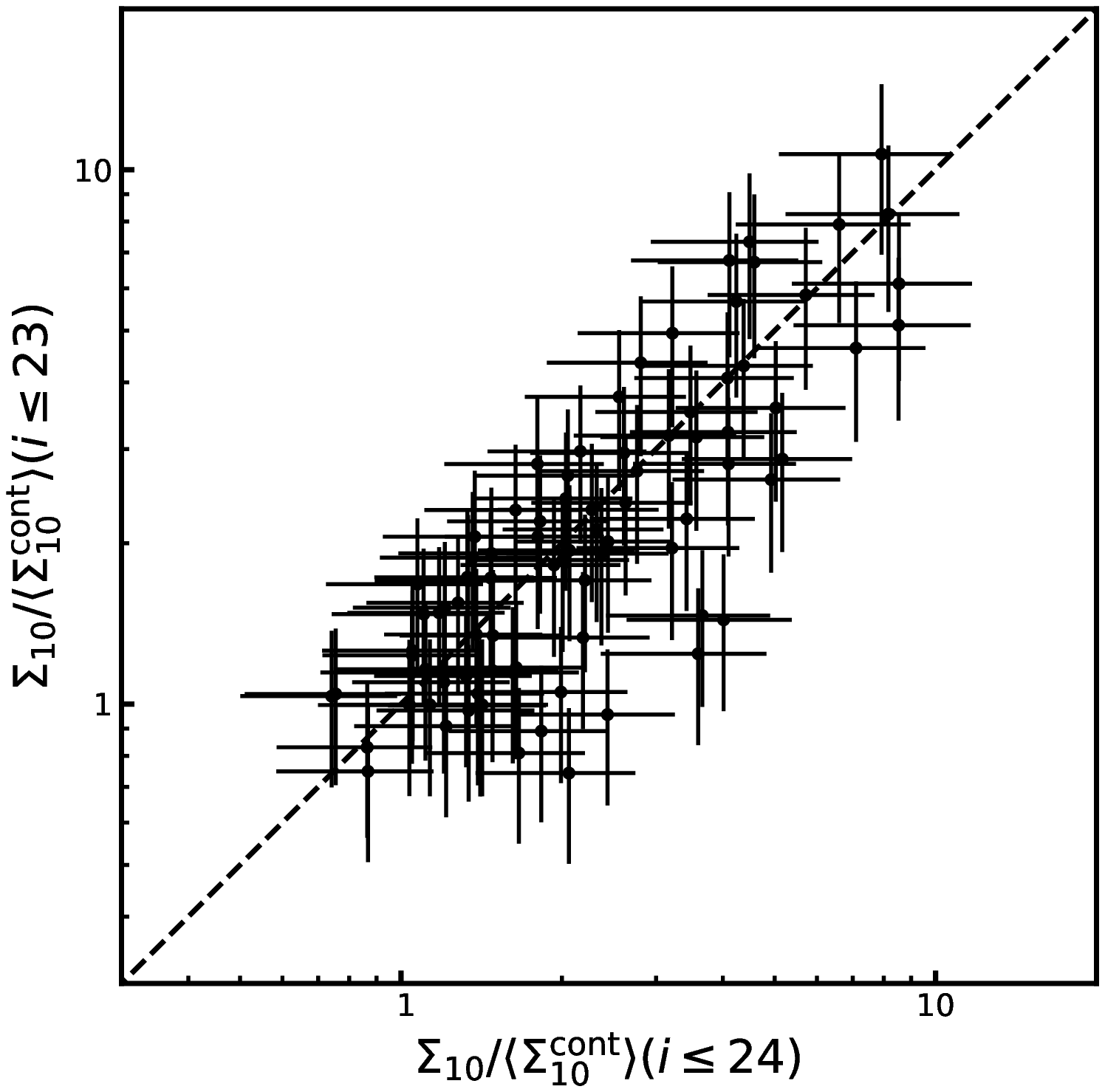}
\caption{
Normalized $\Sigma_{10}$ distributions for individual SuGOHI-g lenses calculated using galaxies brighter than $i \leq 23$ as a function of the same quantity calculated using galaxies brighter than $i \leq 24$.  Most of the normalized $\Sigma_{10}$ values remain consistent within the uncertainties, and there is no systematic trend toward higher overdensities.  Therefore, there does not appear to be a bias arising from our choice of magnitude cut.
\label{fig:sig10_maglim}}
\end{figure}

In Figure~\ref{fig:sig10_maglim}, we plot the relative overdensities of our lens sample against the same quantity calculated using galaxies at $i \leq 23$.  When calculated with the brighter magnitude limit, most of the relative overdensities remain consistent within the uncertainties, and there is no systematic trend toward higher overdensities.  Therefore, there does not appear to be a bias arising from our choice of magnitude cut.

\section{Conclusions} \label{sec:conclusion}
We have presented newly-discovered SuGOHI-g lens candidates from Data Release 2 of the HSC SSP.  In total, we find 41 promising new candidates, comprising 7 ``definite" (grade A) lenses and 34 ``probable" (grade B) lenses.

Using these data, along with previously known and discovered lenses in the survey, we have studied lines of sight of \nmain~strong gravitational lens galaxies to investigate whether they are overdense in comparison to random lines of sight in the Universe.  This study improves on a past analysis by \citet{fassnacht+2011} by taking advantage of the substantial area and multiband nature of the HSC SSP, which allows us to use photometric redshifts to accurately distinguish galaxies in the local lens environment from LOS galaxies in projection.  Our results are in agreement with \citet{fassnacht+2011} in that lens galaxies lie in overdense LOS compared to random, but this effect can be explained by the overdense local environments of lenses.  Once the galaxies in the lens plane are removed, the lens LOS overdensities are consistent with the control sample.  There may be a magnification bias due to the overdense lens environments, but the effect is likely small and is not strong enough to influence our results.  Photometric redshift scatter can lead to a slight overestimate of the lens fields' LOS number counts, but this is likely only a few percent effect.

We have also investigated the local environments of lens galaxies by comparing them to both a random control sample and a sample of ``twin" galaxies that are matched in redshift and velocity dispersion.  This extends the work of \citet{treu+2009}, who used the SLACS sample that had a mean lens redshift of $z_{\mathrm{L}} = 0.2$, out to a redshift of $z_{\mathrm{L}} = 0.8$.  We find that lenses have denser local environments than random fields at the same redshift.  There is indication that their environments are slightly more dense than the twin galaxies, although the statistical significance is marginal, in agreement with \citet{treu+2009}.  A bias in the velocity dispersion measurements of the lens galaxies may explain this result, but higher-precision velocity dispersion measurements are needed for confirmation.  There is no evidence for redshift evolution in the relative overdensity of lens environments.

We have calculated the local and LOS relative overdensities for known galaxy-scale lenses in the HSC SSP through Data Release 2.  These quantities will be valuable for evaluating the potential effects of external structure and biases that can arise in lens modeling from those effects, as well as for studying the relationship between the environments of lens galaxies and their physical properties as inferred from lens modeling.  This work will also inform future studies of this lens sample in terms of follow-up strategies and prioritizing optimal targets.

\acknowledgments
We thank the referee, whose feedback and suggestions were helpful in improving this paper.
We thank Jean Coupon for help with the HSC SSP bright star masks and random object catalog.
We thank Chris Fassnacht for providing us with his galaxy catalog to directly compare number counts in overlapping fields.
The Hyper Suprime-Cam (HSC) collaboration includes the astronomical
communities of Japan and Taiwan, and Princeton University. The HSC instrumentation
and software were developed by the National Astronomical
Observatory of Japan (NAOJ), the Kavli Institute for the Physics and
Mathematics of the Universe (Kavli IPMU), the University of Tokyo, the
High Energy Accelerator Research Organization (KEK), the Academia
Sinica Institute for Astronomy and Astrophysics in Taiwan (ASIAA), and
Princeton University. Funding was contributed by the FIRST program
from Japanese Cabinet Office, the Ministry of Education, Culture, Sports,
Science and Technology (MEXT), the Japan Society for the Promotion of
Science (JSPS), Japan Science and Technology Agency (JST), the Toray
Science Foundation, NAOJ, Kavli IPMU, KEK, ASIAA, and Princeton
University.
Based in part on data collected at the Subaru Telescope and retrieved from the HSC data archive system, which is operated by the Subaru Telescope and Astronomy Data Center at National Astronomical Observatory of Japan.
Funding for SDSS-III has been provided by the Alfred P. Sloan Foundation, the Participating Institutions, the National Science Foundation, and the U.S. Department of Energy Office of Science. The SDSS-III web site is http://www.sdss3.org/.
SDSS-III is managed by the Astrophysical Research Consortium for the Participating Institutions of the SDSS-III Collaboration including the University of Arizona, the Brazilian Participation Group, Brookhaven National Laboratory, Carnegie Mellon University, University of Florida, the French Participation Group, the German Participation Group, Harvard University, the Instituto de Astrofisica de Canarias, the Michigan State/Notre Dame/JINA Participation Group, Johns Hopkins University, Lawrence Berkeley National Laboratory, Max Planck Institute for Astrophysics, Max Planck Institute for Extraterrestrial Physics, New Mexico State University, New York University, Ohio State University, Pennsylvania State University, University of Portsmouth, Princeton University, the Spanish Participation Group, University of Tokyo, University of Utah, Vanderbilt University, University of Virginia, University of Washington, and Yale University. 
K.C.W. is supported by an EACOA Fellowship awarded by the East Asia Core Observatories Association, which consists of the Academia Sinica Institute of Astronomy and Astrophysics, the National Astronomical Observatory of Japan, the National Astronomical Observatories of the Chinese Academy of Sciences, and the Korea Astronomy and Space Science Institute.
This work was supported in part by World Premier International Research Center Initiative (WPI Initiative), MEXT, Japan, and JSPS KAKENHI Grant Number JP15H05892, and JP18K03693.
S.H.S. thanks the Max Planck Society for support through the Max Planck Research Group.

\appendix
\section{Lenses Not In the Main Sample and Comparison with Previous Studies} \label{app:badvd}
Some known lenses are not included in our main sample because they either do not have BOSS spectroscopy or do not have reliable velocity dispersion measurements from BOSS (Section~\ref{subsec:lens}), so we cannot select a sample of twin galaxies for them.  There may also be previously known lenses that are contained within the survey area, but that were not in the lens selection function.  These lenses are listed in Table~\ref{tab:lens_summary_badvd}.  However, we can still calculate $N_{\mathrm{eff}}$ and the local environment overdensity for these systems.  For completeness, we present these results in Table~\ref{tab:overdens_badvd}.  Although these systems could have been used in the parts of our analysis that did not make a direct comparison to the twin sample, we exclude them in order to maintain a consistent main sample throughout.  Some of these lenses (as well as some lenses in the main sample) may also have independent velocity dispersion measurements, but for consistency, we only use the velocity dispersions from BOSS.

We note that two lenses outside of our main sample, SDSS J0924+0219 and SDSS J1226$-$0006, are also in the \citet{fassnacht+2011} sample.  Our calculated relative LOS overdensity is consistent with theirs for SDSS J1226$-$0006, but slightly lower than their value (1.20 in a 45\arcsec~radius aperture) for SDSS J0924+0219, although their errors are likely large due to their small control sample.  The \citet{fassnacht+2011} study goes to a deeper limiting magnitude of $m_{\mathrm{F814W}} \leq 24.25$ on the Vega magnitude system, which could also affect the number counts (C. Fassnacht, private communication).

Among our entire sample (including those outside the main sample), there are eight lenses that overlap the \citet{treu+2009} SLACS sample.  Our calculated relative local overdensities for six of those systems are consistent with their results within the $1\sigma$ uncertainties.  Of the two systems where our results do not agree, SDSS J0935$-$0003 is consistent within $2\sigma$, and SDSS J1250$-$0135 is in a dense cluster environment at a low redshift of $z = 0.087$ where the {\sc mizuki} photometric redshifts have a higher catastrophic outlier rate \citep{tanaka+2018}.

\renewcommand*\arraystretch{1.2}
\begin{table*}
\caption{Additional Lenses not in Main Sample\label{tab:lens_summary_badvd}}
\begin{ruledtabular}
\begin{tabular}{l|ccccc}
Lens &
$\alpha$ (J2000) &
$\delta$ (J2000) &
$z_{\mathrm{L}}$ &
$z_{\mathrm{S}}$ &
Reference\tablenotemark{a}
\\
\tableline
SL2SJ021247$-$055552 & 
 33.1993 & 
 $-$5.9312 & 
0.750 & 
2.74 &
\citet{sonnenfeld+2013} 
\\
SL2SJ021411$-$040502 & 
 33.5467 & 
 $-$4.0841 & 
0.608 & 
1.880 &
\citet{sonnenfeld+2013} 
\\
SL2SJ022056$-$063934 & 
 35.2358 & 
 $-$6.6595 & 
0.330 & 
$-$ &
\citet{sonnenfeld+2013} 
\\
SL2SJ022459$-$040104 & 
 36.2469 & 
 $-$4.0177 & 
0.800 & 
$-$ &
\citet{more+2012} 
\\
SL2SJ022648$-$040610 & 
 36.7016 & 
 $-$4.1029 & 
0.766 & 
$-$ &
\citet{sonnenfeld+2013} 
\\
SL2SJ023251$-$040823 & 
 38.2149 & 
 $-$4.1399 & 
0.352 & 
2.344 &
\citet{sonnenfeld+2013} 
\\
SDSSJ023740.63$-$064112.9 & 
 39.4193 & 
 $-$6.6869 & 
0.486 & 
2.2491 &
\citet{shu+2016} 
\\
SL2SJ085540$-$014730 & 
133.9171 & 
 $-$1.7917 & 
0.365 & 
3.39 &
\citet{sonnenfeld+2013} 
\\
SL2SJ085826$-$014300 & 
134.6108 & 
 $-$1.7168 & 
0.580 & 
$-$ &
\citet{sonnenfeld+2013} 
\\
SL2SJ090408$-$005953 & 
136.0332 & 
 $-$0.9980 & 
0.611 & 
2.36 &
\citet{sonnenfeld+2013} 
\\
HSCJ090613+032939 & 
136.5548 & 
  3.4944 & 
0.617 & 
$-$ &
SuGOHI I 
\\
H-ATLASJ090740.0$-$004200 & 
136.9167 & 
 $-$0.7000 & 
0.613 & 
1.5747 &
\citet{negrello+2010,wong+2017a} 
\\
SDSSJ0912+0029 & 
138.0221 & 
  0.4837 & 
0.164 & 
0.324 &
\citet{bolton+2006} 
\\
SDSSJ0924+0219 & 
141.2325 & 
  2.3236 & 
0.394 & 
1.524 &
\citet{inada+2003,eigenbrod+2006} 
\\
SDSSJ0935$-$0003 & 
143.9331 & 
 $-$0.0597 & 
0.347 & 
0.467 &
\citet{bolton+2008} 
\\
BRI0952$-$0115 & 
148.7500 & 
 $-$1.5014 & 
0.632 & 
4.5 &
\citet{mcmahon+1992,eigenbrod+2007} 
\\
SDSSJ0955+0101 & 
148.8322 & 
  1.0290 & 
0.111 & 
0.316 &
\citet{bolton+2008} 
\\
COSMOS5914+1219 & 
149.7696 & 
  2.2053 & 
1.130 & 
$-$ &
\citet{faure+2008} 
\\
COSMOS5921+0638 & 
149.8407 & 
  2.1107 & 
0.552 & 
3.35 &
\citet{faure+2008,sonnenfeld+2013} 
\\
J095930.93+023427.7 & 
149.8789 & 
  2.5744 & 
0.892 & 
$-$ &
\citet{faure+2011} 
\\
COSMOS5939+3044 & 
149.9132 & 
  2.5122 & 
0.740 & 
$-$ &
\citet{faure+2008,more+2012} 
\\
COSMOS0012+2015 & 
150.0525 & 
  2.3375 & 
0.378 & 
$-$ &
\citet{faure+2011} 
\\
COSMOS0018+3845 & 
150.0767 & 
  2.6458 & 
0.976 & 
3.96 &
\citet{faure+2011} 
\\
COSMOS0038+4133 & 
150.1595 & 
  2.6927 & 
0.738 & 
$-$ &
\citet{faure+2011} 
\\
COSMOS0047+5023 & 
150.1983 & 
  1.8397 & 
0.870 & 
$-$ &
\citet{faure+2011} 
\\
COSMOS0049+5128 & 
150.2053 & 
  1.8578 & 
0.337 & 
0.524 &
\citet{faure+2011} 
\\
COSMOS0050+4901 & 
150.2110 & 
  2.8172 & 
0.960 & 
$-$ &
\citet{faure+2011} 
\\
J100140.12+020040.9 & 
150.4172 & 
  2.0114 & 
0.879 & 
$-$ &
\citet{faure+2011} 
\\
SL2SJ100148+022207 & 
150.4491 & 
  2.3685 & 
0.690 & 
$-$ &
\citet{more+2012} 
\\
COSMOS0211+1139 & 
150.5467 & 
  2.1943 & 
0.920 & 
$-$ &
\citet{faure+2011,more+2012} 
\\
SL2SJ100212+022955 & 
150.5486 & 
  2.4987 & 
0.770 & 
$-$ &
\citet{more+2012} 
\\
SL2SJ100215+023736 & 
150.5619 & 
  2.6268 & 
0.650 & 
$-$ &
\citet{more+2012} 
\\
COSMOS0254+1430 & 
150.7253 & 
  2.2417 & 
0.417 & 
0.779 &
\citet{faure+2011} 
\\
SDSSJ1143$-$0144 & 
175.8735 & 
 $-$1.7417 & 
0.106 & 
0.402 &
\citet{bolton+2008} 
\\
HSTJ114331.46$-$014508.0 & 
175.8811 & 
 $-$1.7522 & 
0.104 & 
0.91 &
\citet{newton+2009} 
\\
SDSSJ1159$-$0007 & 
179.9360 & 
 $-$0.1245 & 
0.579 & 
1.346 &
\citet{brownstein+2012} 
\\
SDSSJ1215+0047 & 
183.7685 & 
  0.7906 & 
0.642 & 
1.297 &
\citet{brownstein+2012} 
\\
SDSSJ1226$-$0006 & 
186.5334 & 
 $-$0.1006 & 
0.516 & 
1.1229 &
\citet{eigenbrod+2006,inada+2008} 
\\
SDSSJ1250$-$0135 & 
192.7105 & 
 $-$1.5921 & 
0.087 & 
0.353 &
\citet{bolton+2008} 
\\
SDSSJ1347$-$0101 & 
206.7707 & 
 $-$1.0176 & 
0.397 & 
0.63 &
\citet{auger+2011} 
\\
SDSSJ1403+0006 & 
210.8729 & 
  0.1115 & 
0.189 & 
0.473 &
\citet{bolton+2008} 
\\
HSCJ141635+010128 & 
214.1476 & 
  1.0247 & 
0.700 & 
$-$ &
SuGOHI I 
\\
HSCJ142053+005620 & 
215.2234 & 
  0.9391 & 
0.616 & 
$-$ &
SuGOHI I 
\\
SDSSJ1436$-$0000 & 
219.1148 & 
 $-$0.0081 & 
0.285 & 
0.805 &
\citet{bolton+2008} 
\\
SDSSJ1524+4409 & 
231.1901 & 
 44.1638 & 
0.320 & 
1.21 &
\citet{oguri+2008} 
\\
HSCJ155319+431824 & 
238.3308 & 
 43.3068 & 
0.629 & 
$-$ &
SuGOHI I 
\\
B1600+434 & 
240.4167 & 
 43.2799 & 
0.414 & 
1.589 &
\citet{jackson+1995,fassnacht+1998} 
\\
SL2SJ220629+005728 & 
331.6225 & 
  0.9580 & 
0.704 & 
$-$ &
\citet{sonnenfeld+2013} 
\\
SL2SJ221326$-$000946 & 
333.3591 & 
 $-$0.1629 & 
0.338 & 
3.447 &
\citet{sonnenfeld+2013} 
\\
SL2SJ221929$-$001743 & 
334.8725 & 
 $-$0.2954 & 
0.289 & 
1.023 &
\citet{sonnenfeld+2013} 
\\
SL2SJ222148+011542 & 
335.4534 & 
  1.2619 & 
0.325 & 
2.35 &
\citet{sonnenfeld+2013} 
\\
SL2SJ222217+001202 & 
335.5735 & 
  0.2008 & 
0.436 & 
1.36 &
\citet{sonnenfeld+2013} 
\\
Q2237+0305 & 
340.1250 & 
  3.3580 & 
0.039 & 
1.695 &
\citet{huchra+1985} 
\\
SDSSJ2300+0022 & 
345.2214 & 
  0.3772 & 
0.228 & 
0.463 &
\citet{bolton+2006} 
\\
HSCJ233230+003821 & 
353.1289 & 
  0.6394 & 
0.623 & 
$-$ &
SuGOHI II 
\\
\end{tabular}
\end{ruledtabular}
\tablecomments{Lenses listed here either do not have velocity dispersion measurements from BOSS or have unreliable velocity dispersion measurements, so they are not included in our main analysis.}
\tablenotetext{1}{SuGOHI I = \citet{sonnenfeld+2018}; SuGOHI II = this work}
\end{table*}

\begin{table*}
\caption{Local and LOS Overdensity of Individual Lens Fields Not in Main Sample\label{tab:overdens_badvd}}
\begin{ruledtabular}
\begin{tabular}{l|cccc|c}
\multirow{2}{*}{Lens} &
$N_{\mathrm{eff}}/\langle N_{\mathrm{eff}}^{\mathrm{cont}} \rangle$ & 
$N_{\mathrm{eff}}/\langle N_{\mathrm{eff}}^{\mathrm{cont}} \rangle$ & 
$N_{\mathrm{eff}}/\langle N_{\mathrm{eff}}^{\mathrm{cont}} \rangle$ & 
$N_{\mathrm{eff}}/\langle N_{\mathrm{eff}}^{\mathrm{cont}} \rangle$ & 
\multirow{2}{*}{$\Sigma_{10}/\langle \Sigma_{10}^{\mathrm{cont}} \rangle$}
\\
&
($r \leq 30\arcsec$) & 
($r \leq 60\arcsec$) & 
($r \leq 90\arcsec$) & 
($r \leq 120\arcsec$) & 
\\
\tableline
SL2SJ021247$-$055552 & 
1.30 $\pm$ 0.32 & 
0.95 $\pm$ 0.13 & 
1.11 $\pm$ 0.10 & 
1.25 $\pm$ 0.08 & 
2.74 $\pm$ 0.92
\\
SL2SJ021411$-$040502 & 
1.85 $\pm$ 0.41 & 
1.34 $\pm$ 0.17 & 
1.20 $\pm$ 0.10 & 
1.13 $\pm$ 0.08 & 
4.18 $\pm$ 1.44
\\
SL2SJ022056$-$063934 & 
0.30 $\pm$ 0.14 & 
0.73 $\pm$ 0.11 & 
0.77 $\pm$ 0.08 & 
0.88 $\pm$ 0.06 & 
0.68 $\pm$ 0.22
\\
SL2SJ022459$-$040104 & 
0.84 $\pm$ 0.25 & 
0.97 $\pm$ 0.13 & 
0.98 $\pm$ 0.09 & 
0.94 $\pm$ 0.07 & 
0.78 $\pm$ 0.25
\\
SL2SJ022648$-$040610 & 
1.40 $\pm$ 0.34 & 
0.97 $\pm$ 0.14 & 
0.94 $\pm$ 0.09 & 
1.05 $\pm$ 0.07 & 
1.94 $\pm$ 0.65
\\
SL2SJ023251$-$040823 & 
1.80 $\pm$ 0.40 & 
1.08 $\pm$ 0.14 & 
1.03 $\pm$ 0.09 & 
1.05 $\pm$ 0.07 & 
1.58 $\pm$ 0.52
\\
SDSSJ023740.63$-$064112.9 & 
0.77 $\pm$ 0.23 & 
0.75 $\pm$ 0.12 & 
0.85 $\pm$ 0.08 & 
0.82 $\pm$ 0.06 & 
0.95 $\pm$ 0.31
\\
SL2SJ085540$-$014730 & 
1.30 $\pm$ 0.32 & 
1.17 $\pm$ 0.15 & 
1.25 $\pm$ 0.11 & 
1.02 $\pm$ 0.07 & 
1.34 $\pm$ 0.44
\\
SL2SJ085826$-$014300 & 
1.35 $\pm$ 0.33 & 
1.36 $\pm$ 0.17 & 
1.22 $\pm$ 0.10 & 
1.15 $\pm$ 0.08 & 
1.95 $\pm$ 0.64
\\
SL2SJ090408$-$005953 & 
1.27 $\pm$ 0.31 & 
1.29 $\pm$ 0.16 & 
1.08 $\pm$ 0.10 & 
0.98 $\pm$ 0.07 & 
0.61 $\pm$ 0.20
\\
HSCJ090613+032939 & 
1.81 $\pm$ 0.40 & 
1.47 $\pm$ 0.18 & 
1.30 $\pm$ 0.11 & 
1.15 $\pm$ 0.08 & 
1.25 $\pm$ 0.41
\\
H$-$ATLASJ090740.0$-$004200 & 
1.01 $\pm$ 0.27 & 
1.05 $\pm$ 0.14 & 
1.12 $\pm$ 0.10 & 
1.03 $\pm$ 0.07 & 
1.04 $\pm$ 0.34
\\
SDSSJ0912+0029 & 
1.73 $\pm$ 0.39 & 
1.53 $\pm$ 0.18 & 
1.40 $\pm$ 0.11 & 
1.25 $\pm$ 0.08 & 
2.28 $\pm$ 0.74
\\
SDSSJ0924+0219 & 
1.14 $\pm$ 0.30 & 
0.91 $\pm$ 0.13 & 
0.99 $\pm$ 0.09 & 
1.00 $\pm$ 0.07 & 
2.12 $\pm$ 0.70
\\
SDSSJ0935$-$0003 & 
2.48 $\pm$ 0.50 & 
2.23 $\pm$ 0.23 & 
1.64 $\pm$ 0.13 & 
1.46 $\pm$ 0.09 & 
5.64 $\pm$ 1.93
\\
BRI0952$-$0115 & 
0.41 $\pm$ 0.17 & 
0.75 $\pm$ 0.12 & 
0.80 $\pm$ 0.08 & 
0.88 $\pm$ 0.06 & 
0.77 $\pm$ 0.25
\\
SDSSJ0955+0101 & 
1.21 $\pm$ 0.31 & 
1.10 $\pm$ 0.15 & 
0.93 $\pm$ 0.09 & 
0.99 $\pm$ 0.07 & 
1.23 $\pm$ 0.40
\\
COSMOS5914+1219 & 
0.62 $\pm$ 0.21 & 
1.00 $\pm$ 0.14 & 
0.99 $\pm$ 0.09 & 
1.03 $\pm$ 0.07 & 
0.92 $\pm$ 0.30
\\
COSMOS5921+0638 & 
0.98 $\pm$ 0.28 & 
1.10 $\pm$ 0.15 & 
1.06 $\pm$ 0.10 & 
1.06 $\pm$ 0.07 & 
0.74 $\pm$ 0.24
\\
J095930.93+023427.7 & 
1.75 $\pm$ 0.39 & 
1.45 $\pm$ 0.17 & 
1.20 $\pm$ 0.10 & 
1.19 $\pm$ 0.08 & 
0.53 $\pm$ 0.17
\\
COSMOS5939+3044 & 
1.81 $\pm$ 0.39 & 
1.89 $\pm$ 0.21 & 
1.79 $\pm$ 0.13 & 
1.68 $\pm$ 0.10 & 
5.77 $\pm$ 2.05
\\
COSMOS0012+2015 & 
1.90 $\pm$ 0.42 & 
1.53 $\pm$ 0.18 & 
1.27 $\pm$ 0.11 & 
1.30 $\pm$ 0.08 & 
0.79 $\pm$ 0.26
\\
COSMOS0018+3845 & 
1.37 $\pm$ 0.33 & 
1.60 $\pm$ 0.18 & 
1.46 $\pm$ 0.12 & 
1.38 $\pm$ 0.09 & 
1.39 $\pm$ 0.46
\\
COSMOS0038+4133 & 
2.80 $\pm$ 0.54 & 
1.77 $\pm$ 0.19 & 
1.45 $\pm$ 0.12 & 
1.39 $\pm$ 0.09 & 
1.61 $\pm$ 0.53
\\
COSMOS0047+5023 & 
1.44 $\pm$ 0.35 & 
0.95 $\pm$ 0.13 & 
1.09 $\pm$ 0.10 & 
1.09 $\pm$ 0.07 & 
4.62 $\pm$ 1.60
\\
COSMOS0049+5128 & 
0.80 $\pm$ 0.24 & 
0.91 $\pm$ 0.13 & 
1.04 $\pm$ 0.10 & 
1.05 $\pm$ 0.07 & 
0.62 $\pm$ 0.20
\\
COSMOS0050+4901 & 
1.81 $\pm$ 0.38 & 
1.50 $\pm$ 0.18 & 
1.27 $\pm$ 0.11 & 
1.19 $\pm$ 0.08 & 
5.01 $\pm$ 1.73
\\
J100140.12+020040.9 & 
2.14 $\pm$ 0.44 & 
1.55 $\pm$ 0.18 & 
1.32 $\pm$ 0.11 & 
1.27 $\pm$ 0.08 & 
1.65 $\pm$ 0.54
\\
SL2SJ100148+022207 & 
2.07 $\pm$ 0.44 & 
1.48 $\pm$ 0.18 & 
1.26 $\pm$ 0.11 & 
1.22 $\pm$ 0.08 & 
1.78 $\pm$ 0.59
\\
COSMOS0211+1139 & 
1.97 $\pm$ 0.42 & 
1.61 $\pm$ 0.19 & 
1.63 $\pm$ 0.13 & 
1.36 $\pm$ 0.08 & 
5.04 $\pm$ 1.74
\\
SL2SJ100212+022955 & 
1.24 $\pm$ 0.31 & 
1.35 $\pm$ 0.17 & 
1.28 $\pm$ 0.11 & 
1.24 $\pm$ 0.08 & 
1.35 $\pm$ 0.45
\\
SL2SJ100215+023736 & 
1.10 $\pm$ 0.29 & 
1.21 $\pm$ 0.15 & 
1.23 $\pm$ 0.10 & 
1.27 $\pm$ 0.08 & 
1.08 $\pm$ 0.35
\\
COSMOS0254+1430 & 
1.38 $\pm$ 0.34 & 
1.08 $\pm$ 0.14 & 
0.98 $\pm$ 0.09 & 
0.92 $\pm$ 0.07 & 
0.45 $\pm$ 0.15
\\
SDSSJ1143$-$0144 & 
1.04 $\pm$ 0.28 & 
1.24 $\pm$ 0.16 & 
1.08 $\pm$ 0.10 & 
1.09 $\pm$ 0.07 & 
3.34 $\pm$ 1.09
\\
HSTJ114331.46$-$014508.0 & 
1.02 $\pm$ 0.28 & 
1.02 $\pm$ 0.14 & 
1.08 $\pm$ 0.10 & 
1.05 $\pm$ 0.07 & 
3.51 $\pm$ 1.15
\\
SDSSJ1159$-$0007 & 
2.00 $\pm$ 0.43 & 
1.45 $\pm$ 0.17 & 
1.14 $\pm$ 0.10 & 
1.18 $\pm$ 0.08 & 
2.24 $\pm$ 0.74
\\
SDSSJ1215+0047 & 
1.55 $\pm$ 0.36 & 
1.23 $\pm$ 0.16 & 
1.15 $\pm$ 0.10 & 
1.12 $\pm$ 0.08 & 
2.00 $\pm$ 0.67
\\
SDSSJ1226$-$0006 & 
1.19 $\pm$ 0.31 & 
1.08 $\pm$ 0.15 & 
1.23 $\pm$ 0.11 & 
1.19 $\pm$ 0.08 & 
0.43 $\pm$ 0.14
\\
SDSSJ1250$-$0135 & 
1.43 $\pm$ 0.35 & 
1.14 $\pm$ 0.15 & 
1.29 $\pm$ 0.11 & 
1.34 $\pm$ 0.08 & 
0.49 $\pm$ 0.16
\\
SDSSJ1347$-$0101 & 
1.06 $\pm$ 0.28 & 
0.88 $\pm$ 0.13 & 
0.97 $\pm$ 0.09 & 
0.98 $\pm$ 0.07 & 
0.54 $\pm$ 0.18
\\
SDSSJ1403+0006 & 
1.96 $\pm$ 0.42 & 
1.47 $\pm$ 0.18 & 
1.38 $\pm$ 0.11 & 
1.25 $\pm$ 0.08 & 
2.12 $\pm$ 0.69
\\
HSCJ141635+010128 & 
2.09 $\pm$ 0.45 & 
1.63 $\pm$ 0.19 & 
1.44 $\pm$ 0.12 & 
1.26 $\pm$ 0.08 & 
3.68 $\pm$ 1.27
\\
HSCJ142053+005620 & 
1.50 $\pm$ 0.36 & 
1.28 $\pm$ 0.16 & 
1.28 $\pm$ 0.11 & 
1.23 $\pm$ 0.08 & 
2.34 $\pm$ 0.78
\\
SDSSJ1436$-$0000 & 
1.46 $\pm$ 0.34 & 
1.14 $\pm$ 0.15 & 
0.99 $\pm$ 0.09 & 
1.07 $\pm$ 0.07 & 
1.47 $\pm$ 0.47
\\
SDSSJ1524+4409 & 
1.15 $\pm$ 0.30 & 
1.45 $\pm$ 0.17 & 
1.37 $\pm$ 0.11 & 
1.24 $\pm$ 0.08 & 
1.60 $\pm$ 0.52
\\
HSCJ155319+431824 & 
1.27 $\pm$ 0.32 & 
1.17 $\pm$ 0.15 & 
1.16 $\pm$ 0.10 & 
1.10 $\pm$ 0.07 & 
1.95 $\pm$ 0.65
\\
B1600+434 & 
0.91 $\pm$ 0.26 & 
1.01 $\pm$ 0.14 & 
1.07 $\pm$ 0.10 & 
0.97 $\pm$ 0.07 & 
0.51 $\pm$ 0.16
\\
SL2SJ220629+005728 & 
0.92 $\pm$ 0.27 & 
1.02 $\pm$ 0.14 & 
0.91 $\pm$ 0.09 & 
0.93 $\pm$ 0.07 & 
0.58 $\pm$ 0.19
\\
SL2SJ221326$-$000946 & 
2.35 $\pm$ 0.48 & 
1.64 $\pm$ 0.19 & 
1.41 $\pm$ 0.11 & 
1.27 $\pm$ 0.08 & 
7.47 $\pm$ 2.59
\\
SL2SJ221929$-$001743 & 
1.80 $\pm$ 0.40 & 
1.24 $\pm$ 0.16 & 
1.11 $\pm$ 0.10 & 
1.10 $\pm$ 0.07 & 
0.78 $\pm$ 0.25
\\
SL2SJ222148+011542 & 
1.47 $\pm$ 0.35 & 
1.46 $\pm$ 0.18 & 
1.31 $\pm$ 0.11 & 
1.10 $\pm$ 0.07 & 
1.44 $\pm$ 0.47
\\
SL2SJ222217+001202 & 
1.09 $\pm$ 0.29 & 
1.02 $\pm$ 0.14 & 
1.00 $\pm$ 0.09 & 
0.90 $\pm$ 0.07 & 
1.42 $\pm$ 0.46
\\
Q2237+0305 & 
1.35 $\pm$ 0.33 & 
1.05 $\pm$ 0.14 & 
0.99 $\pm$ 0.09 & 
0.90 $\pm$ 0.07 & 
0.76 $\pm$ 0.24
\\
SDSSJ2300+0022 & 
2.10 $\pm$ 0.43 & 
2.05 $\pm$ 0.22 & 
1.66 $\pm$ 0.13 & 
1.40 $\pm$ 0.09 & 
2.91 $\pm$ 0.95
\\
HSCJ233230+003821 & 
1.25 $\pm$ 0.31 & 
1.18 $\pm$ 0.15 & 
1.19 $\pm$ 0.10 & 
1.13 $\pm$ 0.08 & 
1.31 $\pm$ 0.43
\\
\end{tabular}
\end{ruledtabular}
\tablecomments{Lenses listed here either do not have velocity dispersion measurements from BOSS or have unreliable velocity dispersion measurements, so they are not included in our main analysis.  We list their local and LOS overdensities here for completeness.}
\end{table*}

\renewcommand*\arraystretch{1.0}

\section{Results Without Including Poisson Error on Number Counts} \label{app:no_poisson}
Tables~\ref{tab:overdens} and~\ref{tab:overdens_badvd} present the LOS and local overdensities of individual lens systems and their associated uncertainties.  Those uncertainties are very conservative since we include Poisson error on the raw counts, $N$, which are the main contributor to the error budget.  However, there may be certain applications of these overdensities where greater precision is desired, so we reproduce those values with uncertainties that do not include Poisson error on $N$ in Tables~\ref{tab:overdens_nopoisson} and~\ref{tab:overdens_badvd_nopoisson}.  We still account for Poisson error in $N_{\mathrm{rand}}$ and the error on the mean of the control sample.

\renewcommand*\arraystretch{0.7}
\begin{table*}
\caption{Local and LOS Overdensity of Individual Lens Fields (No Poisson Error)\label{tab:overdens_nopoisson}}
\begin{ruledtabular}
\begin{tabular}{l|cccc|c}
\multirow{2}{*}{Lens} &
$N_{\mathrm{eff}}/\langle N_{\mathrm{eff}}^{\mathrm{cont}} \rangle$ & 
$N_{\mathrm{eff}}/\langle N_{\mathrm{eff}}^{\mathrm{cont}} \rangle$ & 
$N_{\mathrm{eff}}/\langle N_{\mathrm{eff}}^{\mathrm{cont}} \rangle$ & 
$N_{\mathrm{eff}}/\langle N_{\mathrm{eff}}^{\mathrm{cont}} \rangle$ & 
\multirow{2}{*}{$\Sigma_{10}/\langle \Sigma_{10}^{\mathrm{cont}} \rangle$}
\\
&
($r \leq 30\arcsec$) & 
($r \leq 60\arcsec$) & 
($r \leq 90\arcsec$) & 
($r \leq 120\arcsec$) & 
\\
\tableline
HSCJ015731$-$033057 & 
1.33 $\pm$ 0.16 & 
1.48 $\pm$ 0.09 & 
1.45 $\pm$ 0.06 & 
1.26 $\pm$ 0.04 & 
2.37 $\pm$ 0.27
\\
HSCJ015756$-$021809 & 
1.68 $\pm$ 0.21 & 
1.10 $\pm$ 0.07 & 
1.02 $\pm$ 0.04 & 
0.95 $\pm$ 0.03 & 
0.87 $\pm$ 0.07
\\
SDSSJ0157$-$0056 & 
0.75 $\pm$ 0.09 & 
1.00 $\pm$ 0.06 & 
1.12 $\pm$ 0.05 & 
1.20 $\pm$ 0.04 & 
1.32 $\pm$ 0.11
\\
HSCJ020141$-$030946 & 
1.54 $\pm$ 0.19 & 
1.15 $\pm$ 0.07 & 
0.96 $\pm$ 0.04 & 
1.02 $\pm$ 0.03 & 
2.61 $\pm$ 0.26
\\
HSCJ020241$-$064611 & 
1.28 $\pm$ 0.15 & 
0.95 $\pm$ 0.06 & 
1.03 $\pm$ 0.05 & 
1.13 $\pm$ 0.04 & 
1.05 $\pm$ 0.08
\\
HSCJ020846$-$032727 & 
1.71 $\pm$ 0.20 & 
1.36 $\pm$ 0.08 & 
1.30 $\pm$ 0.06 & 
1.33 $\pm$ 0.04 & 
3.42 $\pm$ 0.44
\\
SL2SJ021737$-$051329 & 
1.90 $\pm$ 0.22 & 
1.53 $\pm$ 0.10 & 
1.19 $\pm$ 0.05 & 
1.10 $\pm$ 0.04 & 
4.49 $\pm$ 0.63
\\
HSCJ022140$-$021020 & 
1.26 $\pm$ 0.14 & 
0.82 $\pm$ 0.05 & 
0.86 $\pm$ 0.04 & 
0.86 $\pm$ 0.03 & 
5.02 $\pm$ 0.75
\\
SL2SJ022315$-$062906 & 
1.90 $\pm$ 0.25 & 
1.16 $\pm$ 0.07 & 
1.01 $\pm$ 0.04 & 
0.91 $\pm$ 0.03 & 
4.24 $\pm$ 0.55
\\
SL2SJ022346$-$053418 & 
1.80 $\pm$ 0.22 & 
1.16 $\pm$ 0.07 & 
1.06 $\pm$ 0.05 & 
1.09 $\pm$ 0.04 & 
2.21 $\pm$ 0.22
\\
SL2SJ022357$-$065142 & 
1.29 $\pm$ 0.14 & 
0.98 $\pm$ 0.06 & 
1.05 $\pm$ 0.04 & 
1.04 $\pm$ 0.03 & 
1.37 $\pm$ 0.11
\\
SL2SJ022439$-$040045 & 
1.35 $\pm$ 0.18 & 
0.87 $\pm$ 0.05 & 
0.96 $\pm$ 0.04 & 
1.05 $\pm$ 0.03 & 
2.06 $\pm$ 0.20
\\
SL2SJ022511$-$045433 & 
1.40 $\pm$ 0.16 & 
1.20 $\pm$ 0.07 & 
1.15 $\pm$ 0.05 & 
1.10 $\pm$ 0.04 & 
1.05 $\pm$ 0.06
\\
SL2SJ022610$-$042011 & 
1.56 $\pm$ 0.17 & 
0.93 $\pm$ 0.06 & 
0.91 $\pm$ 0.04 & 
0.97 $\pm$ 0.03 & 
1.13 $\pm$ 0.09
\\
HSCJ023217$-$021703 & 
1.21 $\pm$ 0.14 & 
1.14 $\pm$ 0.07 & 
1.15 $\pm$ 0.05 & 
1.02 $\pm$ 0.03 & 
1.11 $\pm$ 0.09
\\
SL2SJ023307$-$043838 & 
1.41 $\pm$ 0.17 & 
1.22 $\pm$ 0.08 & 
1.19 $\pm$ 0.05 & 
1.11 $\pm$ 0.04 & 
1.33 $\pm$ 0.13
\\
HSCJ023538$-$063406 & 
1.57 $\pm$ 0.18 & 
1.06 $\pm$ 0.06 & 
0.93 $\pm$ 0.04 & 
0.87 $\pm$ 0.03 & 
1.05 $\pm$ 0.07
\\
HSCJ023637$-$033220 & 
1.86 $\pm$ 0.22 & 
1.57 $\pm$ 0.10 & 
1.49 $\pm$ 0.06 & 
1.36 $\pm$ 0.04 & 
4.08 $\pm$ 0.39
\\
HSCJ023655$-$023656 & 
1.04 $\pm$ 0.13 & 
0.97 $\pm$ 0.06 & 
0.94 $\pm$ 0.04 & 
0.97 $\pm$ 0.03 & 
2.03 $\pm$ 0.20
\\
HSCJ083943+004740 & 
1.05 $\pm$ 0.13 & 
1.03 $\pm$ 0.06 & 
1.01 $\pm$ 0.04 & 
1.10 $\pm$ 0.04 & 
1.48 $\pm$ 0.14
\\
HSCJ085855$-$010208 & 
2.35 $\pm$ 0.29 & 
1.22 $\pm$ 0.08 & 
1.16 $\pm$ 0.05 & 
1.01 $\pm$ 0.03 & 
1.20 $\pm$ 0.10
\\
HSCJ090507$-$001030 & 
1.26 $\pm$ 0.15 & 
0.96 $\pm$ 0.06 & 
0.89 $\pm$ 0.04 & 
0.93 $\pm$ 0.03 & 
1.48 $\pm$ 0.13
\\
HSCJ090709+005648 & 
1.45 $\pm$ 0.17 & 
0.92 $\pm$ 0.06 & 
0.98 $\pm$ 0.04 & 
0.99 $\pm$ 0.03 & 
1.21 $\pm$ 0.10
\\
SDSSJ0915$-$0055 & 
2.61 $\pm$ 0.35 & 
1.96 $\pm$ 0.13 & 
1.52 $\pm$ 0.07 & 
1.38 $\pm$ 0.05 & 
3.22 $\pm$ 0.36
\\
HSCJ091608+034710 & 
1.13 $\pm$ 0.15 & 
1.21 $\pm$ 0.08 & 
1.09 $\pm$ 0.05 & 
1.06 $\pm$ 0.04 & 
3.57 $\pm$ 0.43
\\
HSCJ091904+033638 & 
1.47 $\pm$ 0.17 & 
1.24 $\pm$ 0.08 & 
1.09 $\pm$ 0.05 & 
1.01 $\pm$ 0.03 & 
4.57 $\pm$ 0.57
\\
HSCJ092101+035521 & 
1.59 $\pm$ 0.20 & 
1.00 $\pm$ 0.06 & 
0.83 $\pm$ 0.04 & 
0.81 $\pm$ 0.03 & 
1.64 $\pm$ 0.14
\\
HSCJ093506$-$020031 & 
1.42 $\pm$ 0.18 & 
0.85 $\pm$ 0.06 & 
0.72 $\pm$ 0.03 & 
0.76 $\pm$ 0.03 & 
1.21 $\pm$ 0.10
\\
HSCJ094123+000446 & 
1.32 $\pm$ 0.16 & 
1.00 $\pm$ 0.06 & 
0.94 $\pm$ 0.04 & 
0.88 $\pm$ 0.03 & 
1.80 $\pm$ 0.17
\\
SDSSJ0944$-$0147 & 
1.79 $\pm$ 0.22 & 
1.64 $\pm$ 0.10 & 
1.30 $\pm$ 0.06 & 
1.12 $\pm$ 0.04 & 
2.01 $\pm$ 0.19
\\
HSCJ095750+014507 & 
0.97 $\pm$ 0.12 & 
0.86 $\pm$ 0.05 & 
0.87 $\pm$ 0.04 & 
0.91 $\pm$ 0.03 & 
1.10 $\pm$ 0.09
\\
COSMOS0013+2249 & 
0.93 $\pm$ 0.11 & 
1.09 $\pm$ 0.07 & 
1.12 $\pm$ 0.05 & 
1.15 $\pm$ 0.04 & 
4.10 $\pm$ 0.48
\\
COSMOS0056+1226 & 
1.12 $\pm$ 0.12 & 
0.99 $\pm$ 0.06 & 
1.00 $\pm$ 0.04 & 
0.99 $\pm$ 0.03 & 
2.16 $\pm$ 0.20
\\
HSCJ100659+024735 & 
1.95 $\pm$ 0.24 & 
1.28 $\pm$ 0.08 & 
1.11 $\pm$ 0.05 & 
1.00 $\pm$ 0.03 & 
7.93 $\pm$ 1.32
\\
HSCJ114311$-$013935 & 
0.92 $\pm$ 0.11 & 
0.96 $\pm$ 0.06 & 
1.07 $\pm$ 0.05 & 
1.14 $\pm$ 0.04 & 
1.47 $\pm$ 0.14
\\
HSCJ115653$-$003948 & 
1.19 $\pm$ 0.14 & 
1.36 $\pm$ 0.08 & 
1.38 $\pm$ 0.06 & 
1.43 $\pm$ 0.05 & 
1.18 $\pm$ 0.09
\\
HSCJ120623+001507 & 
1.41 $\pm$ 0.17 & 
1.13 $\pm$ 0.07 & 
1.14 $\pm$ 0.05 & 
1.22 $\pm$ 0.04 & 
3.66 $\pm$ 0.45
\\
HSCJ121052$-$011905 & 
0.85 $\pm$ 0.11 & 
1.06 $\pm$ 0.07 & 
1.03 $\pm$ 0.04 & 
1.19 $\pm$ 0.04 & 
1.49 $\pm$ 0.15
\\
HSCJ122048+002145 & 
3.11 $\pm$ 0.38 & 
2.06 $\pm$ 0.13 & 
1.62 $\pm$ 0.07 & 
1.52 $\pm$ 0.05 & 
5.71 $\pm$ 0.78
\\
HSCJ122314$-$002939 & 
1.25 $\pm$ 0.18 & 
1.32 $\pm$ 0.09 & 
1.19 $\pm$ 0.05 & 
1.07 $\pm$ 0.04 & 
1.66 $\pm$ 0.15
\\
HSCJ123825+003212 & 
2.00 $\pm$ 0.25 & 
1.37 $\pm$ 0.08 & 
1.26 $\pm$ 0.05 & 
1.18 $\pm$ 0.04 & 
2.56 $\pm$ 0.27
\\
HSCJ124320$-$004517 & 
2.10 $\pm$ 0.27 & 
1.40 $\pm$ 0.09 & 
1.27 $\pm$ 0.05 & 
1.14 $\pm$ 0.04 & 
4.09 $\pm$ 0.56
\\
HSCJ125254+004356 & 
1.57 $\pm$ 0.19 & 
1.43 $\pm$ 0.09 & 
1.25 $\pm$ 0.05 & 
1.08 $\pm$ 0.04 & 
2.43 $\pm$ 0.28
\\
HSCJ134351+010817 & 
1.67 $\pm$ 0.20 & 
1.51 $\pm$ 0.10 & 
1.33 $\pm$ 0.06 & 
1.18 $\pm$ 0.04 & 
2.76 $\pm$ 0.31
\\
HSCJ135038+002550 & 
1.49 $\pm$ 0.19 & 
1.40 $\pm$ 0.09 & 
1.19 $\pm$ 0.05 & 
1.03 $\pm$ 0.03 & 
2.32 $\pm$ 0.24
\\
HSCJ135138+002839 & 
2.71 $\pm$ 0.35 & 
1.46 $\pm$ 0.09 & 
1.46 $\pm$ 0.06 & 
1.35 $\pm$ 0.05 & 
8.16 $\pm$ 1.37
\\
HSCJ135242$-$002614 & 
0.70 $\pm$ 0.08 & 
0.77 $\pm$ 0.05 & 
0.84 $\pm$ 0.04 & 
0.93 $\pm$ 0.03 & 
0.75 $\pm$ 0.05
\\
HSCJ135853$-$021525 & 
2.03 $\pm$ 0.26 & 
1.37 $\pm$ 0.09 & 
1.21 $\pm$ 0.05 & 
1.22 $\pm$ 0.04 & 
3.60 $\pm$ 0.47
\\
HSCJ140929$-$011410 & 
1.36 $\pm$ 0.16 & 
1.66 $\pm$ 0.10 & 
1.34 $\pm$ 0.06 & 
1.27 $\pm$ 0.04 & 
2.28 $\pm$ 0.23
\\
HSCJ141001+012956 & 
1.19 $\pm$ 0.14 & 
1.12 $\pm$ 0.07 & 
1.09 $\pm$ 0.05 & 
1.17 $\pm$ 0.04 & 
1.82 $\pm$ 0.17
\\
HSCJ141300$-$012608 & 
1.10 $\pm$ 0.13 & 
0.93 $\pm$ 0.06 & 
0.92 $\pm$ 0.04 & 
0.97 $\pm$ 0.03 & 
0.74 $\pm$ 0.06
\\
HSCJ141831$-$000052 & 
1.63 $\pm$ 0.19 & 
1.59 $\pm$ 0.10 & 
1.15 $\pm$ 0.05 & 
1.15 $\pm$ 0.04 & 
2.44 $\pm$ 0.20
\\
HSCJ142353+013446 & 
1.30 $\pm$ 0.15 & 
1.31 $\pm$ 0.08 & 
1.08 $\pm$ 0.05 & 
1.11 $\pm$ 0.04 & 
2.19 $\pm$ 0.22
\\
HSCJ142449$-$005321 & 
3.20 $\pm$ 0.42 & 
1.79 $\pm$ 0.11 & 
1.74 $\pm$ 0.08 & 
1.68 $\pm$ 0.06 & 
8.54 $\pm$ 1.64
\\
HSCJ142720+001916 & 
1.55 $\pm$ 0.18 & 
1.49 $\pm$ 0.09 & 
1.38 $\pm$ 0.06 & 
1.31 $\pm$ 0.04 & 
3.21 $\pm$ 0.37
\\
HSCJ142748+000958 & 
1.30 $\pm$ 0.15 & 
1.36 $\pm$ 0.09 & 
1.17 $\pm$ 0.05 & 
1.27 $\pm$ 0.04 & 
2.05 $\pm$ 0.20
\\
HSCJ144307$-$004056 & 
0.76 $\pm$ 0.10 & 
1.06 $\pm$ 0.07 & 
0.99 $\pm$ 0.04 & 
0.96 $\pm$ 0.03 & 
1.34 $\pm$ 0.11
\\
HSCJ144428$-$005142 & 
1.96 $\pm$ 0.24 & 
1.26 $\pm$ 0.08 & 
1.09 $\pm$ 0.05 & 
1.09 $\pm$ 0.04 & 
8.53 $\pm$ 1.53
\\
HSCJ145236$-$002142 & 
2.63 $\pm$ 0.33 & 
1.61 $\pm$ 0.10 & 
1.35 $\pm$ 0.06 & 
1.18 $\pm$ 0.04 & 
6.60 $\pm$ 1.13
\\
HSCJ145732$-$015917 & 
2.61 $\pm$ 0.34 & 
1.48 $\pm$ 0.09 & 
1.34 $\pm$ 0.06 & 
1.16 $\pm$ 0.04 & 
4.93 $\pm$ 0.69
\\
HSCJ145759+423019 & 
1.56 $\pm$ 0.17 & 
1.10 $\pm$ 0.07 & 
1.13 $\pm$ 0.05 & 
1.11 $\pm$ 0.04 & 
4.11 $\pm$ 0.57
\\
HSCJ145902$-$012351 & 
1.00 $\pm$ 0.11 & 
1.01 $\pm$ 0.06 & 
0.88 $\pm$ 0.04 & 
1.00 $\pm$ 0.03 & 
1.38 $\pm$ 0.12
\\
HSCJ151336+433251 & 
1.12 $\pm$ 0.13 & 
1.40 $\pm$ 0.09 & 
1.21 $\pm$ 0.05 & 
1.13 $\pm$ 0.04 & 
2.63 $\pm$ 0.28
\\
HSCJ155826+432830 & 
1.35 $\pm$ 0.17 & 
1.07 $\pm$ 0.07 & 
0.93 $\pm$ 0.04 & 
1.02 $\pm$ 0.03 & 
1.04 $\pm$ 0.08
\\
SL2SJ220202+014710 & 
0.78 $\pm$ 0.10 & 
0.67 $\pm$ 0.04 & 
0.84 $\pm$ 0.04 & 
0.88 $\pm$ 0.03 & 
0.86 $\pm$ 0.06
\\
SL2SJ220506+014703 & 
2.22 $\pm$ 0.29 & 
1.69 $\pm$ 0.12 & 
1.45 $\pm$ 0.07 & 
1.35 $\pm$ 0.05 & 
1.83 $\pm$ 0.16
\\
HSCJ220550+041524 & 
1.81 $\pm$ 0.22 & 
1.38 $\pm$ 0.09 & 
1.27 $\pm$ 0.06 & 
1.20 $\pm$ 0.04 & 
2.07 $\pm$ 0.16
\\
SL2SJ220642+041131 & 
1.79 $\pm$ 0.22 & 
1.34 $\pm$ 0.08 & 
1.07 $\pm$ 0.05 & 
1.00 $\pm$ 0.03 & 
5.17 $\pm$ 0.79
\\
HSCJ221726+000350 & 
1.16 $\pm$ 0.14 & 
1.09 $\pm$ 0.07 & 
1.01 $\pm$ 0.04 & 
0.91 $\pm$ 0.03 & 
7.10 $\pm$ 1.05
\\
SL2SJ221852+014038 & 
1.06 $\pm$ 0.13 & 
1.04 $\pm$ 0.06 & 
0.93 $\pm$ 0.04 & 
1.00 $\pm$ 0.03 & 
1.62 $\pm$ 0.15
\\
HSCJ222801+012805 & 
1.10 $\pm$ 0.13 & 
1.00 $\pm$ 0.06 & 
1.02 $\pm$ 0.04 & 
1.11 $\pm$ 0.04 & 
1.36 $\pm$ 0.13
\\
HSCJ223518$-$004747 & 
1.16 $\pm$ 0.13 & 
1.14 $\pm$ 0.07 & 
1.13 $\pm$ 0.05 & 
1.09 $\pm$ 0.04 & 
4.38 $\pm$ 0.62
\\
HSCJ223733+005015 & 
0.83 $\pm$ 0.10 & 
1.06 $\pm$ 0.07 & 
0.95 $\pm$ 0.04 & 
0.96 $\pm$ 0.03 & 
1.33 $\pm$ 0.12
\\
HSCJ224201+022810 & 
1.51 $\pm$ 0.19 & 
0.98 $\pm$ 0.06 & 
0.90 $\pm$ 0.04 & 
1.05 $\pm$ 0.03 & 
1.28 $\pm$ 0.10
\\
HSCJ224221+001144 & 
1.14 $\pm$ 0.14 & 
0.95 $\pm$ 0.06 & 
1.21 $\pm$ 0.05 & 
1.18 $\pm$ 0.04 & 
2.81 $\pm$ 0.29
\\
HSCJ224454+031551 & 
1.29 $\pm$ 0.15 & 
1.11 $\pm$ 0.07 & 
1.01 $\pm$ 0.04 & 
1.03 $\pm$ 0.03 & 
1.39 $\pm$ 0.13
\\
HSCJ224800$-$010259 & 
1.34 $\pm$ 0.16 & 
1.19 $\pm$ 0.07 & 
1.22 $\pm$ 0.05 & 
1.04 $\pm$ 0.04 & 
1.64 $\pm$ 0.12
\\
HSCJ225800+004533 & 
1.02 $\pm$ 0.12 & 
1.30 $\pm$ 0.08 & 
1.19 $\pm$ 0.05 & 
1.14 $\pm$ 0.04 & 
1.93 $\pm$ 0.19
\\
SDSSJ2303+0037 & 
1.96 $\pm$ 0.25 & 
1.30 $\pm$ 0.09 & 
1.16 $\pm$ 0.05 & 
1.11 $\pm$ 0.04 & 
3.17 $\pm$ 0.36
\\
HSCJ230521$-$000211 & 
1.59 $\pm$ 0.18 & 
1.08 $\pm$ 0.07 & 
0.89 $\pm$ 0.04 & 
0.87 $\pm$ 0.03 & 
1.42 $\pm$ 0.12
\\
HSCJ231004+024759 & 
3.30 $\pm$ 0.43 & 
2.00 $\pm$ 0.13 & 
1.71 $\pm$ 0.08 & 
1.50 $\pm$ 0.05 & 
10.29 $\pm$ 1.71
\\
HSCJ231145$-$013039 & 
1.61 $\pm$ 0.18 & 
1.14 $\pm$ 0.07 & 
1.02 $\pm$ 0.04 & 
1.02 $\pm$ 0.03 & 
1.99 $\pm$ 0.20
\\
HSCJ232415+011331 & 
1.04 $\pm$ 0.13 & 
0.95 $\pm$ 0.06 & 
1.02 $\pm$ 0.04 & 
1.16 $\pm$ 0.04 & 
1.80 $\pm$ 0.17
\\
HSCJ233130+003733 & 
1.39 $\pm$ 0.16 & 
1.14 $\pm$ 0.07 & 
1.33 $\pm$ 0.06 & 
1.39 $\pm$ 0.05 & 
1.41 $\pm$ 0.12
\\
HSCJ233146+013845 & 
1.44 $\pm$ 0.17 & 
1.15 $\pm$ 0.07 & 
1.11 $\pm$ 0.05 & 
1.05 $\pm$ 0.04 & 
3.48 $\pm$ 0.39
\\
HSCJ233311+022310 & 
1.06 $\pm$ 0.12 & 
0.91 $\pm$ 0.06 & 
0.98 $\pm$ 0.04 & 
0.89 $\pm$ 0.03 & 
1.07 $\pm$ 0.08
\\
HSCJ233528+001355 & 
1.75 $\pm$ 0.20 & 
1.32 $\pm$ 0.08 & 
1.15 $\pm$ 0.05 & 
1.24 $\pm$ 0.04 & 
4.01 $\pm$ 0.51
\\
\end{tabular}
\end{ruledtabular}
\tablecomments{Relative LOS overdensities are calculated for a magnitude limit of $i \leq 24$.  Overdensities are normalized by the mean values across all control fields.  The uncertainties do not include Poisson uncertainties in the calculation of lens field quantites.}
\end{table*}

\renewcommand*\arraystretch{1.2}
\begin{table*}
\caption{Local and LOS Overdensity of Individual Lens Fields Not in Main Sample (No Poisson Error)\label{tab:overdens_badvd_nopoisson}}
\begin{ruledtabular}
\begin{tabular}{l|cccc|c}
\multirow{2}{*}{Lens} &
$N_{\mathrm{eff}}/\langle N_{\mathrm{eff}}^{\mathrm{cont}} \rangle$ & 
$N_{\mathrm{eff}}/\langle N_{\mathrm{eff}}^{\mathrm{cont}} \rangle$ & 
$N_{\mathrm{eff}}/\langle N_{\mathrm{eff}}^{\mathrm{cont}} \rangle$ & 
$N_{\mathrm{eff}}/\langle N_{\mathrm{eff}}^{\mathrm{cont}} \rangle$ & 
\multirow{2}{*}{$\Sigma_{10}/\langle \Sigma_{10}^{\mathrm{cont}} \rangle$}
\\
&
($r \leq 30\arcsec$) & 
($r \leq 60\arcsec$) & 
($r \leq 90\arcsec$) & 
($r \leq 120\arcsec$) & 
\\
\tableline
SL2SJ021247$-$055552 & 
1.30 $\pm$ 0.16 & 
0.95 $\pm$ 0.06 & 
1.11 $\pm$ 0.05 & 
1.25 $\pm$ 0.04 & 
2.74 $\pm$ 0.32
\\
SL2SJ021411$-$040502 & 
1.85 $\pm$ 0.24 & 
1.34 $\pm$ 0.09 & 
1.20 $\pm$ 0.05 & 
1.13 $\pm$ 0.04 & 
4.18 $\pm$ 0.58
\\
SL2SJ022056$-$063934 & 
0.30 $\pm$ 0.04 & 
0.73 $\pm$ 0.05 & 
0.77 $\pm$ 0.03 & 
0.88 $\pm$ 0.03 & 
0.68 $\pm$ 0.05
\\
SL2SJ022459$-$040104 & 
0.84 $\pm$ 0.10 & 
0.97 $\pm$ 0.06 & 
0.98 $\pm$ 0.04 & 
0.94 $\pm$ 0.03 & 
0.78 $\pm$ 0.06
\\
SL2SJ022648$-$040610 & 
1.40 $\pm$ 0.16 & 
0.97 $\pm$ 0.06 & 
0.94 $\pm$ 0.04 & 
1.05 $\pm$ 0.04 & 
1.94 $\pm$ 0.21
\\
SL2SJ023251$-$040823 & 
1.80 $\pm$ 0.23 & 
1.08 $\pm$ 0.07 & 
1.03 $\pm$ 0.04 & 
1.05 $\pm$ 0.04 & 
1.58 $\pm$ 0.14
\\
SDSSJ023740.63$-$064112.9 & 
0.77 $\pm$ 0.09 & 
0.75 $\pm$ 0.05 & 
0.85 $\pm$ 0.04 & 
0.82 $\pm$ 0.03 & 
0.95 $\pm$ 0.07
\\
SL2SJ085540$-$014730 & 
1.30 $\pm$ 0.15 & 
1.17 $\pm$ 0.07 & 
1.25 $\pm$ 0.05 & 
1.02 $\pm$ 0.03 & 
1.34 $\pm$ 0.11
\\
SL2SJ085826$-$014300 & 
1.35 $\pm$ 0.16 & 
1.36 $\pm$ 0.09 & 
1.22 $\pm$ 0.05 & 
1.15 $\pm$ 0.04 & 
1.95 $\pm$ 0.19
\\
SL2SJ090408$-$005953 & 
1.27 $\pm$ 0.14 & 
1.29 $\pm$ 0.08 & 
1.08 $\pm$ 0.05 & 
0.98 $\pm$ 0.03 & 
0.61 $\pm$ 0.05
\\
HSCJ090613+032939 & 
1.81 $\pm$ 0.21 & 
1.47 $\pm$ 0.09 & 
1.30 $\pm$ 0.06 & 
1.15 $\pm$ 0.04 & 
1.25 $\pm$ 0.11
\\
H-ATLASJ090740.0$-$004200 & 
1.01 $\pm$ 0.11 & 
1.05 $\pm$ 0.06 & 
1.12 $\pm$ 0.05 & 
1.03 $\pm$ 0.03 & 
1.04 $\pm$ 0.09
\\
SDSSJ0912+0029 & 
1.73 $\pm$ 0.22 & 
1.53 $\pm$ 0.10 & 
1.40 $\pm$ 0.06 & 
1.25 $\pm$ 0.04 & 
2.28 $\pm$ 0.18
\\
SDSSJ0924+0219 & 
1.14 $\pm$ 0.13 & 
0.91 $\pm$ 0.05 & 
0.99 $\pm$ 0.04 & 
1.00 $\pm$ 0.03 & 
2.12 $\pm$ 0.21
\\
SDSSJ0935$-$0003 & 
2.48 $\pm$ 0.31 & 
2.23 $\pm$ 0.14 & 
1.64 $\pm$ 0.07 & 
1.46 $\pm$ 0.05 & 
5.64 $\pm$ 0.73
\\
BRI0952$-$0115 & 
0.41 $\pm$ 0.05 & 
0.75 $\pm$ 0.05 & 
0.80 $\pm$ 0.03 & 
0.88 $\pm$ 0.03 & 
0.77 $\pm$ 0.06
\\
SDSSJ0955+0101 & 
1.21 $\pm$ 0.14 & 
1.10 $\pm$ 0.07 & 
0.93 $\pm$ 0.04 & 
0.99 $\pm$ 0.03 & 
1.23 $\pm$ 0.09
\\
COSMOS5914+1219 & 
0.62 $\pm$ 0.07 & 
1.00 $\pm$ 0.06 & 
0.99 $\pm$ 0.04 & 
1.03 $\pm$ 0.03 & 
0.92 $\pm$ 0.06
\\
COSMOS5921+0638 & 
0.98 $\pm$ 0.13 & 
1.10 $\pm$ 0.07 & 
1.06 $\pm$ 0.05 & 
1.06 $\pm$ 0.04 & 
0.74 $\pm$ 0.05
\\
J095930.93+023427.7 & 
1.75 $\pm$ 0.21 & 
1.45 $\pm$ 0.09 & 
1.20 $\pm$ 0.05 & 
1.19 $\pm$ 0.04 & 
0.53 $\pm$ 0.03
\\
COSMOS5939+3044 & 
1.81 $\pm$ 0.20 & 
1.89 $\pm$ 0.12 & 
1.79 $\pm$ 0.08 & 
1.68 $\pm$ 0.06 & 
5.77 $\pm$ 0.93
\\
COSMOS0012+2015 & 
1.90 $\pm$ 0.25 & 
1.53 $\pm$ 0.10 & 
1.27 $\pm$ 0.06 & 
1.30 $\pm$ 0.04 & 
0.79 $\pm$ 0.06
\\
COSMOS0018+3845 & 
1.37 $\pm$ 0.15 & 
1.60 $\pm$ 0.10 & 
1.46 $\pm$ 0.06 & 
1.38 $\pm$ 0.05 & 
1.39 $\pm$ 0.12
\\
COSMOS0038+4133 & 
2.80 $\pm$ 0.34 & 
1.77 $\pm$ 0.11 & 
1.45 $\pm$ 0.06 & 
1.39 $\pm$ 0.05 & 
1.61 $\pm$ 0.16
\\
COSMOS0047+5023 & 
1.44 $\pm$ 0.18 & 
0.95 $\pm$ 0.06 & 
1.09 $\pm$ 0.05 & 
1.09 $\pm$ 0.04 & 
4.62 $\pm$ 0.64
\\
COSMOS0049+5128 & 
0.80 $\pm$ 0.09 & 
0.91 $\pm$ 0.06 & 
1.04 $\pm$ 0.05 & 
1.05 $\pm$ 0.04 & 
0.62 $\pm$ 0.04
\\
COSMOS0050+4901 & 
1.81 $\pm$ 0.19 & 
1.50 $\pm$ 0.09 & 
1.27 $\pm$ 0.06 & 
1.19 $\pm$ 0.04 & 
5.01 $\pm$ 0.69
\\
J100140.12+020040.9 & 
2.14 $\pm$ 0.25 & 
1.55 $\pm$ 0.10 & 
1.32 $\pm$ 0.06 & 
1.27 $\pm$ 0.04 & 
1.65 $\pm$ 0.15
\\
SL2SJ100148+022207 & 
2.07 $\pm$ 0.25 & 
1.48 $\pm$ 0.09 & 
1.26 $\pm$ 0.06 & 
1.22 $\pm$ 0.04 & 
1.78 $\pm$ 0.19
\\
COSMOS0211+1139 & 
1.97 $\pm$ 0.24 & 
1.61 $\pm$ 0.10 & 
1.63 $\pm$ 0.07 & 
1.36 $\pm$ 0.05 & 
5.04 $\pm$ 0.69
\\
SL2SJ100212+022955 & 
1.24 $\pm$ 0.13 & 
1.35 $\pm$ 0.09 & 
1.28 $\pm$ 0.06 & 
1.24 $\pm$ 0.04 & 
1.35 $\pm$ 0.13
\\
SL2SJ100215+023736 & 
1.10 $\pm$ 0.12 & 
1.21 $\pm$ 0.07 & 
1.23 $\pm$ 0.05 & 
1.27 $\pm$ 0.04 & 
1.08 $\pm$ 0.09
\\
COSMOS0254+1430 & 
1.38 $\pm$ 0.17 & 
1.08 $\pm$ 0.07 & 
0.98 $\pm$ 0.04 & 
0.92 $\pm$ 0.03 & 
0.45 $\pm$ 0.03
\\
SDSSJ1143$-$0144 & 
1.04 $\pm$ 0.12 & 
1.24 $\pm$ 0.08 & 
1.08 $\pm$ 0.05 & 
1.09 $\pm$ 0.04 & 
3.34 $\pm$ 0.28
\\
HSTJ114331.46$-$014508.0 & 
1.02 $\pm$ 0.12 & 
1.02 $\pm$ 0.07 & 
1.08 $\pm$ 0.05 & 
1.05 $\pm$ 0.04 & 
3.51 $\pm$ 0.29
\\
SDSSJ1159$-$0007 & 
2.00 $\pm$ 0.24 & 
1.45 $\pm$ 0.09 & 
1.14 $\pm$ 0.05 & 
1.18 $\pm$ 0.04 & 
2.24 $\pm$ 0.23
\\
SDSSJ1215+0047 & 
1.55 $\pm$ 0.19 & 
1.23 $\pm$ 0.08 & 
1.15 $\pm$ 0.05 & 
1.12 $\pm$ 0.04 & 
2.00 $\pm$ 0.21
\\
SDSSJ1226$-$0006 & 
1.19 $\pm$ 0.15 & 
1.08 $\pm$ 0.07 & 
1.23 $\pm$ 0.06 & 
1.19 $\pm$ 0.04 & 
0.43 $\pm$ 0.03
\\
SDSSJ1250$-$0135 & 
1.43 $\pm$ 0.18 & 
1.14 $\pm$ 0.07 & 
1.29 $\pm$ 0.06 & 
1.34 $\pm$ 0.04 & 
0.49 $\pm$ 0.04
\\
SDSSJ1347$-$0101 & 
1.06 $\pm$ 0.11 & 
0.88 $\pm$ 0.05 & 
0.97 $\pm$ 0.04 & 
0.98 $\pm$ 0.03 & 
0.54 $\pm$ 0.04
\\
SDSSJ1403+0006 & 
1.96 $\pm$ 0.23 & 
1.47 $\pm$ 0.09 & 
1.38 $\pm$ 0.06 & 
1.25 $\pm$ 0.04 & 
2.12 $\pm$ 0.15
\\
HSCJ141635+010128 & 
2.09 $\pm$ 0.27 & 
1.63 $\pm$ 0.10 & 
1.44 $\pm$ 0.06 & 
1.26 $\pm$ 0.04 & 
3.68 $\pm$ 0.50
\\
HSCJ142053+005620 & 
1.50 $\pm$ 0.18 & 
1.28 $\pm$ 0.08 & 
1.28 $\pm$ 0.06 & 
1.23 $\pm$ 0.04 & 
2.34 $\pm$ 0.26
\\
SDSSJ1436$-$0000 & 
1.46 $\pm$ 0.17 & 
1.14 $\pm$ 0.07 & 
0.99 $\pm$ 0.04 & 
1.07 $\pm$ 0.04 & 
1.47 $\pm$ 0.10
\\
SDSSJ1524+4409 & 
1.15 $\pm$ 0.14 & 
1.45 $\pm$ 0.09 & 
1.37 $\pm$ 0.06 & 
1.24 $\pm$ 0.04 & 
1.60 $\pm$ 0.13
\\
HSCJ155319+431824 & 
1.27 $\pm$ 0.15 & 
1.17 $\pm$ 0.07 & 
1.16 $\pm$ 0.05 & 
1.10 $\pm$ 0.04 & 
1.95 $\pm$ 0.20
\\
B1600+434 & 
0.91 $\pm$ 0.10 & 
1.01 $\pm$ 0.07 & 
1.07 $\pm$ 0.05 & 
0.97 $\pm$ 0.03 & 
0.51 $\pm$ 0.04
\\
SL2SJ220629+005728 & 
0.92 $\pm$ 0.12 & 
1.02 $\pm$ 0.07 & 
0.91 $\pm$ 0.04 & 
0.93 $\pm$ 0.03 & 
0.58 $\pm$ 0.04
\\
SL2SJ221326$-$000946 & 
2.35 $\pm$ 0.29 & 
1.64 $\pm$ 0.11 & 
1.41 $\pm$ 0.06 & 
1.27 $\pm$ 0.04 & 
7.47 $\pm$ 1.06
\\
SL2SJ221929$-$001743 & 
1.80 $\pm$ 0.23 & 
1.24 $\pm$ 0.08 & 
1.11 $\pm$ 0.05 & 
1.10 $\pm$ 0.04 & 
0.78 $\pm$ 0.05
\\
SL2SJ222148+011542 & 
1.47 $\pm$ 0.17 & 
1.46 $\pm$ 0.09 & 
1.31 $\pm$ 0.06 & 
1.10 $\pm$ 0.04 & 
1.44 $\pm$ 0.12
\\
SL2SJ222217+001202 & 
1.09 $\pm$ 0.13 & 
1.02 $\pm$ 0.06 & 
1.00 $\pm$ 0.04 & 
0.90 $\pm$ 0.03 & 
1.42 $\pm$ 0.12
\\
Q2237+0305 & 
1.35 $\pm$ 0.16 & 
1.05 $\pm$ 0.07 & 
0.99 $\pm$ 0.04 & 
0.90 $\pm$ 0.03 & 
0.76 $\pm$ 0.04
\\
SDSSJ2300+0022 & 
2.10 $\pm$ 0.24 & 
2.05 $\pm$ 0.13 & 
1.66 $\pm$ 0.07 & 
1.40 $\pm$ 0.05 & 
2.91 $\pm$ 0.23
\\
HSCJ233230+003821 & 
1.25 $\pm$ 0.14 & 
1.18 $\pm$ 0.07 & 
1.19 $\pm$ 0.05 & 
1.13 $\pm$ 0.04 & 
1.31 $\pm$ 0.12
\\
\end{tabular}
\end{ruledtabular}
\tablecomments{Lenses listed here either do not have velocity dispersion measurements from BOSS or have unreliable velocity dispersion measurements, so they are not included in our main analysis.  We list their local and LOS overdensities here for completeness.  The uncertainties do not include Poisson uncertainties in the calculation of lens field quantites.}
\end{table*}

\renewcommand*\arraystretch{1.0}


\begin{thebibliography}{}
\expandafter\ifx\csname natexlab\endcsname\relax\def\natexlab#1{#1}\fi

\bibitem[{{Abazajian} {et~al.}(2004){Abazajian}, {Adelman-McCarthy},
  {Ag{\"u}eros}, {Allam}, {Anderson}, {Anderson}, {Annis}, {Bahcall}, {Baldry},
  {Bastian}, {Berlind}, {Bernardi}, {Blanton}, {Bochanski}, {Boroski},
  {Briggs}, {Brinkmann}, {Brunner}, {Budav{\'a}ri}, {Carey}, {Carliles},
  {Castander}, {Connolly}, {Csabai}, {Doi}, {Dong}, {Eisenstein}, {Evans},
  {Fan}, {Finkbeiner}, {Friedman}, {Frieman}, {Fukugita}, {Gal}, {Gillespie},
  {Glazebrook}, {Gray}, {Grebel}, {Gunn}, {Gurbani}, {Hall}, {Hamabe},
  {Harris}, {Harris}, {Harvanek}, {Heckman}, {Hendry}, {Hennessy}, {Hindsley},
  {Hogan}, {Hogg}, {Holmgren}, {Ichikawa}, {Ichikawa}, {Ivezi{\'c}}, {Jester},
  {Johnston}, {Jorgensen}, {Kent}, {Kleinman}, {Knapp}, {Kniazev}, {Kron},
  {Krzesinski}, {Kunszt}, {Kuropatkin}, {Lamb}, {Lampeitl}, {Lee}, {Leger},
  {Li}, {Lin}, {Loh}, {Long}, {Loveday}, {Lupton}, {Malik}, {Margon},
  {Matsubara}, {McGehee}, {McKay}, {Meiksin}, {Munn}, {Nakajima}, {Nash},
  {Neilsen}, {Newberg}, {Newman}, {Nichol}, {Nicinski}, {Nieto-Santisteban},
  {Nitta}, {Okamura}, {O'Mullane}, {Ostriker}, {Owen}, {Padmanabhan},
  {Peoples}, {Pier}, {Pope}, {Quinn}, {Richards}, {Richmond}, {Rix}, {Rockosi},
  {Schlegel}, {Schneider}, {Scranton}, {Sekiguchi}, {Seljak}, {Sergey},
  {Sesar}, {Sheldon}, {Shimasaku}, {Siegmund}, {Silvestri}, {Smith}, {Smol{\v
  c}i{\'c}}, {Snedden}, {Stebbins}, {Stoughton}, {Strauss}, {SubbaRao},
  {Szalay}, {Szapudi}, {Szkody}, {Szokoly}, {Tegmark}, {Teodoro}, {Thakar},
  {Tremonti}, {Tucker}, {Uomoto}, {Vanden Berk}, {Vandenberg}, {Vogeley},
  {Voges}, {Vogt}, {Walkowicz}, {Wang}, {Weinberg}, {West}, {White}, {Wilhite},
  {Xu}, {Yanny}, {Yasuda}, {Yip}, {Yocum}, {York}, {Zehavi}, {Zibetti}, \&
  {Zucker}}]{abazajian+2004}
{Abazajian}, K., {Adelman-McCarthy}, J.~K., {Ag{\"u}eros}, M.~A., {et~al.}
  2004, \aj, 128, 502

\bibitem[{{Abolfathi} {et~al.}(2017){Abolfathi}, {Aguado}, {Aguilar}, {Allende
  Prieto}, {Almeida}, {Tasnim Ananna}, {Anders}, {Anderson}, {Andrews},
  {Anguiano}, \& et~al.}]{abolfathi+2017}
{Abolfathi}, B., {Aguado}, D.~S., {Aguilar}, G., {et~al.} 2017, ArXiv e-prints,
  arXiv:1707.09322

\bibitem[{Aihara {et~al.}(2018{\natexlab{a}})Aihara, Armstrong, Bickerton,
  Bosch, Coupon, Furusawa, Hayashi, Ikeda, Kamata, Karoji, Kawanomoto, Koike,
  Komiyama, Lang, Lupton, Mineo, Miyatake, Miyazaki, Morokuma, Obuchi, Oishi,
  Okura, Price, Takata, Tanaka, Tanaka, Tanaka, Uchida, Uraguchi, Utsumi, Wang,
  Yamada, Yamanoi, Yasuda, Arimoto, Chiba, Finet, Fujimori, Fujimoto, Furusawa,
  Goto, Goulding, Gunn, Harikane, Hattori, Hayashi, He{\l}miniak, Higuchi,
  Hikage, Ho, Hsieh, Huang, Huang, Imanishi, Iwata, Jaelani, Jian, Kashikawa,
  Katayama, Kojima, Konno, Koshida, Kusakabe, Leauthaud, Lee, Lin, Lin,
  Mandelbaum, Matsuoka, Medezinski, Miyama, Momose, More, More, Mukae, Murata,
  Murayama, Nagao, Nakata, Niida, Niikura, Nishizawa, Oguri, Okabe, Ono,
  Onodera, Onoue, Ouchi, Pyo, Shibuya, Shimasaku, Simet, Speagle, Spergel,
  Strauss, Sugahara, Sugiyama, Suto, Suzuki, Tait, Takada, Terai, Toba, Turner,
  Uchiyama, Umetsu, Urata, Usuda, Yeh, \& Yuma}]{aihara+2018a}
Aihara, H., Armstrong, R., Bickerton, S., {et~al.} 2018{\natexlab{a}},
  Publications of the Astronomical Society of Japan, 70, S8

\bibitem[{Aihara {et~al.}(2018{\natexlab{b}})Aihara, Arimoto, Armstrong,
  Arnouts, Bahcall, Bickerton, Bosch, Bundy, Capak, Chan, Chiba, Coupon, Egami,
  Enoki, Finet, Fujimori, Fujimoto, Furusawa, Furusawa, Goto, Goulding, Greco,
  Greene, Gunn, Hamana, Harikane, Hashimoto, Hattori, Hayashi, Hayashi,
  He{\l}miniak, Higuchi, Hikage, Ho, Hsieh, Huang, Huang, Ikeda, Imanishi,
  Inoue, Iwasawa, Iwata, Jaelani, Jian, Kamata, Karoji, Kashikawa, Katayama,
  Kawanomoto, Kayo, Koda, Koike, Kojima, Komiyama, Konno, Koshida, Koyama,
  Kusakabe, Leauthaud, Lee, Lin, Lin, Lupton, Mandelbaum, Matsuoka, Medezinski,
  Mineo, Miyama, Miyatake, Miyazaki, Momose, More, More, Moritani, Moriya,
  Morokuma, Mukae, Murata, Murayama, Nagao, Nakata, Niida, Niikura, Nishizawa,
  Obuchi, Oguri, Oishi, Okabe, Okamoto, Okura, Ono, Onodera, Onoue, Osato,
  Ouchi, Price, Pyo, Sako, Sawicki, Shibuya, Shimasaku, Shimono, Shirasaki,
  Silverman, Simet, Speagle, Spergel, Strauss, Sugahara, Sugiyama, Suto, Suyu,
  Suzuki, Tait, Takada, Takata, Tamura, Tanaka, Tanaka, Tanaka, Tanaka, Terai,
  Terashima, Toba, Tominaga, Toshikawa, Turner, Uchida, Uchiyama, Umetsu,
  Uraguchi, Urata, Usuda, Utsumi, Wang, Wang, Wong, Yabe, Yamada, Yamanoi,
  Yasuda, Yeh, Yonehara, \& Yuma}]{aihara+2018b}
Aihara, H., Arimoto, N., Armstrong, R., {et~al.} 2018{\natexlab{b}},
  Publications of the Astronomical Society of Japan, 70, S4

\bibitem[{{Alam} {et~al.}(2015){Alam}, {Albareti}, {Allende Prieto}, {Anders},
  {Anderson}, {Anderton}, {Andrews}, {Armengaud}, {Aubourg}, {Bailey}, \&
  et~al.}]{alam+2015}
{Alam}, S., {Albareti}, F.~D., {Allende Prieto}, C., {et~al.} 2015, \apjs, 219,
  12

\bibitem[{{Auger} {et~al.}(2010){Auger}, {Treu}, {Bolton}, {Gavazzi},
  {Koopmans}, {Marshall}, {Moustakas}, \& {Burles}}]{auger+2010}
{Auger}, M.~W., {Treu}, T., {Bolton}, A.~S., {et~al.} 2010, \apj, 724, 511

\bibitem[{{Auger} {et~al.}(2011){Auger}, {Treu}, {Brewer}, \&
  {Marshall}}]{auger+2011}
{Auger}, M.~W., {Treu}, T., {Brewer}, B.~J., \& {Marshall}, P.~J. 2011, \mnras,
  411, L6

\bibitem[{{Axelrod} {et~al.}(2010){Axelrod}, {Kantor}, {Lupton}, \&
  {Pierfederici}}]{axelrod+2010}
{Axelrod}, T., {Kantor}, J., {Lupton}, R.~H., \& {Pierfederici}, F. 2010, in
  \procspie, Vol. 7740, Software and Cyberinfrastructure for Astronomy, 774015

\bibitem[{{Bolton} {et~al.}(2006){Bolton}, {Burles}, {Koopmans}, {Treu}, \&
  {Moustakas}}]{bolton+2006}
{Bolton}, A.~S., {Burles}, S., {Koopmans}, L.~V.~E., {Treu}, T., \&
  {Moustakas}, L.~A. 2006, \apj, 638, 703

\bibitem[{{Bolton} {et~al.}(2008){Bolton}, {Treu}, {Koopmans}, {Gavazzi},
  {Moustakas}, {Burles}, {Schlegel}, \& {Wayth}}]{bolton+2008}
{Bolton}, A.~S., {Treu}, T., {Koopmans}, L.~V.~E., {et~al.} 2008, \apj, 684,
  248

\bibitem[{{Bonvin} {et~al.}(2017){Bonvin}, {Courbin}, {Suyu}, {Marshall},
  {Rusu}, {Sluse}, {Tewes}, {Wong}, {Collett}, {Fassnacht}, {Treu}, {Auger},
  {Hilbert}, {Koopmans}, {Meylan}, {Rumbaugh}, {Sonnenfeld}, \&
  {Spiniello}}]{bonvin+2017}
{Bonvin}, V., {Courbin}, F., {Suyu}, S.~H., {et~al.} 2017, \mnras, 465, 4914

\bibitem[{{Bosch} {et~al.}(2018){Bosch}, {Armstrong}, {Bickerton}, {Furusawa},
  {Ikeda}, {Koike}, {Lupton}, {Mineo}, {Price}, {Takata}, {Tanaka}, {Yasuda},
  {AlSayyad}, {Becker}, {Coulton}, {Coupon}, {Garmilla}, {Huang}, {Krughoff},
  {Lang}, {Leauthaud}, {Lim}, {Lust}, {MacArthur}, {Mandelbaum}, {Miyatake},
  {Miyazaki}, {Murata}, {More}, {Okura}, {Owen}, {Swinbank}, {Strauss},
  {Yamada}, \& {Yamanoi}}]{bosch+2018}
{Bosch}, J., {Armstrong}, R., {Bickerton}, S., {et~al.} 2018, \pasj, 70, S5

\bibitem[{{Brownstein} {et~al.}(2012){Brownstein}, {Bolton}, {Schlegel},
  {Eisenstein}, {Kochanek}, {Connolly}, {Maraston}, {Pandey}, {Seitz}, {Wake},
  {Wood-Vasey}, {Brinkmann}, {Schneider}, \& {Weaver}}]{brownstein+2012}
{Brownstein}, J.~R., {Bolton}, A.~S., {Schlegel}, D.~J., {et~al.} 2012, \apj,
  744, 41

\bibitem[{{Chan} {et~al.}(2015){Chan}, {Suyu}, {Chiueh}, {More}, {Marshall},
  {Coupon}, {Oguri}, \& {Price}}]{chan+2015}
{Chan}, J.~H.~H., {Suyu}, S.~H., {Chiueh}, T., {et~al.} 2015, \apj, 807, 138

\bibitem[{{Chiu} {et~al.}(2016){Chiu}, {Dietrich}, {Mohr}, {Applegate},
  {Benson}, {Bleem}, {Bayliss}, {Bocquet}, {Carlstrom}, {Capasso}, {Desai},
  {Gangkofner}, {Gonzalez}, {Gupta}, {Hennig}, {Hoekstra}, {von der Linden},
  {Liu}, {McDonald}, {Reichardt}, {Saro}, {Schrabback}, {Strazzullo}, {Stubbs},
  \& {Zenteno}}]{chiu+2016}
{Chiu}, I., {Dietrich}, J.~P., {Mohr}, J., {et~al.} 2016, \mnras, 457, 3050

\bibitem[{{Collett}(2015)}]{collett2015}
{Collett}, T.~E. 2015, \apj, 811, 20

\bibitem[{{Collett} \& {Cunnington}(2016)}]{collettcunnington2016}
{Collett}, T.~E., \& {Cunnington}, S.~D. 2016, \mnras, 462, 3255

\bibitem[{{Collett} {et~al.}(2013){Collett}, {Marshall}, {Auger}, {Hilbert},
  {Suyu}, {Greene}, {Treu}, {Fassnacht}, {Koopmans}, {Brada{\v c}}, \&
  {Blandford}}]{collett+2013}
{Collett}, T.~E., {Marshall}, P.~J., {Auger}, M.~W., {et~al.} 2013, \mnras,
  432, 679

\bibitem[{{Cooper} {et~al.}(2005){Cooper}, {Newman}, {Madgwick}, {Gerke},
  {Yan}, \& {Davis}}]{cooper+2005}
{Cooper}, M.~C., {Newman}, J.~A., {Madgwick}, D.~S., {et~al.} 2005, \apj, 634,
  833

\bibitem[{{Cooper} {et~al.}(2006){Cooper}, {Newman}, {Croton}, {Weiner},
  {Willmer}, {Gerke}, {Madgwick}, {Faber}, {Davis}, {Coil}, {Finkbeiner},
  {Guhathakurta}, \& {Koo}}]{cooper+2006}
{Cooper}, M.~C., {Newman}, J.~A., {Croton}, D.~J., {et~al.} 2006, \mnras, 370,
  198

\bibitem[{Coupon {et~al.}(2018)Coupon, Czakon, Bosch, Komiyama, Medezinski,
  Miyazaki, \& Oguri}]{coupon+2018}
Coupon, J., Czakon, N., Bosch, J., {et~al.} 2018, Publications of the
  Astronomical Society of Japan, 70, S7

\bibitem[{{Dawson} {et~al.}(2013){Dawson}, {Schlegel}, {Ahn}, {Anderson},
  {Aubourg}, {Bailey}, {Barkhouser}, {Bautista}, {Beifiori}, {Berlind},
  {Bhardwaj}, {Bizyaev}, {Blake}, {Blanton}, {Blomqvist}, {Bolton}, {Borde},
  {Bovy}, {Brandt}, {Brewington}, {Brinkmann}, {Brown}, {Brownstein}, {Bundy},
  {Busca}, {Carithers}, {Carnero}, {Carr}, {Chen}, {Comparat}, {Connolly},
  {Cope}, {Croft}, {Cuesta}, {da Costa}, {Davenport}, {Delubac}, {de Putter},
  {Dhital}, {Ealet}, {Ebelke}, {Eisenstein}, {Escoffier}, {Fan}, {Filiz Ak},
  {Finley}, {Font-Ribera}, {G{\'e}nova-Santos}, {Gunn}, {Guo}, {Haggard},
  {Hall}, {Hamilton}, {Harris}, {Harris}, {Ho}, {Hogg}, {Holder}, {Honscheid},
  {Huehnerhoff}, {Jordan}, {Jordan}, {Kauffmann}, {Kazin}, {Kirkby}, {Klaene},
  {Kneib}, {Le Goff}, {Lee}, {Long}, {Loomis}, {Lundgren}, {Lupton}, {Maia},
  {Makler}, {Malanushenko}, {Malanushenko}, {Mandelbaum}, {Manera}, {Maraston},
  {Margala}, {Masters}, {McBride}, {McDonald}, {McGreer}, {McMahon}, {Mena},
  {Miralda-Escud{\'e}}, {Montero-Dorta}, {Montesano}, {Muna}, {Myers},
  {Naugle}, {Nichol}, {Noterdaeme}, {Nuza}, {Olmstead}, {Oravetz}, {Oravetz},
  {Owen}, {Padmanabhan}, {Palanque-Delabrouille}, {Pan}, {Parejko},
  {P{\^a}ris}, {Percival}, {P{\'e}rez-Fournon}, {P{\'e}rez-R{\`a}fols},
  {Petitjean}, {Pfaffenberger}, {Pforr}, {Pieri}, {Prada}, {Price-Whelan},
  {Raddick}, {Rebolo}, {Rich}, {Richards}, {Rockosi}, {Roe}, {Ross}, {Ross},
  {Rossi}, {Rubi{\~n}o-Martin}, {Samushia}, {S{\'a}nchez}, {Sayres}, {Schmidt},
  {Schneider}, {Sc{\'o}ccola}, {Seo}, {Shelden}, {Sheldon}, {Shen}, {Shu},
  {Slosar}, {Smee}, {Snedden}, {Stauffer}, {Steele}, {Strauss}, {Streblyanska},
  {Suzuki}, {Swanson}, {Tal}, {Tanaka}, {Thomas}, {Tinker}, {Tojeiro},
  {Tremonti}, {Vargas Maga{\~n}a}, {Verde}, {Viel}, {Wake}, {Watson}, {Weaver},
  {Weinberg}, {Weiner}, {West}, {White}, {Wood-Vasey}, {Yeche}, {Zehavi},
  {Zhao}, \& {Zheng}}]{dawson+2013}
{Dawson}, K.~S., {Schlegel}, D.~J., {Ahn}, C.~P., {et~al.} 2013, \aj, 145, 10

\bibitem[{{Dressler}(1980)}]{dressler1980}
{Dressler}, A. 1980, \apj, 236, 351

\bibitem[{{Eigenbrod} {et~al.}(2006){Eigenbrod}, {Courbin}, {Dye}, {Meylan},
  {Sluse}, {Vuissoz}, \& {Magain}}]{eigenbrod+2006}
{Eigenbrod}, A., {Courbin}, F., {Dye}, S., {et~al.} 2006, \aap, 451, 747

\bibitem[{{Eigenbrod} {et~al.}(2007){Eigenbrod}, {Courbin}, \&
  {Meylan}}]{eigenbrod+2007}
{Eigenbrod}, A., {Courbin}, F., \& {Meylan}, G. 2007, \aap, 465, 51

\bibitem[{{Eisenstein} {et~al.}(2011){Eisenstein}, {Weinberg}, {Agol},
  {Aihara}, {Allende Prieto}, {Anderson}, {Arns}, {Aubourg}, {Bailey},
  {Balbinot}, \& et~al.}]{eisenstein+2011}
{Eisenstein}, D.~J., {Weinberg}, D.~H., {Agol}, E., {et~al.} 2011, \aj, 142, 72

\bibitem[{{Falco} {et~al.}(1985){Falco}, {Gorenstein}, \&
  {Shapiro}}]{falco+1985}
{Falco}, E.~E., {Gorenstein}, M.~V., \& {Shapiro}, I.~I. 1985, \apjl, 289, L1

\bibitem[{{Fassnacht} \& {Cohen}(1998)}]{fassnacht+1998}
{Fassnacht}, C.~D., \& {Cohen}, J.~G. 1998, \aj, 115, 377

\bibitem[{{Fassnacht} {et~al.}(2011){Fassnacht}, {Koopmans}, \&
  {Wong}}]{fassnacht+2011}
{Fassnacht}, C.~D., {Koopmans}, L.~V.~E., \& {Wong}, K.~C. 2011, \mnras, 410,
  2167

\bibitem[{{Faure} {et~al.}(2008){Faure}, {Kneib}, {Covone}, {Tasca},
  {Leauthaud}, {Capak}, {Jahnke}, {Smolcic}, {de la Torre}, {Ellis},
  {Finoguenov}, {Koekemoer}, {Le Fevre}, {Massey}, {Mellier}, {Refregier},
  {Rhodes}, {Scoville}, {Schinnerer}, {Taylor}, {Van Waerbeke}, \&
  {Walcher}}]{faure+2008}
{Faure}, C., {Kneib}, J.-P., {Covone}, G., {et~al.} 2008, \apjs, 176, 19

\bibitem[{{Faure} {et~al.}(2011){Faure}, {Anguita}, {Alloin}, {Bundy},
  {Finoguenov}, {Leauthaud}, {Knobel}, {Kneib}, {Jullo}, {Ilbert}, {Koekemoer},
  {Capak}, {Scoville}, \& {Tasca}}]{faure+2011}
{Faure}, C., {Anguita}, T., {Alloin}, D., {et~al.} 2011, \aap, 529, A72

\bibitem[{{Freedman} {et~al.}(2012){Freedman}, {Madore}, {Scowcroft}, {Burns},
  {Monson}, {Persson}, {Seibert}, \& {Rigby}}]{freedman+2012}
{Freedman}, W.~L., {Madore}, B.~F., {Scowcroft}, V., {et~al.} 2012, \apj, 758,
  24

\bibitem[{{Furusawa} {et~al.}(2018){Furusawa}, {Koike}, {Takata}, {Okura},
  {Miyatake}, {Lupton}, {Bickerton}, {Price}, {Bosch}, {Yasuda}, {Mineo},
  {Yamada}, {Miyazaki}, {Nakata}, {Koshida}, {Komiyama}, {Utsumi},
  {Kawanomoto}, {Jeschke}, {Noumaru}, {Schubert}, {Iwata}, {Finet},
  {Fujiyoshi}, {Tajitsu}, {Terai}, \& {Lee}}]{furusawa+2018}
{Furusawa}, H., {Koike}, M., {Takata}, T., {et~al.} 2018, \pasj, 70, S3

\bibitem[{{Gorenstein} {et~al.}(1988){Gorenstein}, {Falco}, \&
  {Shapiro}}]{gorenstein+1988}
{Gorenstein}, M.~V., {Falco}, E.~E., \& {Shapiro}, I.~I. 1988, \apj, 327, 693

\bibitem[{{Greene} {et~al.}(2013){Greene}, {Suyu}, {Treu}, {Hilbert}, {Auger},
  {Collett}, {Marshall}, {Fassnacht}, {Blandford}, {Brada{\v c}}, \&
  {Koopmans}}]{greene+2013}
{Greene}, Z.~S., {Suyu}, S.~H., {Treu}, T., {et~al.} 2013, \apj, 768, 39

\bibitem[{{Hilbert} {et~al.}(2007){Hilbert}, {White}, {Hartlap}, \&
  {Schneider}}]{hilbert+2007}
{Hilbert}, S., {White}, S.~D.~M., {Hartlap}, J., \& {Schneider}, P. 2007,
  \mnras, 382, 121

\bibitem[{{Huchra} {et~al.}(1985){Huchra}, {Gorenstein}, {Kent}, {Shapiro},
  {Smith}, {Horine}, \& {Perley}}]{huchra+1985}
{Huchra}, J., {Gorenstein}, M., {Kent}, S., {et~al.} 1985, \aj, 90, 691

\bibitem[{{Inada} {et~al.}(2003){Inada}, {Becker}, {Burles}, {Castander},
  {Eisenstein}, {Hall}, {Johnston}, {Pindor}, {Richards}, {Schechter},
  {Sekiguchi}, {White}, {Brinkmann}, {Frieman}, {Kleinman}, {Krzesi{\'n}ski},
  {Long}, {Neilsen}, {Newman}, {Nitta}, {Schneider}, {Snedden}, \&
  {York}}]{inada+2003}
{Inada}, N., {Becker}, R.~H., {Burles}, S., {et~al.} 2003, \aj, 126, 666

\bibitem[{{Inada} {et~al.}(2008){Inada}, {Oguri}, {Becker}, {Shin}, {Richards},
  {Hennawi}, {White}, {Pindor}, {Strauss}, {Kochanek}, {Johnston}, {Gregg},
  {Kayo}, {Eisenstein}, {Hall}, {Castander}, {Clocchiatti}, {Anderson},
  {Schneider}, {York}, {Lupton}, {Chiu}, {Kawano}, {Scranton}, {Frieman},
  {Keeton}, {Morokuma}, {Rix}, {Turner}, {Burles}, {Brunner}, {Sheldon},
  {Bahcall}, \& {Masataka}}]{inada+2008}
{Inada}, N., {Oguri}, M., {Becker}, R.~H., {et~al.} 2008, \aj, 135, 496

\bibitem[{{Jackson} {et~al.}(1995){Jackson}, {de Bruyn}, {Myers}, {Bremer},
  {Miley}, {Schilizzi}, {Browne}, {Nair}, {Wilkinson}, {Blandford}, {Pearson},
  \& {Readhead}}]{jackson+1995}
{Jackson}, N., {de Bruyn}, A.~G., {Myers}, S., {et~al.} 1995, \mnras, 274, L25

\bibitem[{{Juri{\'c}} {et~al.}(2015){Juri{\'c}}, {Kantor}, {Lim}, {Lupton},
  {Dubois-Felsmann}, {Jenness}, {Axelrod}, {Aleksi{\'c}}, {Allsman},
  {AlSayyad}, {Alt}, {Armstrong}, {Basney}, {Becker}, {Becla}, {Bickerton},
  {Biswas}, {Bosch}, {Boutigny}, {Carrasco Kind}, {Ciardi}, {Connolly},
  {Daniel}, {Daues}, {Economou}, {Chiang}, {Fausti}, {Fisher-Levine},
  {Freemon}, {Gee}, {Gris}, {Hernandez}, {Hoblitt}, {Ivezi{\'c}}, {Jammes},
  {Jevremovi{\'c}}, {Jones}, {Bryce Kalmbach}, {Kasliwal}, {Krughoff}, {Lang},
  {Lurie}, {Lust}, {Mullally}, {MacArthur}, {Melchior}, {Moeyens}, {Nidever},
  {Owen}, {Parejko}, {Peterson}, {Petravick}, {Pietrowicz}, {Price}, {Reiss},
  {Shaw}, {Sick}, {Slater}, {Strauss}, {Sullivan}, {Swinbank}, {Van Dyk},
  {Vuj{\v c}i{\'c}}, {Withers}, {Yoachim}, \& {LSST Project}}]{juric+2015}
{Juri{\'c}}, M., {Kantor}, J., {Lim}, K., {et~al.} 2015, ArXiv e-prints,
  arXiv:1512.07914

\bibitem[{Kawanomoto {et~al.}(2018)Kawanomoto, Uraguchi, Komiyama, Miyazaki,
  Furusawa, Finet, Hattori, Wang, Yasuda, \& Suzuki}]{kawanomoto+2018}
Kawanomoto, S., Uraguchi, F., Komiyama, Y., {et~al.} 2018, Publications of the
  Astronomical Society of Japan, psy056

\bibitem[{{Keeton}(2001)}]{keeton2001}
{Keeton}, C.~R. 2001, ArXiv Astrophysics e-prints, astro-ph/0102340

\bibitem[{{Kelly}(2007)}]{kelly2007}
{Kelly}, B.~C. 2007, \apj, 665, 1489

\bibitem[{{Komiyama} {et~al.}(2018){Komiyama}, {Obuchi}, {Nakaya}, {Kamata},
  {Kawanomoto}, {Utsumi}, {Miyazaki}, {Uraguchi}, {Furusawa}, {Morokuma},
  {Uchida}, {Miyatake}, {Mineo}, {Fujimori}, {Aihara}, {Karoji}, {Gunn}, \&
  {Wang}}]{komiyama+2018}
{Komiyama}, Y., {Obuchi}, Y., {Nakaya}, H., {et~al.} 2018, \pasj, 70, S2

\bibitem[{{Koopmans} {et~al.}(2009){Koopmans}, {Bolton}, {Treu}, {Czoske},
  {Auger}, {Barnab{\`e}}, {Vegetti}, {Gavazzi}, {Moustakas}, \&
  {Burles}}]{koopmans+2009}
{Koopmans}, L.~V.~E., {Bolton}, A., {Treu}, T., {et~al.} 2009, \apjl, 703, L51

\bibitem[{{McCully} {et~al.}(2014){McCully}, {Keeton}, {Wong}, \&
  {Zabludoff}}]{mccully+2014}
{McCully}, C., {Keeton}, C.~R., {Wong}, K.~C., \& {Zabludoff}, A.~I. 2014,
  \mnras, 443, 3631

\bibitem[{{McCully} {et~al.}(2017){McCully}, {Keeton}, {Wong}, \&
  {Zabludoff}}]{mccully+2017}
---. 2017, \apj, 836, 141

\bibitem[{{McMahon} {et~al.}(1992){McMahon}, {Irwin}, \&
  {Hazard}}]{mcmahon+1992}
{McMahon}, R., {Irwin}, M., \& {Hazard}, C. 1992, GEMINI Newsletter Royal
  Greenwich Observatory, 36, 1

\bibitem[{{Miyazaki} {et~al.}(2012){Miyazaki}, {Komiyama}, {Nakaya}, {Kamata},
  {Doi}, {Hamana}, {Karoji}, {Furusawa}, {Kawanomoto}, {Morokuma}, {Ishizuka},
  {Nariai}, {Tanaka}, {Uraguchi}, {Utsumi}, {Obuchi}, {Okura}, {Oguri},
  {Takata}, {Tomono}, {Kurakami}, {Namikawa}, {Usuda}, {Yamanoi}, {Terai},
  {Uekiyo}, {Yamada}, {Koike}, {Aihara}, {Fujimori}, {Mineo}, {Miyatake},
  {Yasuda}, {Nishizawa}, {Saito}, {Tanaka}, {Uchida}, {Katayama}, {Wang},
  {Chen}, {Lupton}, {Loomis}, {Bickerton}, {Price}, {Gunn}, {Suzuki},
  {Miyazaki}, {Muramatsu}, {Yamamoto}, {Endo}, {Ezaki}, {Itoh}, {Miwa},
  {Yokota}, {Matsuda}, {Ebinuma}, \& {Takeshi}}]{miyazaki+2012}
{Miyazaki}, S., {Komiyama}, Y., {Nakaya}, H., {et~al.} 2012, in \procspie, Vol.
  8446, Ground-based and Airborne Instrumentation for Astronomy IV, 84460Z

\bibitem[{{Miyazaki} {et~al.}(2018){Miyazaki}, {Komiyama}, {Kawanomoto}, {Doi},
  {Furusawa}, {Hamana}, {Hayashi}, {Ikeda}, {Kamata}, {Karoji}, {Koike},
  {Kurakami}, {Miyama}, {Morokuma}, {Nakata}, {Namikawa}, {Nakaya}, {Nariai},
  {Obuchi}, {Oishi}, {Okada}, {Okura}, {Tait}, {Takata}, {Tanaka}, {Tanaka},
  {Terai}, {Tomono}, {Uraguchi}, {Usuda}, {Utsumi}, {Yamada}, {Yamanoi},
  {Aihara}, {Fujimori}, {Mineo}, {Miyatake}, {Oguri}, {Uchida}, {Tanaka},
  {Yasuda}, {Takada}, {Murayama}, {Nishizawa}, {Sugiyama}, {Chiba}, {Futamase},
  {Wang}, {Chen}, {Ho}, {Liaw}, {Chiu}, {Ho}, {Lai}, {Lee}, {Jeng}, {Iwamura},
  {Armstrong}, {Bickerton}, {Bosch}, {Gunn}, {Lupton}, {Loomis}, {Price},
  {Smith}, {Strauss}, {Turner}, {Suzuki}, {Miyazaki}, {Muramatsu}, {Yamamoto},
  {Endo}, {Ezaki}, {Ito}, {Kawaguchi}, {Sofuku}, {Taniike}, {Akutsu}, {Dojo},
  {Kasumi}, {Matsuda}, {Imoto}, {Miwa}, {Suzuki}, {Takeshi}, \&
  {Yokota}}]{miyazaki+2018}
{Miyazaki}, S., {Komiyama}, Y., {Kawanomoto}, S., {et~al.} 2018, \pasj, 70, S1

\bibitem[{{Momcheva} {et~al.}(2006){Momcheva}, {Williams}, {Keeton}, \&
  {Zabludoff}}]{momcheva+2006}
{Momcheva}, I., {Williams}, K., {Keeton}, C., \& {Zabludoff}, A. 2006, \apj,
  641, 169

\bibitem[{{More} {et~al.}(2012){More}, {Cabanac}, {More}, {Alard}, {Limousin},
  {Kneib}, {Gavazzi}, \& {Motta}}]{more+2012}
{More}, A., {Cabanac}, R., {More}, S., {et~al.} 2012, \apj, 749, 38

\bibitem[{{Navarro} {et~al.}(1997){Navarro}, {Frenk}, \&
  {White}}]{navarro+1997}
{Navarro}, J.~F., {Frenk}, C.~S., \& {White}, S.~D.~M. 1997, \apj, 490, 493

\bibitem[{{Negrello} {et~al.}(2010){Negrello}, {Hopwood}, {De Zotti}, {Cooray},
  {Verma}, {Bock}, {Frayer}, {Gurwell}, {Omont}, {Neri}, {Dannerbauer},
  {Leeuw}, {Barton}, {Cooke}, {Kim}, {da Cunha}, {Rodighiero}, {Cox},
  {Bonfield}, {Jarvis}, {Serjeant}, {Ivison}, {Dye}, {Aretxaga}, {Hughes},
  {Ibar}, {Bertoldi}, {Valtchanov}, {Eales}, {Dunne}, {Driver}, {Auld},
  {Buttiglione}, {Cava}, {Grady}, {Clements}, {Dariush}, {Fritz}, {Hill},
  {Hornbeck}, {Kelvin}, {Lagache}, {Lopez-Caniego}, {Gonzalez-Nuevo}, {Maddox},
  {Pascale}, {Pohlen}, {Rigby}, {Robotham}, {Simpson}, {Smith}, {Temi},
  {Thompson}, {Woodgate}, {York}, {Aguirre}, {Beelen}, {Blain}, {Baker},
  {Birkinshaw}, {Blundell}, {Bradford}, {Burgarella}, {Danese}, {Dunlop},
  {Fleuren}, {Glenn}, {Harris}, {Kamenetzky}, {Lupu}, {Maddalena}, {Madore},
  {Maloney}, {Matsuhara}, {Micha{\l}owski}, {Murphy}, {Naylor}, {Nguyen},
  {Popescu}, {Rawlings}, {Rigopoulou}, {Scott}, {Scott}, {Seibert}, {Smail},
  {Tuffs}, {Vieira}, {van der Werf}, \& {Zmuidzinas}}]{negrello+2010}
{Negrello}, M., {Hopwood}, R., {De Zotti}, G., {et~al.} 2010, Science, 330, 800

\bibitem[{{Newton} {et~al.}(2009){Newton}, {Marshall}, \& {Treu}}]{newton+2009}
{Newton}, E.~R., {Marshall}, P.~J., \& {Treu}, T. 2009, \apj, 696, 1125

\bibitem[{{Oguri} {et~al.}(2005){Oguri}, {Keeton}, \& {Dalal}}]{oguri+2005}
{Oguri}, M., {Keeton}, C.~R., \& {Dalal}, N. 2005, \mnras, 364, 1451

\bibitem[{{Oguri} {et~al.}(2008){Oguri}, {Inada}, {Clocchiatti}, {Kayo},
  {Shin}, {Hennawi}, {Strauss}, {Morokuma}, {Schneider}, \&
  {York}}]{oguri+2008}
{Oguri}, M., {Inada}, N., {Clocchiatti}, A., {et~al.} 2008, \aj, 135, 520

\bibitem[{{Refsdal}(1964)}]{refsdal1964}
{Refsdal}, S. 1964, \mnras, 128, 307

\bibitem[{{Riess} {et~al.}(2016){Riess}, {Macri}, {Hoffmann}, {Scolnic},
  {Casertano}, {Filippenko}, {Tucker}, {Reid}, {Jones}, {Silverman},
  {Chornock}, {Challis}, {Yuan}, {Brown}, \& {Foley}}]{riess+2016}
{Riess}, A.~G., {Macri}, L.~M., {Hoffmann}, S.~L., {et~al.} 2016, \apj, 826, 56

\bibitem[{{Rusu} {et~al.}(2017){Rusu}, {Fassnacht}, {Sluse}, {Hilbert}, {Wong},
  {Huang}, {Suyu}, {Collett}, {Marshall}, {Treu}, \& {Koopmans}}]{rusu+2017}
{Rusu}, C.~E., {Fassnacht}, C.~D., {Sluse}, D., {et~al.} 2017, \mnras, 467,
  4220

\bibitem[{{Saha}(2000)}]{saha2000}
{Saha}, P. 2000, \aj, 120, 1654

\bibitem[{{Schneider} \& {Sluse}(2013)}]{schneider+2013}
{Schneider}, P., \& {Sluse}, D. 2013, \aap, 559, A37

\bibitem[{{Shu} {et~al.}(2016){Shu}, {Bolton}, {Kochanek}, {Oguri},
  {P{\'e}rez-Fournon}, {Zheng}, {Mao}, {Montero-Dorta}, {Brownstein},
  {Marques-Chaves}, \& {M{\'e}nard}}]{shu+2016}
{Shu}, Y., {Bolton}, A.~S., {Kochanek}, C.~S., {et~al.} 2016, \apj, 824, 86

\bibitem[{{Sluse} {et~al.}(2017){Sluse}, {Sonnenfeld}, {Rumbaugh}, {Rusu},
  {Fassnacht}, {Treu}, {Suyu}, {Wong}, {Auger}, {Bonvin}, {Collett}, {Courbin},
  {Hilbert}, {Koopmans}, {Marshall}, {Meylan}, {Spiniello}, \&
  {Tewes}}]{sluse+2017}
{Sluse}, D., {Sonnenfeld}, A., {Rumbaugh}, N., {et~al.} 2017, \mnras, 470, 4838

\bibitem[{{Sonnenfeld} {et~al.}(2013{\natexlab{a}}){Sonnenfeld}, {Gavazzi},
  {Suyu}, {Treu}, \& {Marshall}}]{sonnenfeld+2013}
{Sonnenfeld}, A., {Gavazzi}, R., {Suyu}, S.~H., {Treu}, T., \& {Marshall},
  P.~J. 2013{\natexlab{a}}, \apj, 777, 97

\bibitem[{{Sonnenfeld} {et~al.}(2013{\natexlab{b}}){Sonnenfeld}, {Treu},
  {Gavazzi}, {Suyu}, {Marshall}, {Auger}, \& {Nipoti}}]{sonnenfeld+2013b}
{Sonnenfeld}, A., {Treu}, T., {Gavazzi}, R., {et~al.} 2013{\natexlab{b}}, \apj,
  777, 98

\bibitem[{Sonnenfeld {et~al.}(2018)Sonnenfeld, Chan, Shu, More, Oguri, Suyu,
  Wong, Lee, Coupon, Yonehara, Bolton, Jaelani, Tanaka, Miyazaki, \&
  Komiyama}]{sonnenfeld+2018}
Sonnenfeld, A., Chan, J. H.~H., Shu, Y., {et~al.} 2018, Publications of the
  Astronomical Society of Japan, 70, S29

\bibitem[{{Suyu} {et~al.}(2017){Suyu}, {Bonvin}, {Courbin}, {Fassnacht},
  {Rusu}, {Sluse}, {Treu}, {Wong}, {Auger}, {Ding}, {Hilbert}, {Marshall},
  {Rumbaugh}, {Sonnenfeld}, {Tewes}, {Tihhonova}, {Agnello}, {Blandford},
  {Chen}, {Collett}, {Koopmans}, {Liao}, {Meylan}, \& {Spiniello}}]{suyu+2017}
{Suyu}, S.~H., {Bonvin}, V., {Courbin}, F., {et~al.} 2017, \mnras, 468, 2590

\bibitem[{{Tanaka}(2015)}]{tanaka2015}
{Tanaka}, M. 2015, \apj, 801, 20

\bibitem[{{Tanaka} {et~al.}(2016){Tanaka}, {Wong}, {More}, {Dezuka}, {Egami},
  {Oguri}, {Suyu}, {Sonnenfeld}, {Higuchi}, {Komiyama}, {Miyazaki}, {Onoue},
  {Oyamada}, \& {Utsumi}}]{tanaka+2016}
{Tanaka}, M., {Wong}, K.~C., {More}, A., {et~al.} 2016, \apjl, 826, L19

\bibitem[{Tanaka {et~al.}(2018)Tanaka, Coupon, Hsieh, Mineo, Nishizawa,
  Speagle, Furusawa, Miyazaki, \& Murayama}]{tanaka+2018}
Tanaka, M., Coupon, J., Hsieh, B.-C., {et~al.} 2018, Publications of the
  Astronomical Society of Japan, 70, S9

\bibitem[{{Treu} {et~al.}(2009){Treu}, {Gavazzi}, {Gorecki}, {Marshall},
  {Koopmans}, {Bolton}, {Moustakas}, \& {Burles}}]{treu+2009}
{Treu}, T., {Gavazzi}, R., {Gorecki}, A., {et~al.} 2009, \apj, 690, 670

\bibitem[{{Treu} {et~al.}(2006){Treu}, {Koopmans}, {Bolton}, {Burles}, \&
  {Moustakas}}]{treu+2006}
{Treu}, T., {Koopmans}, L.~V., {Bolton}, A.~S., {Burles}, S., \& {Moustakas},
  L.~A. 2006, \apj, 640, 662

\bibitem[{{Wilson} {et~al.}(2016){Wilson}, {Zabludoff}, {Ammons}, {Momcheva},
  {Williams}, \& {Keeton}}]{wilson+2016}
{Wilson}, M.~L., {Zabludoff}, A.~I., {Ammons}, S.~M., {et~al.} 2016, \apj, 833,
  194

\bibitem[{{Wilson} {et~al.}(2017){Wilson}, {Zabludoff}, {Keeton}, {Wong},
  {Williams}, {French}, \& {Momcheva}}]{wilson+2017}
{Wilson}, M.~L., {Zabludoff}, A.~I., {Keeton}, C.~R., {et~al.} 2017, \apj, 850,
  94

\bibitem[{{Wong} {et~al.}(2017{\natexlab{a}}){Wong}, {Ishida}, {Tamura},
  {Suyu}, {Oguri}, \& {Matsushita}}]{wong+2017a}
{Wong}, K.~C., {Ishida}, T., {Tamura}, Y., {et~al.} 2017{\natexlab{a}}, \apjl,
  843, L35

\bibitem[{{Wong} {et~al.}(2011){Wong}, {Keeton}, {Williams}, {Momcheva}, \&
  {Zabludoff}}]{wong+2011}
{Wong}, K.~C., {Keeton}, C.~R., {Williams}, K.~A., {Momcheva}, I.~G., \&
  {Zabludoff}, A.~I. 2011, \apj, 726, 84

\bibitem[{{Wong} {et~al.}(2017{\natexlab{b}}){Wong}, {Suyu}, {Auger}, {Bonvin},
  {Courbin}, {Fassnacht}, {Halkola}, {Rusu}, {Sluse}, {Sonnenfeld}, {Treu},
  {Collett}, {Hilbert}, {Koopmans}, {Marshall}, \& {Rumbaugh}}]{wong+2017}
{Wong}, K.~C., {Suyu}, S.~H., {Auger}, M.~W., {et~al.} 2017{\natexlab{b}},
  \mnras, 465, 4895

\bibitem[{{Xu} {et~al.}(2016){Xu}, {Sluse}, {Schneider}, {Springel},
  {Vogelsberger}, {Nelson}, \& {Hernquist}}]{xu+2016}
{Xu}, D., {Sluse}, D., {Schneider}, P., {et~al.} 2016, \mnras, 456, 739

\bibitem[{{Zehavi} {et~al.}(2005){Zehavi}, {Zheng}, {Weinberg}, {Frieman},
  {Berlind}, {Blanton}, {Scoccimarro}, {Sheth}, {Strauss}, {Kayo}, {Suto},
  {Fukugita}, {Nakamura}, {Bahcall}, {Brinkmann}, {Gunn}, {Hennessy},
  {Ivezi{\'c}}, {Knapp}, {Loveday}, {Meiksin}, {Schlegel}, {Schneider},
  {Szapudi}, {Tegmark}, {Vogeley}, {York}, \& {SDSS
  Collaboration}}]{zehavi+2005}
{Zehavi}, I., {Zheng}, Z., {Weinberg}, D.~H., {et~al.} 2005, \apj, 630, 1

\bibitem[{{Zhao} {et~al.}(2009){Zhao}, {Jing}, {Mo}, \&
  {B{\"o}rner}}]{zhao+2009}
{Zhao}, D.~H., {Jing}, Y.~P., {Mo}, H.~J., \& {B{\"o}rner}, G. 2009, \apj, 707,
  354

\end{thebibliography}
\end{document}